%% file: main.tex
\RequirePackage{snapshot}

\documentclass[twoside, 11pt, a4paper]{article}

\usepackage[utf8]{inputenc}
\usepackage{pgfplots}
\usepgfplotslibrary{groupplots}
\usepackage{epsfig}
\usepackage{subfigure}
\usepackage{caption}
\usepackage{psfrag}
\usepackage{color}
\usepackage[intlimits]{amsmath}
\usepackage{amsfonts}
\usepackage{amssymb}
\usepackage{mathrsfs}
\usepackage[explicit]{titlesec}
\usepackage[top=3.5cm, bottom=4cm, left=3.5cm]{geometry}
\usepackage{tikz}
\usepackage{natbib}
\usepackage{pgf}
\usetikzlibrary{arrows,cd}
\usepackage{fancyhdr}

\newcommand{\articleTitle}[0]{Adjoint chaos via cumulant truncation}

\renewcommand{\subsectionmark}[1]{}

\titleformat{\subsubsection}[runin]{}{\thesubsubsection\quad\caps{#1}.}{1em}{}
\titleformat{\subsection}{\filcenter}{\thesubsection\quad\emph{#1}\indent}{1em}{}
\titleformat{\section}{\filcenter\bf}{\thesection\quad#1}{1em}{}
\titleformat{name=\section,numberless}[block]{\filcenter\bf}{}{1em}{#1}

\DeclareTextFontCommand\textsfi{\usefont{OT1}{cmss}{m}{sl}}
\DeclareMathAlphabet\mathsfi            {OT1}{cmss}{m}{sl}
\DeclareTextFontCommand\textsfb{\usefont{OT1}{cmss}{bx}{n}}
\DeclareMathAlphabet\mathsfb            {OT1}{cmss}{bx}{n}
\DeclareTextFontCommand\textsfbi{\usefont{OT1}{cmss}{m}{sl}}
\DeclareMathAlphabet\mathsfbi            {OT1}{cmss}{m}{sl}

\begin{document}

\newcommand{\av}[1]{\overline{#1}}
\newcommand{\mathsbv}[1]{\boldsymbol{#1}}
\newcommand{\bfs}[1]{\textsf{\textbf{#1}}}
\newcommand{\ad}[1]{#1^{\dagger}}
\newcommand{\od}[2]{\dfrac{\mathrm{d} #1}{\mathrm{d} #2}}
\newcommand{\pd}[2]{\dfrac{\partial #1}{\partial #2}}
\newcommand{\dispbinom}[2]{\displaystyle\binom{#1}{#2}}
\newcommand{\rev}[1]{#1}

\setlength{\abovecaptionskip}{-5pt plus 3pt minus 2pt} 

\begin{center}
\vspace{-1cm}
\noindent{\Large\articleTitle}\\ \medskip
{\small John Craske$^{1}$ \\
  $^{1}$Department of Civil and Environmental Engineering, Imperial College London,\\
  London SW7 2AZ, UK \\
  (Last updated: \today)}
\end{center}

\thispagestyle{empty}

\begin{quote}
  \small We describe a simple \rev{and systematic} method for
  obtaining approximate sensitivity information from a chaotic
  dynamical system using a hierarchy of cumulant equations. The
  resulting forward and adjoint systems yield information about
  \rev{gradients of functionals} of the system and do not suffer
  from the convergence issues that are associated with the tangent
  linear representation of chaotic systems. The functionals on
  which we focus are ensemble-averaged quantities, whose dynamics
  are not necessarily chaotic; hence we analyse the system's
  statistical state dynamics, rather than individual
  trajectories. The approach is designed for extracting parameter
  sensitivity information from the detailed statistics that can be
  obtained from direct numerical simulation or experiments. We
  advocate a data-driven approach that incorporates observations
  of a system's cumulants to determine an optimal closure for a
  hierarchy of cumulants that does not require the specification
  of model parameters. Whilst the sensitivity information from the
  resulting surrogate model is approximate, the approach is
  designed to be used in the analysis of turbulence, whose number
  of degrees of freedom \rev{and complexity} currently prohibits
  the use of more accurate techniques. Here we apply the method to
  obtain functional gradients from low-dimensional representations
  of Rayleigh-B\'{e}nard convection.
\end{quote}

\section{Introduction}

Complete information about a particular solution \rev{of} an
engineering problem is often less useful than knowledge of the way
in which a small number of functionals of the solution change with
respect to input parameters. An example in fluid mechanics is the
effect that a body's shape has on the drag to which it is
subjected \citep{PirOjfm1974a, JamAjsc1988a}. Further examples can
be found in the fields of data assimilation \citep{DimFtel1986a},
uncertainty quantification \citep{CacDboo2003a}, stability
analysis \citep{LucPafm2014a, FarPsam2014a}, flow reconstruction
\citep{FouDjfm2014a} and flow optimisation more generally
\citep{LioJboo1971a}. In these situations it is natural to focus
on adjoint variables, which represent the derivative of a given
functional with respect to the problem's constraints or governing
equations.  With adjoint variables the derivative of the
functional with respect to any combination of input parameters can
be readily computed with a single dot product, alleviating the
need to run a large ensemble of simulations to obtain gradients in
different directions. For a general introduction to the theory the
reader is referred to \citet{MarGboo1995a} and
\citet{GilMftc2000a}.

Whilst adjoint analysis is well established and used successfully
in many fields, the problem of obtaining functional gradients from
chaotic dynamical systems, such as turbulence, is an open question
\citep{VisRjcp2015a}. Whether such gradients are well-defined
depends on the properties of the dynamical system. For example, if
the system is uniformly hyperbolic \citep{SmaSams1967a,
  EckJrmp1985a} then linear response theory provides the required
formula \citep{EyiGnon2004a, RueDnon2009a}. In all chaotic
systems, however, the linearised description, on which both
forward and adjoint analysis \rev{is} based, produces divergent
trajectories that make it impossible to compute accurate gradients
over large times in the conventional way \citep{LeaDtel2000a}. A
variety of different methods have been proposed to overcome this
practical difficulty. A possible approximation is to obtain an
estimation of a system's linear response by taking finite
differences \citep{RusSjfm2016a}. An approach employing adjoint
formalism is to limit the duration over which sensitivity
information is obtained \citep{VisRjcp2015a}, or to collect an
ensemble of gradients to compensate for the short time intervals
to which the adjoint equations are otherwise restricted
\citep{LeaDtel2000a, EyiGnon2004a}. In addition to the requirement
of having to obtain a potentially large ensemble, the difficulty
of the latter approach is in determining an appropriate time
interval \emph{a priori}. Consequently, probability density
functions have also received attention as a reliable source of
gradient information. \citet{ThuJqms2005a} proposed solving an
adjoint Fokker-Planck equation, which, though capable of producing
accurate derivatives, is computationally expensive and involves
approximation in the selection of stochastic forcing
terms. Related work uses ideas from the fluctuation-dissipation
theorem \citep[][]{MarUphr2008a} to determine sensitivities
\citep[e.g.][]{CooFjas2011a}, and typically relies on an
assumption about the underlying probability density function.


Recent efforts to reconcile adjoint techniques and chaotic systems
have focused on deriving sensitivities from shadow trajectories,
which are defined as remaining uniformly close to a given
trajectory of the system over time \citep{WanQjcp2013a}, and can
therefore yield meaningful sensitivity information. An improvement
of the method proposed by \citet{WanQjcp2013a}, which relied on
the calculation of Lyapunov exponents and was therefore restricted
to low-dimensional dynamical systems, is the least-squares
shadowing method proposed by \citet{WanQjcp2014a}. The
least-squares shadowing method involves solving an optimisation
problem to determine a perturbed trajectory that is closest to
\rev{the} chosen reference trajectory. Notably, the least-squares
shadowing method has been applied to the Kuramoto-Sivashinsky
equation, and yields accurate gradient information for certain
states \citep{BloPcsf2014a}.

\rev{The issue regarding divergent trajectories in tangent and
  adjoint systems can be circumvented altogether by computing
  sensitivities of unstable periodic orbits
  \citep{LasDsam2018a}. Perturbations of unstable periodic orbits,
  which behave like a skeleton around chaotic orbits \citep[see e.g.][]{AueDprl1987a},
  provide a proxy for the latter's sensitivity. In general, each
  unstable period orbit returns a different sensitivity. In
  certain cases, however, the sensitivities are closely aligned and
  exhibit a good agreement with that of the underlying chaotic
  orbit \citep{LasDsam2018a}. Principal among the challenges
  associated with this technique is the difficulty of finding
  unstable periodic orbits in chaotic systems of high dimension,
  such as turbulence at high Reynolds number \citep[see
  e.g.][]{LucDjfm2017a}.}


The need to overcome the sensitive dependence on perturbations
inherent in chaotic systems might be regarded as unnecessary, in
view of the fact that one is often interested in gradients of
\emph{ensemble-averaged} quantities. Indeed, following
\citet{HopEarm1952a} and \citet{LorEboo1967a}, it is possible to
directly simulate a system's statistics or cumulant dynamics. With
the use of the original governing equations, the cumulant
equations can be derived from a single flow functional
\citep{HopEarm1952a} and provide a direct means of understanding
the behaviour of a flow's statistics. In addition to their
evolution being slower and not necessarily chaotic, the cumulant
equations can be used to investigate statistically unsteady
problems, statistical stability and to provide an analytical means
of determining the linear response of a system
\citep{FarBarx2014a}. The evolution of a finite set of dependent
variables corresponds to an infinite hierarchy of cumulant
equations. The benefits of focusing on the evolution of statistics
are therefore offset by the requirement of finding a suitable
closure \citep[][]{RotAprs1993a}. Fortunately, heterogeneous flows
that are dominated by the interaction of eddies with a mean shear
are amenable to relatively simple closures, because the evolution
of third-order cumulants, describing eddy-eddy interactions, can
sometimes be neglected \citep{FarBarx2014a}. Statistical state
dynamics, or direct statistical simulation \citep{TobSast2011a,
  ChaFnjp2016a} has therefore been applied with success in
simulations of planetary jets \citep{MarJjas2008a, TobSprl2013a}
and wall-bounded shear-flow \citep{FarBjfm2016a}. Whilst strongly
nonlinear systems, such as the model for Rayleigh-B\'{e}nard
convection given by the Lorenz equations \citep{LorEjas1963a},
require a more sophisticated treatment that accounts for the role
of cumulants beyond second order, direct statistical simulation
can nevertheless produce accurate predictions
\citep{AllApre2016a}.

The approach that we describe combines the desirable features of
the statistical state equations with observations from direct
simulation and classical adjoint techniques. In \S\ref{sec:prob}
we describe the problems associated with the adjoint analysis of
chaotic systems, before deriving a well-conditioned adjoint
operator from a system's cumulant equations in
\S\ref{sec:cumulant}. In \S\ref{sec:res} we apply the approach to
the sensitivity analysis of thermal convection via the Lorenz
equations, and consider their extension to a 9-dimensional phase
space \rev{in \S\ref{sec:rd9}}. Conclusions and suggestions for
further work are made in \S\ref{sec:con}.

\section{The problem}

\label{sec:prob}

Consider a dynamical system whose state, \rev{$\mathsbv{Q}(t)$}, evolves according to

\begin{equation}
\od{\mathsbv{Q}}{t}=\mathsbv{F}(\mathsbv{Q},\mathsbv{m}),
\label{eq:ode}
\end{equation}

\noindent where $\mathsbv{m}$ is a vector of system parameters. If
the dynamical system \eqref{eq:ode} is chaotic then an
understanding of the system's statistics becomes
crucial. \rev{Fortunately, engineers and scientists are typically
  interested in a small subset of the possible statistics that can
  be obtained from \eqref{eq:ode}. Unfortunately, they typically
  wish to understand how sensitively such statistics depend on
  \emph{each} element of the parameter vector $\mathsbv{m}$}.

We focus our attention on the Lorenz equations as a specific
example. \citet{LorEjas1963a} derived the following system of
equations from a truncated description of Rayleigh-B\'{e}nard
convection between hot and cold horizontal surfaces:

\begin{equation}
\od{X}{t} = s(Y-X),\quad\quad
\od{Y}{t} = rX-Y-XZ,\quad\quad
\od{Z}{t} = XY - bZ.
\label{eq:lorenz}
\end{equation}

\noindent The state $\mathsbv{Q}=(X,Y,Z)$ describes the strength
of the velocity field, the difference in temperature between
ascending and descending fluid, and the strength of the
horizontally averaged temperature with respect to a state of pure
conduction, respectively. The parameters $\mathsbv{m}=(s,b,r)$ are
the Prandtl number, the aspect ratio of the resulting convection
rolls and the Rayleigh number, respectively, the latter normalised
with respect to a critical Rayleigh number.

\begin{figure}
  \setlength{\abovecaptionskip}{-10pt plus 3pt minus 2pt}
\begin{center}
  \input{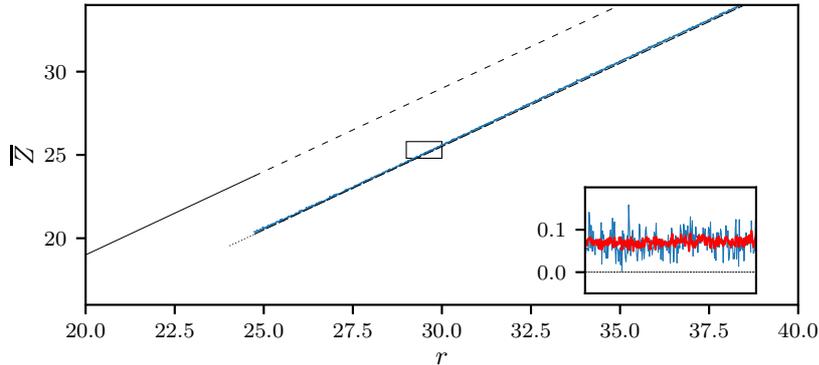}
  \end{center}
  \caption{The dependence of the statistic $J=\av{Z}$, \rev{as
      defined in \eqref{eq:J}}, on the renormalised Rayleigh
    number $r$ from a solution of the Lorenz equations
    \eqref{eq:lorenz} with $(s,b)=(10, 8/3)$. The results were
    obtained from integration \rev{along a statistically stationary}
    trajectory for \rev{$\tau=1000$} time units (blue/dark lines)
    and \rev{$\tau =10000$} time units (red/light lines). The thin
    solid black line corresponds to the location $Z=r-1$ of the
    stable fixed points of the system for $r<24.7$ \rev{and to an
      upper bound of $\av{Z}$ for $r\geq 24.7$}. \rev{The dashed black
    straight line $\av{Z}=r-4.50$ is provided for reference}. The
    dependence \rev{$\av{Z}\propto r$} has been subtracted from
    the data that are displayed in the inset window.}
  \label{fig:01}
\end{figure}

\rev{We will focus on statistics $\mathsbv{J}$ that correspond to a finite time
  average $\av{\mathsbv{L}}$ of a function $\mathsbv{L}(\mathsbv{Q})$:
\begin{equation}
 \mathsbv{J}[\mathsbv{Q}]\equiv \av{\mathsbv{L}}=\frac{1}{\tau}\int_{0}^{\tau}\mathsbv{L}(\mathsbv{Q})\mathrm{d}t,
\label{eq:J}
\end{equation}
}
\rev{
\noindent which depends implicitly on the parameters $\mathsbv{m}$
via $\mathsbv{Q}(t)$.} \rev{Under the assumption of ergodicity,
the estimator $J$ \rev{using} the finite time average in \eqref{eq:J} 
corresponds to a phase-average of the function $L$ when $\tau\rightarrow\infty$.}
\rev{Following previous work on the sensitivity analysis of
  chaotic systems \citep{LeaDtel2000a, WanQjcp2013a}, we will
  focus on \rev{$L\equiv Z$, such that $J\equiv\av{Z}$} estimates
  the average amplitude of the horizontally-averaged temperature
  fluctuations. Figure \ref{fig:01} displays $\av{Z}$ computed
  numerically from simulations of \eqref{eq:lorenz} with $s=10$
  and $b=8/3$. In spite of the chaotic dynamics described by
  equation \eqref{eq:lorenz}, $\av{Z}$ appears to vary linearly
  with respect to $r>r_{c}$, for the values of $r$ considered,
  where $r_{c}\approx 24.74$ is a critical value of $r$. \rev{At
    $r=r_{c}$} the two stable fixed points of the system, for the
  given values of $s$ and $b$, become unstable. The value $r_{c}$
  marks the threshold of sustained chaotic behaviour, on which we
  focus, for almost all initial conditions, in contrast to the
  transient chaos that can be observed on an unstable chaotic set
  for $13.93<r<24.06$ \citep{YorJjsp1979a}. The oscillations in
  $\av{Z}$ in figure \ref{fig:01} are due to the fact that
  $\av{Z}$ is an estimator obtained from a finite time
  interval. Indeed, comparison of the statistic obtained over
  $\tau=1000$ with that obtained over $\tau=10000$ in figure
  \ref{fig:01}, indicates that the oscillations reduce in
  amplitude as the length of the time interval increases}.

\subsection{A finite difference approach}

\noindent An estimation of the G\^{a}teaux derivative of a single functional $J_{j}$ with respect to the $i$th component $m_{i}$ of $\mathsbv{m}$ is
\rev{
\begin{equation}
  \pd{J_{j}}{m_{i}} \approx \dfrac{\delta J_{j}}{\delta m_{i}}\equiv \dfrac{ J_{j}[\mathsbv{Q}(t|\mathsbv{m}+\epsilon\mathsbv{e}_{i})]-J_{j}[\mathsbv{Q}(t|\mathsbv{m})]}{\epsilon},
\label{eq:fd}
\end{equation}
}
\noindent \rev{in which all elements of $\mathsbv{e}_{i}$ are
  equal to zero, with the exception of the $i$th element, which is
  equal to $1$.} If $\epsilon$ is relatively large, then
$\delta J_{j}/\delta m_{i}$ will not provide an accurate
approximation to the local derivative. If, on the other hand,
$\epsilon$ is relatively \emph{small}, the non-smooth behaviour of
$J_{j}$ for finite-time averages evident in figure \ref{fig:01}
suggests that we would need to obtain statistics over a
correspondingly \emph{large} time to obtain meaningful results
\citep[see e.g.][]{RusSjfm2016a}. \rev{Moreover}, the use of such
an approach to obtain the sensitivity of $J_{j}$ with respect to
other system parameters requires \rev{the entire simulation to be
  run at least twice for each parameter $m_{i}$}.

\subsection{The tangent linear equations}

\rev{Instead of by looking at finite differences between independent
simulations}, a functional's gradient can, in theory, be calculated
\rev{exactly} using information from a single simulation. We will
outline a naive version of the method before explaining \rev{its
  problems} in the case of chaotic systems. Assuming that it is
well-defined, the derivative of a given functional $J_{j}$ with
respect to a given component $m_{i}$ of $\mathsbv{m}$ for a known
trajectory $\mathsbv{Q}(t)$ can be evaluated as

%
\rev{
\begin{equation}
  \pd{J_{j}}{m_{i}}= \frac{1}{\tau}\int_{0}^{\tau}\pd{L_{j}}{\mathsbv{Q}}\cdot\mathsbv{q}_{i}\,\mathrm{d}t, \label{eq:dJdm}
\end{equation}
}
\rev{
\noindent where
$\mathsbv{q}_{i}\equiv \mathrm{d}\mathsbv{Q}/\mathrm{d}m_{i}$. By
augmenting the time-dependent functions with boundary values:

\renewcommand{\arraystretch}{1.2}

\begin{equation}
  \tilde{\mathsbv{q}}_{i}\equiv
  \begin{bmatrix}
    \mathsbv{q}_{i}(t) \\
    \mathsbv{q}_{i}(\tau)
  \end{bmatrix},\quad\quad
    \tilde{\mathsbv{g}}_{j}\equiv
  \begin{bmatrix}
    \pd{L_{j}}{\mathsbv{Q}}(t) \\
 \mathsbv{0}
  \end{bmatrix},
  \label{eq:Telem}
\end{equation}

\noindent and defining an inner product $(\cdot,\cdot)$, based on
\eqref{eq:dJdm}, for the space to which the elements
$\tilde{\mathsbv{q}}_{i}$ and $\tilde{\mathsbv{g}}_{j}$ belong,
the functional's derivative can be expressed as \citep[see
e.g.][]{SewMboo1987a}

\begin{equation}
  \pd{J_{j}}{m_{i}}=\left(\tilde{\mathsbv{g}_{j}}, \tilde{\mathsbv{q}_{i}}\right).
\label{eq:dJdmT}
\end{equation}
}
\noindent In general, a penalty term that depends on the
  state trajectory's end point $\mathsbv{Q}(\tau)$ can be added to
  \eqref{eq:J}, which would modify the $\mathsbv{0}$ that appears in \eqref{eq:Telem}
for $\tilde{\mathsbv{g}}_{j}$.  The perturbation $\mathsbv{q}_{i}$
  satisfies the tangent linear equations, which are obtained by
  differentiating $\mathsbv{F}$ with respect to $m_{i}$:

\begin{equation}
  \mathsfbi{T}\tilde{\mathsbv{q}}_{i}\equiv
  \begin{bmatrix}
    \od{\mathsbv{q}_{i}}{t}-\left(\pd{\mathsbv{F}}{\mathsbv{Q}}\right)\mathsbv{q}_{i}\\
\mathsbv{q}_{i}(0)    
  \end{bmatrix}
=
  \begin{bmatrix}
   \pd{\mathsbv{F}}{m_{i}} \\
    \mathsbv{0}
  \end{bmatrix}\equiv \tilde{\mathsbv{f}}_{i},
  \label{eq:tang}
\end{equation}

\noindent where $\mathsbv{q}_{i}(0)$ sets the perturbation of
  the initial condition to be zero. For the Lorenz equations the
tangent linear system for $m_{i}=r$ is equivalent to

\begin{equation}
  \od{}{t}
  \begin{pmatrix}
    x \\
    y \\
    z \\
  \end{pmatrix}
  =
  \begin{pmatrix}
    -s & s & 0 \\
    r-Z & -1 & -X \\
    Y & X & -b \\
  \end{pmatrix}
    \begin{pmatrix}
    x \\
    y \\
    z \\
  \end{pmatrix}
  +
    \begin{pmatrix}
    0 \\
    X \\
    0 \\
  \end{pmatrix}
,
\label{eq:tang_lorenz}
\end{equation}
\rev{
  \noindent in addition to the initial condition
  $x(0)=y(0)=z(0)=0$. Note that the $X$, $Y$ and $Z$ appearing in
  \eqref{eq:tang_lorenz} are known, albeit time-dependent,
  variables.} Once the perturbed trajectory
  $\mathsbv{q}_{i}\equiv (x,y,z)$ is known, the derivative
  of \emph{any} functional $J_{j}$ with respect to $m_{i}$ can in
  principle be calculated using \eqref{eq:dJdmT} by changing
  $\mathsbv{g}_{j}$. Calculation of the derivative of a given $J$
  with respect to a different parameter $m$ is more difficult,
  because it requires us to find a different perturbed trajectory
  from \eqref{eq:tang} for use in \eqref{eq:dJdm}. This motivates
  an alternative way of factorising \eqref{eq:dJdm}, to obtain
  adjoint variables that describe the change in $J$ with respect
  to a change in the constraints $\mathsbv{F}$.

\subsection{The adjoint equations}
\label{sec:adj}

Introducing the adjoint variables $\mathsbv{p}_{j}$ to enforce the
the equations of motion $\mathsbv{F}$, \rev{which act as
  constraints}, results in
\rev{
\begin{equation}
  \pd{J_{j}}{m_{i}}=\left(\tilde{\mathsbv{g}}_{j}, \tilde{\mathsbv{q}}_{i}\right)+\langle\tilde{\mathsbv{p}}_{j}, \tilde{\mathsbv{f}}_{i}-\mathsfbi{T}\tilde{\mathsbv{q}}_{i}\rangle,
\label{eq:dJdm2}
\end{equation}
}
\rev{
\noindent where $\langle\cdot,\cdot\rangle$ is an inner product for the dual space containing the elements

\begin{equation}
  \tilde{\mathsbv{p}}_{j}=
  \begin{bmatrix}
    \mathsbv{p}_{j}(t) \\
    \mathsbv{p}_{j}(0)
  \end{bmatrix},\quad\quad
    \tilde{\mathsbv{f}}_{i}=
  \begin{bmatrix}
    \pd{\mathsbv{F}}{m_{i}}(t) \\
 \mathsbv{0}
  \end{bmatrix},
  \label{eq:Delem}
\end{equation}
\noindent If $\tilde{\mathsbv{p}}_{j}$ satisfies the adjoint equations:
\begin{equation}
  \mathsfbi{T}^{\dagger}\tilde{\mathsbv{p}}_{j}\equiv
  \begin{bmatrix}
    -\od{\mathsbv{p}_{j}}{t}-\left(\pd{\mathsbv{F}}{\mathsbv{Q}}\right)^{\dagger}\mathsbv{p}_{j}\\
\mathsbv{p}_{j}(\tau)    
  \end{bmatrix}
  =
    \begin{bmatrix}
    \pd{L_{j}}{\mathsbv{Q}}(t) \\
 \mathsbv{0}
  \end{bmatrix}\equiv \tilde{\mathsbv{g}}_{j},
  \label{eq:adjoint}
\end{equation}

\noindent where $\dagger$ denotes the adjoint/transpose of an
operator, then integration by parts of \eqref{eq:dJdm2} results in
\begin{equation}
  \pd{J_{j}}{m_{i}}=\langle\tilde{\mathsbv{p}}_{j}, \tilde{\mathsbv{f}_{i}}\rangle.
\label{eq:dJdmD}
\end{equation}
}
\noindent \rev{Note that the penalty term $\mathsbv{0}$ in
  $\tilde{\mathsbv{g}}_{j}$ corresponds to the initial condition
  $\mathsbv{p}_{j}(\tau)$, and that the penalty term on
  $\mathsbv{p}_{j}(0)$ in \eqref{eq:dJdmD} corresponds to the
  initial condition $\mathsbv{q}_{i}(0)$. Since $\mathsbv{p}_{j}$
  is equal to the derivative of $J_{j}$ with respect to a change
  in the constraints, equation \eqref{eq:dJdmD} enables us to
  readily compute the sensitivity of a given functional $J_{j}$
  with respect to \emph{any} parameter $m_{i}$. To appreciate
  this, observe that \eqref{eq:dJdmD} contains $i$, whereas
  \eqref{eq:adjoint} does not, and that the converse statement is
  true for \eqref{eq:dJdmT} and \eqref{eq:tang}.} On the other
hand, calculation of the the sensitivity of a different functional
is difficult using the adjoint approach, because it would require
a new solution of \eqref{eq:adjoint}, which, like obtaining a
solution to \eqref{eq:tang} is computationally demanding in
comparison with the evaluation of \eqref{eq:dJdmT} or
\eqref{eq:dJdmD}.

\rev{Solutions to both the tangent system \eqref{eq:tang} and the
  adjoint system \eqref{eq:adjoint} grow without bound as the time
  $\tau$ in \eqref{eq:J} increases.} As pointed out by
\citet{ThuJqms2005a}, the cause of the difference between the
actual gradient and a gradient obtained from \rev{either} the
tangent or adjoint system is the fact that the operation of time
averaging over $\tau\rightarrow\infty$ does not, in general,
commute with the finite difference of an infinite time average
over an interval of $\epsilon\rightarrow 0$:

\begin{equation}
  \pd{J}{m_{i}}=\lim_{\epsilon\rightarrow 0} \lim_{\tau\rightarrow\infty} \frac{\delta J}{\delta m_{i}} \neq  \lim_{\tau\rightarrow\infty} \lim_{\epsilon\rightarrow 0} \frac{\delta J}{\delta m_{i}}.
\label{eq:comm}
\end{equation}

\noindent \rev{The finite difference of the functional $J_{j}$
  does not converge uniformly to the sought-after derivative for
  all integration times $\tau$ and, therefore, neither do
  \eqref{eq:dJdmT} nor \eqref{eq:dJdmD}.}

An approximation to $\partial_{m_{i}}J_{j}$ can be obtained if
\eqref{eq:adjoint} is integrated over relatively short time intervals
\citep{LeaDtel2000a}. On the other hand, if finite differences are
employed using equation \eqref{eq:fd}, then the minimal time $\tau$
over which accurate statistics can be obtained is determined by
$\epsilon \ll 1$. An accurate finite difference approximation
requires a small value of $\epsilon$, which requires a large value
of $\tau$ \citep[see e.g.][]{RusSjfm2016a}. Thus, approximate gradients
can be obtained by using the tangent linear equations or finite
differences, provided that small or large time intervals are used,
respectively.

That it is crucial to take the limit $\tau \rightarrow \infty$
\emph{before} analysing derivatives suggests that sensitivity
analysis of the equations governing the statistics of the process
might result in a more tractable problem.  At the expense of
introducing additional unknowns, we therefore focus on obtaining
adjoint information for the equations satisfied by the system's
cumulants.

\section{The cumulant equations and their closure}

\label{sec:cumulant}

The equations that govern the behaviour of cumulants provide a
means of establishing the leading-order \rev{relationships between
  the statistics of a chaotic attractor. These relationships
  constrain the response of statistics to changes in parameters.}
The cumulants and their dynamics have a natural hierarchy and can
be derived in a systematic way from the equations that govern
individual trajectories.

\subsection{The cumulant generating functional}
\label{sec:cumgen}

The cumulants $\mathsbv{U}$ of a dynamical system can be defined
in terms of a cumulant generating functional $\mathrm{log}\av{\psi}$:

\begin{equation}
  U_{\alpha_{1}\alpha_{2}\ldots\alpha_{d}} = (-\imath)^{|\alpha|}\left.\pd{^{\alpha}}{\mathsbv{P}^{\alpha}}\log\,\av{\psi}\right|_{\mathsbv{P}=0},
\label{eq:cumgen}
\end{equation}

\noindent where $\alpha=(\alpha_{1},\alpha_{2}\ldots,\alpha_{d})$ is a multi-index
for the system of $d$ \rev{time-dependent variables} $\mathsbv{Q}(t)$, and

\rev{
\begin{equation}
  \psi\equiv \mathrm{exp}\left( \imath\,P_{i}Q_{i}(t) \right),
\label{eq:cumgen2}
\end{equation}
}

\noindent where $\psi$ is the Hopf generating \rev{functional}
\citep[see e.g.][]{HopEarm1952a, FriUboo1995a} and
\rev{$\imath=\sqrt{-1}$}. \rev{The over-bar in \eqref{eq:cumgen}
  denotes the finite time average defined in \eqref{eq:J}, which
  we assume converges to a phase average when the duration the
  averaging interval $\tau\rightarrow\infty$.} Due to the
  logarithm in \eqref{eq:cumgen}, a cumulant, unlike a moment,
  derived from the sum of two independent random variables, is
  equal to the sum of their respective cumulants. This
  commutativity is related to the fact that cumulants isolate the
  interdependence of random variables without including the
  effects of correlations between statistics of lower order. For
  example, according to \eqref{eq:cumgen} and \eqref{eq:cumgen2},
  \begin{equation}
    \av{Z^{3}}=U_{003}+3U_{002}U_{001}+U_{001}^{3},
  \end{equation}
  \noindent where the coefficients of the three terms on the
  right-hand side correspond to the number of ways of partitioning
  a multiset of three (identical) elements into (a) a single
  multiset of three; (b) a multiset of two and a set of one; (c)
  three sets of one. In this respect, cumulants are the atoms of
  which moments are comprised, and therefore have simpler
  algebraic properties than the latter. Further examples of the
  decomposition of moments into cumulants include
  $\av{XZ}=U_{101}+U_{100}U_{001}$, and

  \begin{equation}
  \begin{aligned}
      \av{XYZ^{2}}=&\overbrace{U_{112}}^{\text{4th order}}+\overbrace{2U_{111}U_{001}+U_{100}U_{012}+U_{010}U_{102}}^{\text{3rd and 1st order}}+\overbrace{U_{110}U_{002}+2U_{101}U_{011}}^{\text{2nd order}}\\
            &+\underbrace{U_{100}U_{010}U_{002} +2U_{100}U_{001}U_{011}+2U_{010}U_{001}U_{101}}_{\text{2nd and 1st order}}+\underbrace{U_{100}U_{010}U_{001}^{2}}_{\text{1st order}},
            \label{eq:cumulants_eg}
          \end{aligned}
  \end{equation}

\noindent in which the grouped terms correspond to a summation over the
  different ways that the multiset of four elements
  $\{X,Y,Z,Z\}$ can be partitioned into subsets of a given
  cardinality. The connection between cumulants and moments is discussed in
  more detail in appendix \ref{sec:app1}, in which it is helpful to compare \eqref{eq:cumulants_eg} with \eqref{eq:cumulants_eg2}.

Noting from \eqref{eq:cumgen2} that $\mathsbv{Q}$
plays the role of $-\imath \partial_{\mathsbv{P}}$, the Hopf
function $\psi$ satisfies the \emph{linear} equation

\rev{
\begin{equation}
  \imath \pd{\psi}{t} = -P_{i}F_{i}\left(-\imath\pd{}{\mathsbv{P}}, \mathsbv{m}\right)\psi.
\label{eq:dpsidt}
\end{equation}

\noindent The original $d$ nonlinear equations $\mathsbv{F}$ from
\eqref{eq:ode}} are recovered by differentiating \eqref{eq:dpsidt}
with respect to the vector \rev{$\mathsbv{P}$}. Associated with the
original system \eqref{eq:ode} are an infinite hierarchy of
cumulant equations,
\rev{
\begin{equation}
\mathsbv{H}(\mathsbv{U},\mathsbv{m})=0,
  \label{eq:cdyn}
\end{equation}
}
\noindent which are obtained, \rev{under the assumption of
  ergodicity}, by averaging \eqref{eq:dpsidt} to obtain a
\rev{stationary} equation for $\av{\psi}$, in which the equation
for a given cumulant $U_{\alpha}$ corresponds to the coefficient
of \rev{$\mathsbv{P}^{\alpha}$}. Readers are referred to
\citet{FriUboo1995a} for further details.

  Despite the fact that they do not form a closed system, the
cumulant equations provide useful information. For example, as
noted by \citet{KnoEjsp1979a}, the cumulant equations for the
Lorenz system indicate that

\begin{equation}
 \av{Z}=\frac{\av{X^{2}}}{b}=r-1 -\frac{1}{s^{2}\av{Z}}\od{\av{X^{2}}}{t}-\frac{\av{Z^{2}}-\av{Z}^{2}}{\av{Z}},
\label{eq:example1}
\end{equation}

\noindent which, since $\av{Z^{2}}\geq\av{Z}^{2}$, implies that
$0\leq \av{Z}\leq r-1$ in a statistically steady state. 

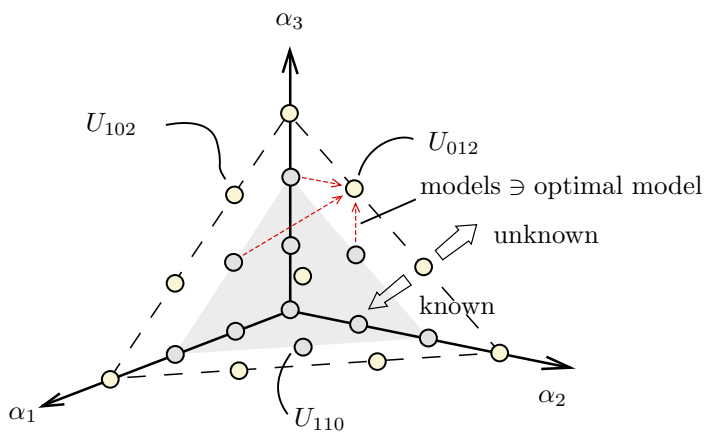
\begin{figure}
  {\small
  \begin{center}
  \input{diag04}
\end{center}
}
\caption{Cumulant space for $d=3$ indexed with
  $\alpha=(\alpha_{1},\alpha_{2},\alpha_{3})$ and truncation at order
  $N=2$. For systems, such as \eqref{eq:lorenz}, with quadratic
  nonlinearities, a model is the specification of the dependence
  of cumulants of order $N+1$ on those of lower order.}
  \label{diag:02}
\end{figure}

If the original system evolves on a $d$-dimensional phase space
then, ignoring symmetries in the governing equations, the number
of cumulants at order $j=|\alpha|$ is equal to the number of ways that $j$
indistinguishable objects can be assigned to $d$ sets; hence the
number of cumulants up to and including those of order $N$ is
\begin{equation}
  \sum_{j=1}^{N}\binom{j+d-1}{d-1},
  \label{eq:binom}
\end{equation}
\noindent as illustrated in figure \ref{diag:02}$(a)$. \rev{Known
  symmetries of a system reduce the number of independent unknown
  cumulants. In the case of the Lorenz equations
  \eqref{eq:lorenz}, for which $\mathsbv{F}$ is invariant under
  the mapping $(X,Y)\mapsto (-X,-Y)$, a cumulant
  $U_{\alpha_{1}\alpha_{2}\alpha_{3}}$ for which
  $\alpha_{1}+\alpha_{2}$ is odd, is necessarily equal to zero. In
  the case of $d=3$, of the $34$ and $55$ available cumulants up
  to order $N=4$ and $N=5$, $18$ and $27$, respectively, are non
  zero.}  Closures of the cumulant hierarchy aim to strike a
balance between the incorporation of \rev{additional} physics
\rev{from} nonlinear interactions and \rev{keeping the number of
  unknowns small}.



\subsection{Adjoint cumulant dynamics}

As described in \S\ref{sec:prob}, if one wishes to differentiate a
vector-valued function with respect to vector-valued input one can
employ one of two dual approaches. Each approach uses the chain
rule: one forwards and the other backwards.  The first approach is to
propagate derivatives with respect to a single input parameter
`up' the computational graph, to find the derivatives of all
output variables. The second approach is to propagate derivatives
of a single output variable `down' the computational graph with
respect to all input variables. The efficiency of the two methods
depends on the number of input parameters compared with the number
of output variables.

\rev{Regarding $\mathsbv{J}[\mathsbv{U}(\mathsbv{m})]$, originally defined in
\eqref{eq:J}, as a functional of an infinite hierarchy of
cumulants, }

\begin{equation}
  \pd{\mathsbv{J}}{\mathsbv{m}}=-\rlap{$\underbrace{\phantom{\pd{\mathsbv{J}}{\mathsbv{U}}\left(\pd{\mathsbv{H}}{\mathsbv{U}}\right)^{-1}}}_{\mathsbv{v}}$}
      \pd{\mathsbv{J}}{\mathsbv{U}}\overbrace{\left(\pd{\mathsbv{H}}{\mathsbv{U}}\right)^{-1}\pd{\mathsbv{H}}{\mathsbv{m}}}^{-\mathsbv{u}},
      \label{eq:dJdm4}
    \end{equation}

    \noindent where the cumulant perturbation $\mathsbv{u}$ and
    the corresponding adjoint variables $\mathsbv{v}$ are defined
    according to

\begin{equation}
  \mathsbv{u}\equiv \pd{\mathsbv{U}}{\mathsbv{m}} = -\left(\pd{\mathsbv{H}}{\mathsbv{U}}\right)^{-1}\pd{\mathsbv{H}}{\mathsbv{m}},\quad\quad
  \mathsbv{v}\equiv \pd{\mathsbv{J}}{\mathsbv{H}} = \pd{\mathsbv{J}}{\mathsbv{U}}\left(\pd{\mathsbv{H}}{\mathsbv{U}}\right)^{-1}.
\end{equation}

\noindent As discussed in \S\ref{sec:prob}, if the problem
\rev{involves} more functionals than parameters, it is
computationally preferable to solve the tangent linear system to
find $\mathsbv{u}$ before evaluating \eqref{eq:dJdm4}. If, on the
other hand, the problem contains more parameters than functionals,
then it is computationally preferable to find the adjoint
variables $\mathsbv{v}$ before evaluating \eqref{eq:dJdm4}. For a
given functional $J_{j}$ and a given parameter $m_{i}$, the two
alternatives can be expressed as

\begin{equation}
  \pd{J_{j}}{m_{i}}=
  \begin{cases}
    (\mathsbv{g}_{j},\mathsbv{u}_{i})\ \ &\mathrm{s.t.}\ \ \mathsfbi{T}\mathsbv{u}_{i}=\mathsbv{f}_{i},\\
    \langle \mathsbv{v}_{j},\mathsbv{f}_{i}\rangle\ \ &\mathrm{s.t.}\ \ \mathsfbi{T}^{\dagger}\mathsbv{v}_{j}=\mathsbv{g}_{j},
  \end{cases}
\end{equation}

\noindent where

\begin{equation}
\mathsfbi{T} = \pd{\mathsbv{H}}{\mathsbv{U}},\quad\quad
\mathsbv{f}_{i} = -\pd{\mathsbv{H}}{m_{i}},\quad\quad
\mathsbv{g}_{j} = \pd{J_{j}}{\mathsbv{U}},
\end{equation}

\noindent We focus on the adjoint problem of determining the
sensitivity of a single functional \rev{$J$ (we omit the subscript
  $j$ hereafter)} with respect to a potentially large number of
unknown parameters. Unlike the systems \eqref{eq:tang} and
\eqref{eq:adjoint}, whose validity relies on the commutation of
time averaging and differentiation with respect to $m_{i}$,
\eqref{eq:dJdm4} works with time averaged variables directly.

\subsection{Building the cumulant operator}
\label{sec:build}

\noindent If the original system of equations \eqref{eq:ode}
contains nonlinear terms then the equations for the cumulants of
order $j$ will depend on cumulants of order $j+1$ and higher,
depending on the degree of nonlinearity. For the Lorenz equations
\eqref{eq:lorenz}, and indeed the quadratic equations governing
fluid mechanics more generally, cumulants of order $j$ do not have
a dependence on cumulants whose order is higher than $j+1$. It is
nevertheless necessary to close the problem, as illustrated by the
shape of the tangent linear operator:

\begin{equation}
  \begin{bmatrix}
    \mathsfbi{T}^{(11)} & \mathsfbi{T}^{(12)} & 0 & 0 & \ldots \\
    \mathsfbi{T}^{(21)} & \mathsfbi{T}^{(22)} & \mathsfbi{T}^{(23)} & 0 & \ldots \\
     \vdots & \vdots   & \vdots    & & \\
  \end{bmatrix}
  \begin{pmatrix}
    \mathsbv{u}_{i}^{(1)} \\
    \mathsbv{u}_{i}^{(2)} \\
    \mathsbv{u}_{i}^{(3)} \\
    \vdots
  \end{pmatrix}
  =
  \mathsbv{f}.
\end{equation}

\noindent Here $\mathsbv{u}_{i}^{(j)}$ represents perturbations
$\partial_{m_{i}}{U}^{(j)}$ to the cumulants of order
$j=|\alpha|$.  \rev{According to \eqref{eq:binom},} each operator
$\mathsfbi{T}^{(ij)}$ has \rev{the shape}

\begin{equation}
  \mathrm{shape}(\mathsfbi{T}^{(ij)}) = \binom{i+d-1}{d-1}\times\binom{j+d-1}{d-1}.
\end{equation}

\noindent For the Lorenz equations,

\begin{equation}
  [\mathsfbi{T}^{(11)}, \mathsfbi{T}^{(12)}] =
  \begin{bmatrix}
    s & -s & 0 & 0 & 0 & 0 & 0 & 0 & 0 \\
    r-\av{Z} & -1 & -\av{X} & 0 & 0 & -1 &0  &0 &  0 \\
    \av{Y} & \av{X} & b & 0 & 1 & 0 & 0  & 0  & 0
  \end{bmatrix}.
  \label{eq:cum1}
\end{equation}

Whilst the tangent linear system is \emph{under determined}, the
adjoint system $\ad{\mathsfbi{T}}\mathsbv{v}=\mathsbv{f}$ is
\emph{over determined}. The overall properties of the system can be
be seen in the self-adjoint problem that combines the tangent
linear and adjoint operators. With the equations for perturbations
to the first order cumulants (e.g. $\av{X}$, $\av{Y}$ and
$\av{Z}$), one finds
\begin{equation}
  \begin{bmatrix}
0 & \mathsfbi{T}^{(11)} & \mathsfbi{T}^{(12)} \\
\mathsfbi{T}^{(11)\dagger} & 0 & 0 \\
\mathsfbi{T}^{(12)\dagger} & 0 & 0 \\
  \end{bmatrix}
\begin{pmatrix}
  \mathsbv{v}^{(1)} \\
  \mathsbv{u}_{i}^{(1)} \\
  \mathsbv{u}_{i}^{(2)}
\end{pmatrix}
=
\begin{pmatrix}
  \mathsbv{f}_{i}^{(1)} \\
  \mathsbv{g}^{(1)} \\
  \mathsbv{g}^{(2)} 
\end{pmatrix}.
\label{eq:extended}
\end{equation}
\noindent If the functional $J$, and therefore the vector
$\partial_{\mathsbv{U}^{(1)}}J\equiv \mathsbv{g}^{(1)}$, is
specified then one can solve for the adjoint variables
$\mathsbv{v}^{(1)}$ according to the second row of
\eqref{eq:extended}. However, a consistency requirement for the
extended system \eqref{eq:extended} to possess a solution is that
$\mathsbv{g}^{(2)}=\mathsfbi{T}^{(12)\dagger}\mathsbv{v}^{(1)}\neq
0$, in general.  We are therefore not at liberty to choose the
functional $J$ arbitrarily, because it will automatically contain
a contribution scaled by $\mathsbv{g}^{(2)}$ from \rev{the}
unclosed \rev{perturbations $\mathsbv{u}^{2}_{i}$}.

\setlength{\abovecaptionskip}{5pt plus 2pt minus 3pt}
\begin{figure}
  \begin{center}
  \input{diag03}
\  \caption{$(a)$ Modification of the functional due to the inner
    product of unclosed cumulants $\mathsbv{u}^{(N+1)}$ and the
    weighting factor $\mathsbv{g}^{(N+1)}$. $(b)$ Local observation
    of \rev{the} underlying functional $J[\mathsbv{U}(\mathsbv{m})]$ (red
    circle) and gradient $\nabla J_{*}$ from the optimal model as
    an approximation of the underlying exact gradient $\nabla J$.}
\label{diag:03}
\end{center}
\end{figure}

The vacuous consequence of using \eqref{eq:extended} is that only
functionals whose value can be determined identically from the
original cumulant equations, such as equation \eqref{eq:example1},
can be determined exactly. For \eqref{eq:extended} to yield novel
information an assumption is required about the response of the
unknown cumulant perturbations $\mathsbv{u}^{(2)}_{i}$. \rev{The
  simplest, albeit naive, approach} is to assume that
$( \mathsbv{g}^{(2)}, \mathsbv{u}_{i}^{(2)})=0$, which corresponds
to the unknown high-order perturbations $\mathsbv{u}^{(2)}_{i}$
being either zero or orthogonal to the weighting vector
$\mathsbv{g}^{(2)}$, as illustrated in figure \ref{diag:03}$(a)$.
\rev{More generally, taking the system \eqref{eq:extended} as an
  example, a closure corresponds to the specification of
  $( \mathsbv{g}^{(2)}, \mathsbv{u}_{i}^{(2)})$ in terms of the
  the retained cumulant sensitivities $\mathsbv{u}_{i}^{(1)}$.  If
  $\mathsbv{m}$ belongs to a three-dimensional parameter space,
  then truncation at order $N$ entails three assumptions,
  determining $(\mathsbv{g}^{(N+1)}, \mathsbv{u}_{i}^{(N+1)})$ for
  $i = 1,2,3$. In this respect, the number of required assumptions
  is independent of the order $N$ at which a closure is invoked,
  which arguably makes finding a suitable closure for sensitivity
  analysis less onerous than finding a suitable closure for the
  original cumulant equations.}

\subsection{Closure}
\label{sec:close}

\rev{As illustrated in figure \ref{diag:02}}, to obtain a
closed system of cumulant equations one needs to make an
assumption about how the highest-order cumulants are related to
those of lower order and, therefore, the way in which they depend
on the problem's parameters. One approach is to assume that
cumulants whose order is higher than $N$ are not affected, or
respond sufficiently slowly, to changes in the problem's
parameters, which is a sufficient condition for
$(\mathsbv{g}^{(N+1)}, \mathsbv{u}^{(N+1)})=0$. For $N=2$ in a
system with quadratic nonlinearities, this approach is consistent
with the assumption that the probability distribution of the
underlying process is Gaussian and is therefore completely
determined by its cumulants \rev{of first and second order}
\citep{FriUboo1995a}. 

\rev{As outlined in section \S\ref{sec:build}}, for sensitivity
analysis the implications of discarding cumulants beyond a certain
order are weaker than those associated with direct simulation of
the truncated equations. For example, truncation of the cumulant
equations at order $N=3$ and assuming that
$\mathsbv{U}^{(4)}\equiv 0$, produces non-realisable statistics
\citep{KraRmis1980a}, leading to a negative energy spectrum in
turbulence \citep{OguYjas1962a}. From the perspective of
sensitivity analysis, however, the orthogonality condition
$(\mathsbv{g}^{(4)}, \mathsbv{u}^{(4)})=0$ does not necessarily
imply that $\mathsbv{U}^{(4)}\equiv 0$. Similarly,
$(\mathsbv{u}^{(3)}, \mathsbv{g}^{(3)})=0$ does not necessarily
imply that the process is Gaussian. \rev{It is nevertheless
  important to note that the behaviour of higher-order cumulants
  in a Gaussian distribution is a special case, because
  probability distributions possessing non-zero cumulants at order
  $N_{*}>2$, followed by zero cumulants at all orders $N>N_{*}$,
  do not exist \citep[][p. 223]{LukEboo1970a}}.

One can discard cumulants of order higher than $N$ and model their
effects with a forcing function such as $\mathsbv{M}^{(N)}$,
which, in general, will depend on a vector $\mathsbv{\mu}$ of
unknown parameters:

\begin{equation}
  \mathsbv{H}^{(N)}(\pi_{N}\mathsbv{U}, \mathsbv{m}) = \mathsbv{R}^{(N)} + \mathsbv{M}^{(N)}(\pi_{N}\mathsbv{U},\mathsbv{\mu}),
\label{eq:closure}
\end{equation}

\noindent where $\mathsbv{R}^{(N)}$ represents the residuals
arising from the truncation and $\pi_{N}\mathsbv{U}$ is the
projection that sets the value of cumulants whose order exceeds
$N$ to zero. Assuming that the residual $\mathsbv{R}^{N}$ can be
made small with a suitable choice of $\mathsbv{M}^{(N)}$, and that
for a given $\mathsbv{M}^{(N)}$, $\mathsbv{R}^{(N)}$ does not
depend on $\mathsbv{m}$, the tangent linear equations \rev{at
  order $N$} are

\begin{equation}
\left(\pd{ \mathsbv{H} }{\mathsbv{U}}^{(N)}-\pd{ \mathsbv{M}}{\mathsbv{U}}^{(N)}\right)\mathsbv{u}^{(N)} = -\pd{ \mathsbv{H}}{\mathsbv{m}}^{(N)}.
  \label{eq:closure2}
\end{equation}

\noindent A key assumption underlying the use of
\eqref{eq:closure2} as a model for the tangent linear behaviour of
the system is that the model parameters $\mathsbv{\mu}$ in
\eqref{eq:closure} exhibit a weak dependence on the problem
parameters $\mathsbv{m}$ (hence
$\partial_{\mathsbv{m}}\mathsbv{M}$ is not included in
\eqref{eq:closure2}), \rev{which is consistent with the assumption
  that $\mathsbv{R}^{(N)}=0$ in the vicinity of
  $\mathsbv{m}$}. Utilising \eqref{eq:closure2} for truncation at
$N=3$ in the sensitivity analysis of a system with quadratic
nonlinearities, under the assumption that $\mathsbv{M}^{(N)}$
depends only on the highest retained cumulants
$\mathsbv{U}^{(N)}$, yields

\begin{equation}
  \begin{bmatrix}
0 & 0 & 0 & \mathsfbi{T}^{(11)} & \mathsfbi{T}^{(12)} & 0 \\
0 & 0 & 0 & \mathsfbi{T}^{(21)} & \mathsfbi{T}^{(22)} & \mathsfbi{T}^{(23)} \\
0 & 0 & 0 & \mathsfbi{T}^{(31)} & \mathsfbi{T}^{(32)} & \mathsfbi{T}^{(33)}-\mathsfbi{M}^{(33)} \\

\mathsfbi{T}^{(11)\dagger} & \mathsfbi{T}^{(21)\dagger} & \mathsfbi{T}^{(31)\dagger} & 0 & 0 & 0 \\
\mathsfbi{T}^{(12)\dagger} & \mathsfbi{T}^{(22)\dagger} & \mathsfbi{T}^{(32)\dagger} & 0 & 0 & 0 \\
0                  & \mathsfbi{T}^{(23)\dagger} & \mathsfbi{T}^{(33)\dagger}-\mathsfbi{M}^{(33)\dagger} & 0 & 0 & 0 \\
  \end{bmatrix}
\begin{pmatrix}
  \mathsbv{v}^{(1)} \\
  \mathsbv{v}^{(2)} \\
  \mathsbv{v}^{(3)} \\
  \mathsbv{u}^{(1)} \\
  \mathsbv{u}^{(2)} \\
  \mathsbv{u}^{(3)}
\end{pmatrix}
=
\begin{pmatrix}
  \mathsbv{f}^{(1)} \\
  \mathsbv{f}^{(2)} \\
  \mathsbv{f}^{(3)} \\
  \mathsbv{g}^{(1)} \\
  \mathsbv{g}^{(2)} \\ 
  \mathsbv{g}^{(3)} \\
\end{pmatrix},
\label{eq:extended_closure}
\end{equation}

\noindent where

\begin{equation}
\mathsfbi{M}^{(ij)} \equiv \pd{\mathsbv{M}^{(i)}}{\mathsbv{U}^{(j)}}.
\end{equation}

\noindent The closed system of extended equations
\eqref{eq:extended_closure} is, in general, invertible and
therefore provides a set of solutions for the adjoint variables
$\mathsbv{v}$ for a specified set of weights
$\mathsbv{g}$. Without selecting the model parameters
$\mathsbv{\mu}$, inversion of the adjoint operator
$(\mathsfbi{T}-\mathsfbi{M})^{\dagger}$ yields a fan of gradients,
as indicated in figure \ref{diag:03}$(b)$. The determination of a
unique gradient \rev{from the fan} requires the selection of an
optimal set of model parameters $\mathsbv{\mu}_{*}$. For example,
the optimal parameters \rev{could} be chosen to minimise
$\mathsbv{R}^{(N)}$ according to statistical observations from a
direct simulation:
\begin{equation}
\mathsbv{\mu}_{*} = \mathrm{arg}\min_{\mathsbv{\mu}}\left\|  \mathsbv{H}^{(N)} - \mathsbv{M}^{(N)}\right\|.
\label{eq:argmin}
\end{equation}
\noindent \rev{Using local observational data} the procedure of
obtaining sensitivity information can \rev{therefore} be freed
from tunable parameters once a suitable class of models has been
selected.

The extent to which it is necessary to include cumulants of order
greater than $N$ for sensitivity calculations depends on \rev{the
  role they play in maintaining the statistical equilibrium
  defined by \eqref{eq:cdyn}.} Although the truncation of the
cumulants at second order yields realisable statistics, the second
order cumulants alone will in general not be capable of describing
the fully nonlinear features of a flow \citep{FriUboo1995a}. As
described above, inclusion of the third-order cumulants (the
quasi-normal approximation) without accounting for the
fourth-order cumulants is problematic in simulations, because the
latter play a crucial role in damping the third-order \rev{cumulants}
\citep{BohTboo2005a}. Therefore, a popular choice, known as the
Eddy Damped Quasi-Normal Markovian approximation \citep[see
e.g.][]{LeiCjas1972a}, is to truncate the cumulants at third order
and to include a damping term to account for the discarded
fourth-order cumulants:

\begin{equation}
  \mathsbv{M}^{(3)} = \mu \mathsbv{U}^{(3)};\quad \mathrm{hence}\ \mathsfbi{M}=
  \begin{bmatrix}
    0, 0, \mu \mathsfbi{I}
    \end{bmatrix}.
\end{equation}

\noindent When $\mu\rightarrow \infty$ the cumulants of order
$N=3$ become increasingly damped and the closure corresponds to a
truncation at $N=2$; when $\mu\rightarrow 0$ the closure
corresponds to truncation at $N=3$ \citep{AllApre2016a}. The
eddy-damping parameter therefore produces a fan of possible
functional gradients, as illustrated in figure \ref{diag:03}$(b)$.

The optimal value of $\mu$ that minimises the size of the normed residual $\|\mathsbv{R}^{(3)}\|$ is

\begin{equation}
  \mu_{*} = \frac{(\mathsbv{U}^{(3)},\mathsbv{H}^{(3)})}{\|\mathsbv{U}^{(3)}\|^{2}},
\end{equation}

\noindent which enables the optimal functional gradient to be
determined according to

\begin{equation}
  \pd{J}{m_{i}} = \Bigg\langle \left(
\begin{bmatrix}
  \mathsfbi{T}^{(11)\dagger} & \mathsfbi{T}^{(21)\dagger} & \mathsfbi{T}^{(31)\dagger} \\
  \mathsfbi{T}^{(12)\dagger} & \mathsfbi{T}^{(22)\dagger} & \mathsfbi{T}^{(32)\dagger} \\
  0 & \mathsfbi{T}^{(23)\dagger} & \mathsfbi{T}^{(33)\dagger} \\
\end{bmatrix}
+
\begin{bmatrix}
  0 & 0 & 0 \\
  0 & 0 & 0 \\
  0 & 0 & -\mu_{*} \mathsfbi{I}
\end{bmatrix}
    \right)^{-1}\mathsbv{g}, \mathsbv{f}_{i} \Bigg\rangle.
\label{eq:dJmu}
\end{equation}

The procedure described in this section consists of identifying
the order $N$ at which the cumulant hierarchy should be truncated,
before selecting a subclass of possible models for the unknown
cumulants. The optimal parameters $\mathsbv{\mu}_{*}$, and
therefore optimal gradient $\nabla J_{*}$ in figure
\ref{diag:03}$(b)$, can be determined by minimising the residual
between statistics from direct simulation and the corresponding
model prediction according to equation \eqref{eq:argmin}.

\setlength{\abovecaptionskip}{-5pt plus 2pt minus 3pt}
{
\begin{figure}
  \begin{center}
    \input{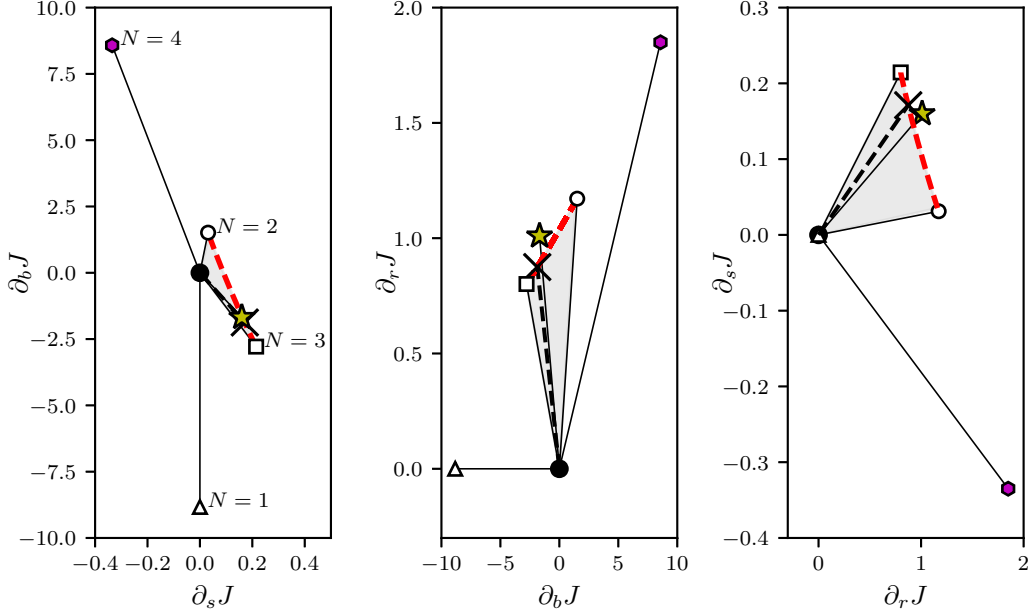}
  \end{center}
  \caption{Direction of the gradient $\nabla J=\nabla\av{Z}$
    evaluated at $m=(10,8/3,28)$ using the hierarchy of cumulant
    equations truncated at order $N=1,2,3,4$. The star corresponds
    to the observed gradient reported in \cite{WanQjcp2013a} and
    $\times$ corresponds to the optimised eddy-damped
    model. Statistics were obtained over $\tau=10^{6}$
    dimensionless time units. Points along the thick dashed red
    line that connects $N=2$ with $N=3$ were obtained by
    determining $\nabla\av{Z}$ for different values of
    $\mu\in (0,\infty)$.}
  \label{fig:main_XYZ}
\end{figure}
}

\section{Two-dimensional convection (the Lorenz equations, $d=3$)}
\label{sec:res}

\subsection{Truncation of the cumulant hierarchy}
\label{sec:rd3}

To test the method for obtaining functional gradients described in
\S \ref{sec:close}, we collect statistics from direct simulations
of the Lorenz equations \eqref{eq:lorenz}. We focus on the
statistically stationary state produced by parameters
$(s,b,r)=(10,8/3, 28)$, which is well documented and was the state
chosen for the sensitivity analysis of \citet{WanQjcp2013a}. The
dynamical equations are integrated using the DOPRI5 explicit
Runge-Kutta method in Python's SciPy library. To check convergence
of the computed cumulants the time \rev{$\tau$} \rev{used to
  define the time average} \eqref{eq:J} was varied from
\rev{$\tau=10^{3}$} to \rev{$\tau=10^{6}$}. To allow for transient
behaviour, the initial time used in the simulations is
$-100$. Integrals such as \eqref{eq:J} were computed using a
trapezium rule over the discrete points obtained from the
simulations.

\rev{Gradients of the functional $J=\av{Z}$ are displayed in
  figure \ref{fig:main_XYZ}, which shows the projection of the
  gradient vector $\nabla J$ onto two-dimensional planes. The
  symbols denote the gradients that are obtained by truncating the
  cumulant hierarchy at order $N=1,2,3,4$, without modelling the
  discarded cumulants.} Truncation of the tangent linear system at
$N=1$ yields an inaccurate representation of the gradient of
$\av{Z}$. The response of the second order cumulants to changes in
the parameters is evidently significant and therefore the
assumption that their dependence on parameters is identically zero
(or, more generally, orthogonal to $\mathsbv{g}^{(2)}$, as
described in \S\ref{sec:build}) produces poor
predictions. Truncation of the tangent linear system at $N=2$ also
yields a poor approximation of $\nabla \av{Z}$, \rev{particularly
  $\partial_{b}\av{Z}$}, but one that is an improvement in
comparison with truncation at $N=1$. As discussed in
\S\ref{sec:close}, in shearless turbulence the effect on eddies of
eddy-eddy interactions, captured by the third order cumulants
\citep{FarBarx2014a}, is expected to play a crucial role in
maintaining statistical equilibrium in the case of the Lorenz
equations. Indeed, the third order cumulants play a dynamically
important role in determining the response of the Lorenz system to
parametric changes, and \rev{figure \ref{fig:main_XYZ} shows that}
their retention yields a reasonable approximation of
$\nabla \av{Z}$.

\begin{figure}
  \begin{center}
    \input{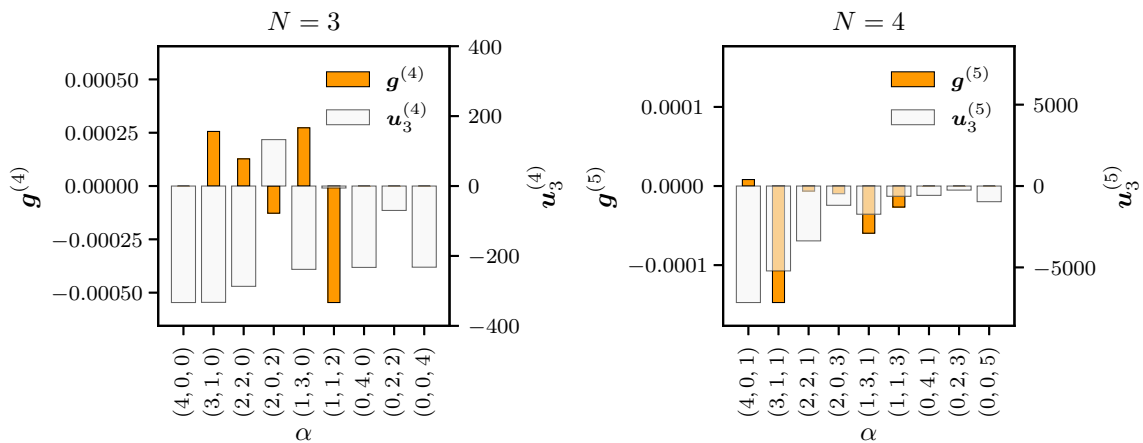}
  \end{center}
  \caption{The constituent parts of the error
  $(\mathsbv{g}^{(N+1)}, \mathsbv{u}^{(N+1)}_{3})$ for truncation at order $N=3$ (left) and $N=4$ (right). The values used to create this figure can be found in tables \ref{tab:error4} and \ref{tab:error5} in appendix \ref{sec:app2}.}
  \label{fig:11b}
\end{figure}

Truncation of the cumulant equations at $N=4$ yields a poor
approximation to $\nabla \av{Z}$, which illustrates the need to
find a compromise between the efficiency and simplicity of
truncation at relatively low order and the additional physics that
is captured by higher-order \rev{cumulants}. \rev{In the absence
  of physical} justification, truncation at higher order, rather
than lower order, does not necessarily imply an improved
estimation of the behaviour of the retained
cumulants. \rev{Indeed, as noted in \S\ref{sec:close},
  distributions with cumulants that are non-zero up to order $N$,
  followed by cumulants that are zero above order $N$, are not
  realisable for $N>2$. In this respect, it is perhaps not
  surprising that the fourth order approximation shown in figure
  \ref{fig:main_XYZ} is inaccurate.}
\rev{ 

  \subsection{Error analysis}
\label{sec:error}

  The difference between the approximation
  $\sum_{j=1}^{N}\langle \mathsbv{v}^{(j)},
  \mathsbv{f}^{(j)}_{i}\rangle$ and the observed gradient
  $\nabla \av{Z}$ that was obtained by truncating the cumulant
  hierarchy (depicted in figure \ref{fig:main_XYZ} with a star)
  can be understood by inspecting the derivatives of the discarded
  cumulants. As discussed at the end of \S\ref{sec:build}, the
  error associated with the $i^{\mathrm{th}}$ component of the
  gradient $\nabla\av{Z}$ for truncation at order $N$ is
  $(\mathsbv{g}^{(N+1)}, \mathsbv{u}^{(N+1)}_{i})$, where
  $\mathsbv{u}^{(N+1)}_{i}$ are the perturbations of the neglected
  cumulants, and $\mathsbv{g}^{(N+1)}$ determines the influence
  they have on the functional in question:

\begin{equation}
\pd{J}{m_{i}} = \underbrace{\sum_{j=1}^{N}\langle \mathsbv{v}^{(j)}, \mathsbv{f}^{(j)}_{i}\rangle}_{\text{approximation}}-\underbrace{\left(\mathsbv{g}^{(N+1)}, \mathsbv{u}^{(N+1)}_{i}\right)}_{\text{error}}.
\end{equation}

We focus on the error associated with the derivative of $\av{Z}$
with respect to $r$ (i.e. $i=3$), and display
$\mathsbv{g}^{(N+1)}$ and $\mathsbv{u}_{3}^{(N+1)}$ for $N=1,2,3$
and $4$ in figures \ref{fig:11a} and \ref{fig:11b}. We restrict
attention to non-zero cumulants using the symmetry arguments made
in \S\ref{sec:cumgen}. The gradients were determined by analysing
statistics from $256$ simulations employing values of $r$
distributed uniformly over a unit interval centred on
$r=28$. Further details are provided in appendix \ref{sec:app2}.}

\begin{figure}
  \begin{center}
    \input{fig_11a.pgf}
  \end{center}
  \caption{The constituent parts of the error
  $(\mathsbv{g}^{(N+1)}, \mathsbv{u}^{(N+1)}_{3})$ for truncation at order $N=1$ (left) and $N=2$ (right). The values used to create this figure can be found in tables \ref{tab:error2} and \ref{tab:error3} in appendix \ref{sec:app2}.}
  \label{fig:11a}
\end{figure}

\rev{The error associated with truncation at $N=1$ in figure
  \ref{fig:11a} is entirely due to the behaviour of the cumulant
  $\av{XY}=U_{110}$. The remaining cumulants, for which the
  corresponding values of $\mathsbv{g}^{(2)}$ are zero, do not
  contribute to the error, as can be seen directly from the second
  row of the cumulant equations in equation \eqref{eq:cum1}. At
  order $N=2$ in figure \ref{fig:11a}, the dominant contribution
  to the error comes from the response of $U_{111}$, and at order
  $N=3$, shown in figure \ref{fig:11b}, it comes predominantly
  from $U_{310}$ and $U_{130}$, which are related to the moments
  $\av{XY^{3}}$ and $\av{X^{3}Y}$. For truncation at order $N=4$,
  the perturbations in the discarded cumulants are large
  $O(10^{3})$, with figure \ref{fig:11b} indicating that the
  dominant contribution to the error comes from $U_{311}$, which
  is related to the moment $\av{X^{3}YZ}$. The effect on the error
  of the growing sensitivity and number of discarded cumulants is,
  to a limited extent, compensated by their diminishing influence
  on the gradient $\partial_{r}\av{Z}$, as evidenced by the
  relatively small values of $\mathsbv{g}^{(5)}$ in figure
  \ref{fig:11b}.}

\rev{A summary of the truncation errors obtained at each order is
  provided in table \ref{tab:errors}. Obtaining accurate
  observations of the sensitivity of fifth-order statistics from
  the Lorenz attractor is challenging, because it requires the use
  of relatively large intervals for time averaging. The
  approximate equality between the third and fourth columns of
  table \ref{tab:errors} nevertheless indicates that the sum of
  the inferred gradient
  $\sum_{j=1}^{N}\langle \mathsbv{v}^{(j)},
  \mathsbv{f}^{(N)}_{i}\rangle$ and the error
  $-(\mathsbv{g}^{(N+1)}, \mathsbv{u}^{(N+1)}_{3})$ agrees with
  $\partial_{r}\av{Z}$, and therefore satisfies the original
  cumulant equations to within $2\%$. At orders $1,2$ and $3$ the
  difference between the third and fourth columns of table
  \ref{tab:errors} implies that the cumulant equations are
  satisfied to within approximately $0.1\%$.}

\renewcommand{\neg}{\phantom{-}}

\begin{table}[b]
\renewcommand{\arraystretch}{1.2}
\rev{
  \begin{center}
  \begin{tabular}{lccccc}
    \hline
    \vspace{0.2em}
    & \hspace{12em} $N=$ & $1$ & $2$ & $3$ & $4$ \\
    \hline
{\small Approximation}  & {\small $\sum_{j=1}^{N}\langle \mathsbv{v}^{(j)}, \mathsbv{f}^{(N)}_{3}\rangle$} & -0.0003 & \neg 1.1715 & \neg 0.8012 & \neg 1.8502 \\ 
{\small Error}  &    {\small $-(\mathsbv{g}^{(N+1)}, \mathsbv{u}^{(N+1)})$}                          &\neg 1.0033 & -0.1691 & \neg 0.2007 & -0.8625 \\ 
 &    {\small $\sum_{j=1}^{N}\langle \mathsbv{v}^{(j)}, \mathsbv{f}^{(N)}_{3}\rangle-(\mathsbv{g}^{(N+1)}, \mathsbv{u}^{(N+1)})$} & \neg 1.0030 & \neg 1.0023 & \neg 1.0019 & \neg 0.9877 \\ 
{\small Observation} & {\small $\partial_{r}\av{Z}$} & \neg 1.0030 & \neg 1.0030 & \neg 1.0030 & \neg 1.0030 \\ 
    \hline
  \end{tabular}
    \vspace{5mm}
    \caption{The error
      $-(\mathsbv{g}^{(N+1)}, \mathsbv{u}_{3}^{(N+1)})$ in the
      estimation of $\partial_{r}\av{Z}$ using a truncation of the
      cumulant hierarchy at order $N$. See appendix \ref{sec:app2} for further details.}
  \label{tab:errors}
\end{center}
}
\end{table}

\subsection{Optimal closure}

In addition to the relatively simple truncations \rev{discussed in
  \S\S\ref{sec:rd3}-\ref{sec:error}}, corresponding to assumption
that $(\mathsbv{g}^{(N+1)}, \mathsbv{u}^{(N+1)})=0$, the
projections in figure \ref{fig:main_XYZ} also display the
gradients that are obtained by varying the eddy-damping parameter
$\mu$ described in \S\ref{sec:close}. The resulting family of
gradients produce a fan of gradient vectors lying between the
limit points associated with second-order truncation
($\mu\rightarrow \infty$) and the third-order truncation
($\mu\rightarrow 0$). A single member of the family corresponds to
the eddy damping that is optimal, in the sense of equation
\eqref{eq:argmin}, with respect to observations. Although the
optimal eddy damping $\mu_{*}$ yields a gradient that is close to
the observed gradient, \rev{figure \ref{fig:main_XYZ} indicates
  that} other values of $\mu$ would yield a slightly improved
prediction. The reason for this is that the parameter that
minimises the residual of the difference between the cumulant
equations and the observations is not necessarily that which
minimises the difference between the predicted and observed
gradients of a given functional, and therefore typifies the
difficulties of deriving gradients from \rev{a single set of
  statistics}.

\begin{figure}
  \begin{center}
  \input{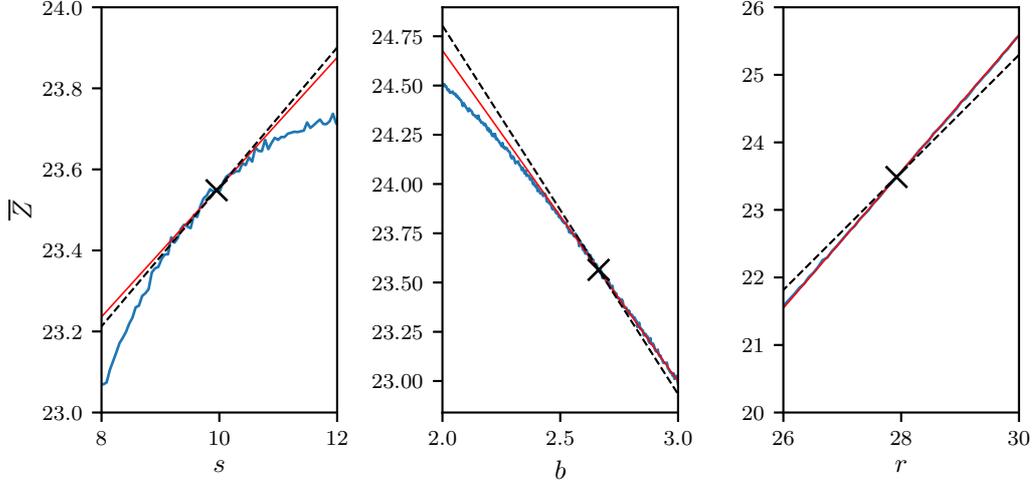}
  \end{center}
  \caption{Local observation and approximation of the derivative
    of $\av{Z}$ at the point $(s,b,r)=(10,8/3,28)$ (marked
    $\times$) with respect to the parameters $s$, $b$ and $r$. The
    thick blue curve corresponds to observations from direct
    simulations of the Lorenz equations, in which one parameter is
    varied and the others are held constant. The straight red line
    corresponds to a local fit to the gradient by
    \citet{WanQjcp2013a} and the dashed line corresponds to the
    gradient obtained from the optimised model approximation
    developed in \S \ref{sec:close}.}
  \label{fig:05}
\end{figure}

\begin{table}
  \renewcommand{\arraystretch}{1.0}
  \begin{center}
    \begin{tabular}{llll}
\hline \vspace{-1em}
\\
    & $\partial_{s}\av{Z}$ & $\partial_{b}\av{Z}$ & $\partial_{r}\av{Z}$ \\
\hline 
      \citet[][regression]{WanQjcp2013a} & \neg 0.16 & -1.68 & \neg 1.01 \\
      \vspace{2mm}
    \citet[]{WanQjcp2013a} & \neg 0.21 & -1.74 & \neg 0.97 \\

1st order  & \neg 0.0000  (\neg 0.0000)& -8.8346      (-8.8327)     & -0.0003     (-0.0001)    \\ 
2nd order  & \neg 0.0312  (\neg 0.0312)& \neg 1.5146  (\neg 1.5145)& \neg 1.1715 (\neg 1.1715)\\ 
3rd order  & \neg 0.2144  (\neg 0.2145)& -2.7844      (-2.7840)     & \neg 0.8012 (\neg 0.8012)\\ 
\vspace{2mm}
4th order  & -0.3353      (-0.3350)     & \neg 8.5854  (\neg 8.5774)& \neg 1.8502 (\neg 1.8495)\\ 
Model $(\tau=10^{3})$ & \neg 0.2186 & -2.5566 & \neg 0.8379 \\ 
Model $(\tau=10^{4})$ & \neg 0.1754 & -1.9172 & \neg 0.8723 \\ 
Model $(\tau=10^{5})$ & \neg 0.1734 & -1.9082 & \neg 0.8730 \\ 
Model $(\tau=10^{6})$ & \neg 0.1712 & -1.8743 & \neg 0.8748 \\ 

    \hline
  \end{tabular}
    \vspace{5mm}
    \caption{Cumulant sensitivities for the Lorenz
      equations. $N$th order corresponds to truncation of the
      cumulant equations at order $N$ (i.e. discarding cumulants
      of order $N+1$, \rev{which is equivalent to assuming that
        $(\mathsbv{u}^{(2)}, \mathsbv{g}^{(2)})=0$}), obtained
      from integrals over $\tau=10^{5}$ dimensionless time units
      (values corresponding to $\tau=10^{6}$ are shown in
      parentheses). The entries marked `Model' correspond to those
      obtained by using an optimal eddy damping parameter
      $\mu_{*}$ in the equations for the third order cumulants.}
  \label{tab:01}
  \end{center}
\end{table}

Figure \ref{fig:05} displays orthogonal slices through the
functional $\av{Z}$ to illustrate its partial dependence on the
parameters $s, b$ and $r$. The gradients that are obtained by
using the optimal model approach described in \S\ref{sec:close}
are displayed in comparison with those that were obtained by
linear regression analysis \citep{WanQjcp2013a}. The optimal model
approach yields a reasonably good agreement with the observed
gradients of $\av{Z}$ at $(s,b,r)=(10,8/3,28)$. The optimal value
of $\mu=\mu_{*}$ was found to be 8.96. A summary of the results,
including the dependence of the computed gradients on the
integration time used to obtain statistics, is provided in table
\ref{tab:01}.

\begin{figure}
  \begin{center}
  \input{fig_07.pgf}
  \caption{The solution to the inverse problem of determining the
    value of the renormalised Reynolds number $r$ from a known observed statistic $\av{Z}_{*}$. The circles denote points at which the functional and its gradients were evaluated to search for the minimum value $(\av{Z}-\av{Z}_{*})^{2}$.}
  \label{fig:07}
    \end{center}
\end{figure}
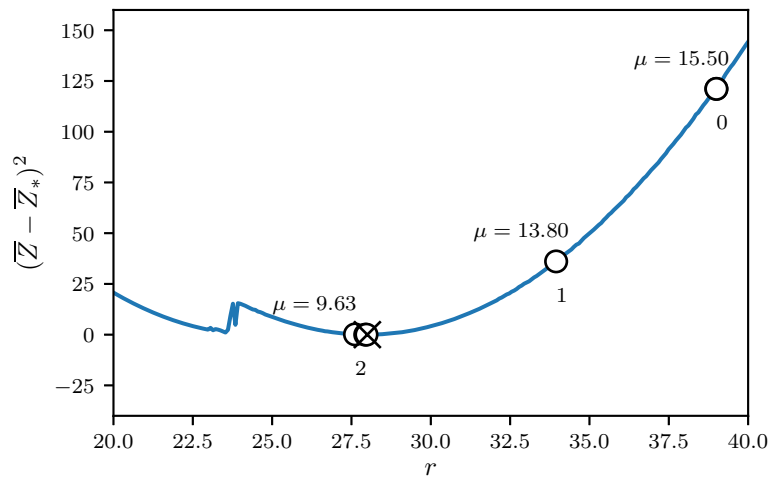

\subsection{Test optimisation problem}

In practice, local gradient information can be used in a
gradient-based optimisation routine. To demonstrate, we define the
functional $(\av{Z}-\av{Z}_{*})^{2}$ where $\av{Z}_{*}$
corresponds to the desired value of $\av{Z}$. For convenience we
define $\av{Z}_{*}$ as the value of $\av{Z}$ corresponding to the
parameters $(s,b,r)=(10,8/3,28)$ and attempt to solve the inverse
problem of determining \rev{an \emph{a priori} unknown $r$ from}
the known value $\av{Z}_{*}$. \rev{During each iteration of the
  optimisation procedure}, we calculate the cumulants
corresponding to a given set of parameters, and therefore the
functional $\av{Z}$. We then find the optimal eddy-damping
parameter $\mu_{*}$, before \rev{approximating} the gradient of
the functional. The optimal eddy-damping can therefore change at
each step of the iteration procedure. We use the BFGS optimisation
routine in the SciPy library and look for the parameter $r_{*}$
corresponding to $\av{Z}_{*}$. We set $s=10$ and $b=8/3$ and
select $r=39$ as an initial guess for $r$. Within four iterations
the optimisation routine finds $r_{*}=27.97$ and $\av{Z}_{*}$ to
within a tolerance of less than $0.001$. This is an interesting,
albeit contrived, example of a problem for which the use of
sub-optimal gradients can nevertheless lead to an optimal solution
because $\av{Z}=\av{Z}_{*}$ implies that $\nabla J=0$, regardless
of whether $\nabla\av{Z}=0$. More general optimisation problems,
for which the value of an extremum might not be known in advance,
will not necessarily share this property. Attempts to use the BFGS
optimisation routine without providing local gradients were
unsuccessful.

\renewcommand{\neg}{\phantom{-}}

\section{Three-dimensional convection ($d=9$)}

\label{sec:rd9}

A logical extension of the model for two-dimensional Boussinesq
convection analysed in the previous section is the model for
three-dimensional Boussinesq convection studied by
\citet{ReiPjpa1998a}. Like its two-dimensional counter part, the
system is a truncated Galerkin representation of the full
dynamics. Unlike its two-dimensional counter part, the system
evolves on a $d=9$ dimensional, rather than $d=3$ dimensional,
phase space and therefore yields statistics that exhibit a more
complicated dependence on the problem's parameters. Expressing
temperature and velocity in terms of a triple Fourier series and
retaining terms up to second order yields the following closed
system of equations \citep{ReiPjpa1998a}:

\setlength{\abovecaptionskip}{15pt plus 3pt minus 2pt}

\rev{
\begin{equation}
    \left.
  \begin{aligned}
    \dot{Q}_{0} &= -s\,b_{1}\,Q_{0} - Q_{1}\,Q_{3} + b_{4}\,Q_{3}^{2} + b_{3}\,Q_{2}\,Q_{4} - s\,b_{2}\,Q_{6},\\
    \dot{Q}_{1} &= -s\,Q_{1} + Q_{0}\,Q_{3} - Q_{1}\,Q_{4} + Q_{3}\,Q_{4} - s\,Q_{8}/2,\\
    \dot{Q}_{2} &= -s\,b_{1}\,Q_{2} + Q_{1}\,Q_{3} - b_{4}\,Q_{1}^{2} - b_{3}\,Q_{0}\,Q_{4} + s\,b_{2}\,Q_{7},\\
    \dot{Q}_{3} &= -s\,Q_{3} - Q_{1}\,Q_{2} - Q_{1}\,Q_{4} + Q_{3}\,Q_{4} + s\,Q_{8}/2,\\
    \dot{Q}_{4} &= -s\,b_{5}\,Q_{4} + Q_{1}^{2}/2 - Q_{3}^{2}/2,\\
    \dot{Q}_{5} &= -b_{6}\,Q_{5} + Q_{1}\,Q_{8} - Q_{3}\,Q_{8},\\
    \dot{Q}_{6} &= -b_{1}\,Q_{6} - r\,Q_{0} + 2\,Q_{4}\,Q_{7} - Q_{3}\,Q_{8},\\
    \dot{Q}_{7} &= -b_{1}\,Q_{7} + r\,Q_{2} - 2\,Q_{4}\,Q_{6}+Q_{1}\,Q_{8},\\
    \dot{Q}_{8} &= -Q_{8} -r\,Q_{1} +r\,Q_{3} -2\,Q_{1}\,Q_{5}+2\,Q_{3}\,Q_{5}+Q_{3}\,Q_{6}-Q_{1}\,Q_{7},
  \end{aligned}
\right\}
    \label{eq:lorenz9}
\end{equation}
\noindent where
\begin{equation}
  \left.
  \begin{aligned}
    b_{1} &= 4\,\frac{1+k^2}{1+2\,k^2},\quad &b_{2} = \frac{1+2\,k^2}{2\,(1+k^2)},\quad & b_{3} = 2\,\frac{1-k^2}{1+k^2},\\
    b_{4} &= \frac{k^2}{1+k^2},\quad         &b_{5} = 8\,\frac{k^2}{1+2\,k^2},\quad     & b_{6} = \frac{4}{1+2\,k^2}.
      \end{aligned}
    \right\}
        \label{eq:bs}
\end{equation}
}

\setlength{\abovecaptionskip}{5pt plus 3pt minus 2pt}
\begin{figure}[t]
  \begin{center}
  \input{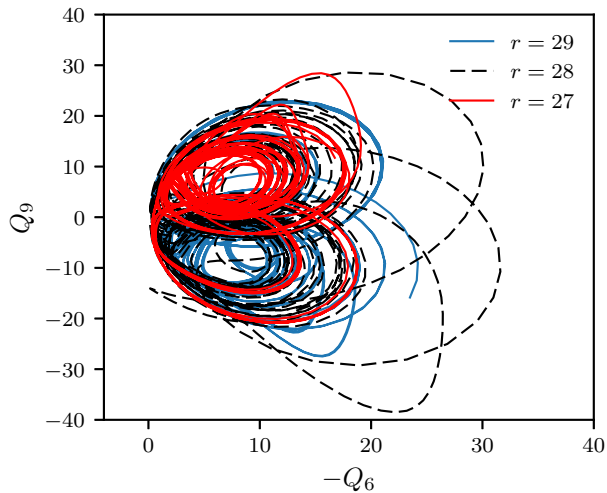}
  \caption{Projection of the chaotic attractor associated with the
    nine-dimensional system \eqref{eq:lorenz9} onto the plane
    describing the mode associated with horizontally average
    temperature $-Q_{6}$ and the difference in temperature between
    ascending and descending fluid $Q_{9}$. The trajectory corresponds to $2000$ dimensionless time units.}
  \label{fig:l9traj}
    \end{center}
\end{figure}

\noindent The parameters $s$ and $r$ continue to represent the
Prandtl number and the renormalised Rayleigh number. In addition,
equation \eqref{eq:bs} defines a set of geometrical parameters, as
a function of the wave number $k$, which correspond to $b$ in the
previous problem. To within constants of proportionality, the
variables $X$, $Y$ and $Z$ in the two-dimensional case correspond
to $Q_{4}$, $Q_{9}$ and $Q_{6}$, respectively. More precisely,
because $\av{Z} = -A\,\av{Q}_{6}$ for $A>0$, we focus on the
dependence of $-\av{Q}_{6}$ on $r$, where $-Q_{6}$ is proportional
to the strength of the horizontal average temperature with respect
to a state of pure conduction. For details pertaining to the
derivation of \eqref{eq:lorenz9}, the reader is referred to
\citet{ReiPjpa1998a}. To aid comparison with the results presented
in \citet{ReiPjpa1998a}, we choose $s=10$, $k=1/2$ and vary
$r$. The statistics were obtained over a dimensionless time
$\tau=10^{4}$.

As described in \citet{ReiPjpa1998a}, when $r>14.17$ for $s=10$
and $k=1/2$, the system is chaotic. When projected onto the
$Q_{6}, Q_{9}$ plane the attractor consists of two lobes either
side of the hyperplane $Q_{9}=0$, as can be seen in figure
\ref{fig:l9traj}. As $r$ increases the deviation of the
horizontally averaged temperature from the linear behaviour
associated with pure conduction increases.

\begin{figure}
  \begin{center}
  \input{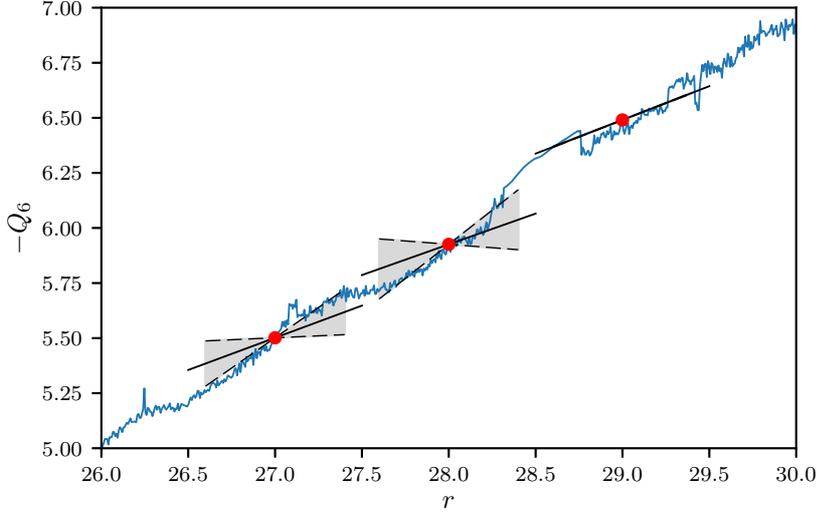}
  \caption{The dependence of $-Q_{6}$ on the system parameter $r$
    (blue line) and approximations to the local derivative using a
    truncated cumulant expansion. The statistics were obtained
    over a dimensionless time $\tau=10^{4}$. The shaded regions
    indicate the linear sensitivity of the computed gradient to
    changes in the eddy-damping parameter $\mu$; the gradient of
    the bounds of the shaded regions are calculated according to
    $\pm\mu_{*}\partial_{\mu}\partial_{r}Q_{6}$.}
  \label{fig:l9grad}
    \end{center}
\end{figure}

The precise relationship between $r$ and $-Q_{6}$ for the
parameters $s=10$ and $k=1/2$ is displayed in figure
\ref{fig:l9grad}. In spite of the discontinuities \rev{resulting
  from the use of a finite time average for each value of $r$}, the
  relationship indicates that $-Q_{6}$ tends to increase as $r$
  increases. At a glance, a linear relationship between $r$ and
  $-Q_{6}$ over $[26,30]\ni r$ appears to provide a reasonable first
  description of the sensitivity. However, closer inspection
  reveals that $-\partial_{r}\av{Q}_{6}$ varies significantly on
  scales of approximately $\Delta r\sim 0.5$, in contrast to the
  equivalent relationship for the Lorenz system (see figure
  \ref{fig:01}), for which $\partial_{r}\av{Z}$ is approximately
  constant over a large range of $r$.

\setlength{\abovecaptionskip}{5pt plus 3pt minus 2pt}

Since the dynamical system has $d=9$ degrees of freedom, the
number of cumulants up to order $N$ is given by equation \eqref{eq:binom}:

\begin{equation}
  \sum_{j=1}^{N}\binom{j+d-1}{d-1}=9+45+165+\ldots+\binom{N+d-1}{d-1}.
\end{equation}

\noindent The derivative of $-\av{Q}_{6}$ with respect to $r$ was
computed by truncating the cumulant equations at $N=3$ and
invoking the optimal eddy-damping closure described in
\S\ref{sec:close}. As is evident from figure \ref{fig:l9grad}, the
computed gradients appear to under estimate the underlying exact
gradients in general, but nevertheless provide a reasonably good
approximation. As pointed out in \S\ref{sec:close}, the
least-squares optimal eddy-damping parameter yields an
approximation to the gradient based on point-wise observations,
rather than the best approximation to the gradient.  It is
therefore useful to consider the sensitivity of the computed
gradient to changes in the eddy-damping parameter $\mu$ by
calculating the derivative
$-\partial_{\mu}\partial_{r}Q_{6}$. Figure \ref{fig:l9grad}
displays gradients corresponding to the optimal eddy damping
parameter $\mu_{*}$, along with lines whose gradients are
$\pm\mu_{*}\partial_{\mu}\partial_{r}Q_{6}(\mu_{*})$ to indicate the
sensitivity of the results to changes in $\mu$. It is interesting that at
$r=29$, we observe that $\partial_{\mu}\partial_{r}Q_{6}=0$, which indicates that
the computed gradient is insensitive to changes in $\mu$.

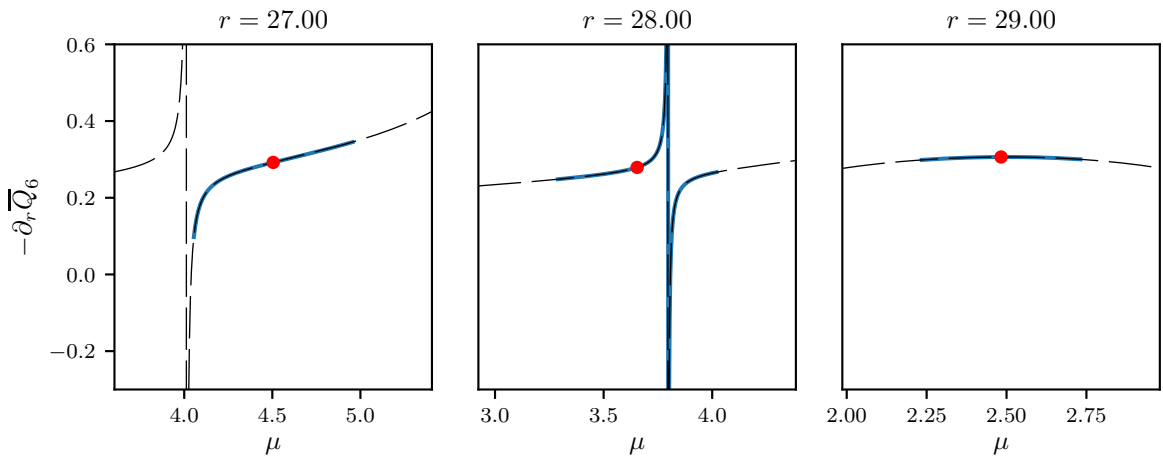
\begin{figure}
  \begin{center}
  \input{fig_l9dJ.pgf}
  \caption{The dependence of the derivative
    $\partial_{r}\av{Q}_{6}$, derived from a truncation of the
    cumulant hierarchy at $N=3$, according to \eqref{eq:dJmu}.}
  \label{fig:l9dJ}
    \end{center}
\end{figure}

As discussed in \S\ref{sec:close}, different values of $\mu$
correspond to different assumptions about the involvement of
third-order cumulants in the statistical
\rev{equilibrium}. Picking an arbitrary value of $\mu$ in equation
\eqref{eq:dJmu} might result in the adjoint operator being close
to singular and therefore yielding gradients that depend
sensitively on the choice of $\mu$. To illustrate this, figure
\ref{fig:l9dJ} shows evaluations of the derivative of $\av{Q}_{6}$
with respect to $r$ using equation \eqref{eq:dJmu} for values of
$\mu$ in the vicinity of the optimal value $\mu_{*}$ as determined
by equation \eqref{eq:argmin}. When $r=27.0$ and $r=28.0$ it is
evident that some choices of $\mu$ result in a singular or
near-singular adjoint operator and, therefore, a large amount of
uncertainty in the resulting gradients. To obtain robust results
in this particular case it is therefore necessary to use an
optimal eddy-damping parameter that is determined systematically,
rather than an estimation that is independent of observations. The
optimal parameter $\mu_{*}$ appears to find a local maximum in the
value of $-\partial_{r}\av{Q}_{6}$ when $r=29.0$, which explains
why the estimated gradient is locally insensitive to changes in
$\mu$.

\section{Conclusions}
\label{sec:con}

We have described a systematic means of obtaining approximate
forward and adjoint sensitivity information from a chaotic system
using a truncated system of cumulant equations. Unlike
linearisation of the underlying evolution equations for individual
trajectories, the cumulant equations yield robust, albeit
approximate, information about functional derivatives. The method
was designed for situations in which one has access to statistical
data from the direct simulation of a potentially high-dimensional
chaotic system and wishes to approximate the gradients of a
functional with respect to many input parameters. In principle
the method could also be applied to obtain gradients of flow
functionals from experimental measurements.

We combined data from direct simulation with tangent linear and
adjoint equations for the system's statistical state
dynamics. These equations can be obtained from the original system
systemically using a cumulant generating function. Whilst the
method is approximate, because it relies on truncation of the
cumulant equations, the incorporation of observations to derive
optimal truncations significantly improves its accuracy and
robustness.  Although the method itself is not restricted to
statistically stationary problems, we expect the acquisition and
incorporation of the corresponding unsteady statistical
observations to be challenging.

The extraction of gradient information from functionals of chaotic
dynamical systems is a stringent test for modelling and closure
schemes. A given model can be tuned to adequately represent a
given problem. However, unless it accurately describes the
underlying physics, it is unlikely to yield accurate information
about how an output functional changes with respect to changes in
the problem's definition. Hence, the class of models from which
one selects a suitable surrogate must be capable of describing the
dynamics correctly. In the absence of shear, Rayleigh B\'{e}nard
convection and, specifically, the Lorenz model, provide a
difficult test for cumulant closures because truncation of the
equations at second order removes interactions that are vital in
determining the response of the system's \rev{statistical
  equilibrium}. In contrast, for problems dominated by mean shear,
such as jets, it is likely that cumulant truncation at second
order would adequately capture the leading-order dynamics and
would significantly simplify the approach to obtaining gradient
information. The basic approach that we have described can be
refined by exploring more appropriate ways of fitting the
surrogate model.

Although we have focused on relatively low-dimensional dynamical
systems, the idea of using cumulant expansions was motivated by
the need to analyse high-dimensional dynamical systems. The
challenge in the successful application of the method to large
systems lies in the acquisition of a large number of accurate high
order cumulants and the systematic derivation and manipulation of
a potentially large number of cumulant equations. In such cases
statistical symmetries of a given problem can be used to
significantly reduce the number of unknowns. An alternative or
complementary approach would be to map the full system onto a
relatively low-order model, for which the cumulants and their
dynamics can be more readily obtained. \rev{Guided by the
  classical moment problem, further work should also incorporate
  restrictions that could be imposed on the gradients of cumulants
  to ensure that they point in a realisable direction.}

\begin{figure}
  \begin{center}
    \input{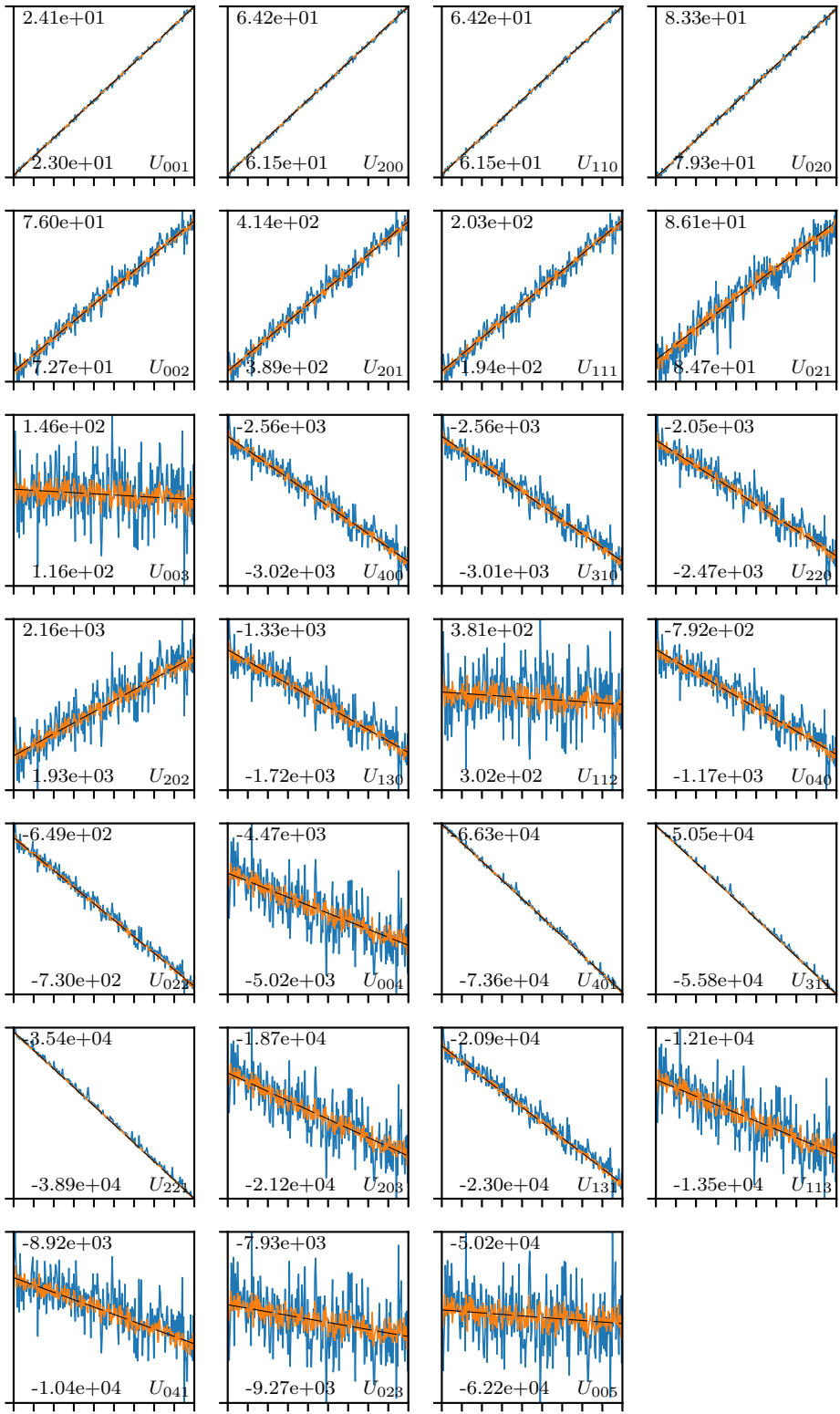}
  \end{center}
  \caption{\rev{Estimators for the non-zero cumulants of the Lorenz
    attractor from simulations of duration $\tau = 10^{5}$ time
    units (blue/dark) and $\tau = 10^{6}$ time units
    (red/light). The gradients of the data were computed from
    simulations of duration $\tau = 10^{6}$ time units. The
    cumulant $U_{\alpha}$ corresponds to the moment
    $\av{\mathsbv{Q}^{\alpha}}$ modulo all combinations of the
    corresponding low-order cumulants, as described in appendix
    \ref{sec:app1}.}}
  \label{fig:08}
\end{figure}

\appendix
\rev{
\section{Derivation of the cumulant equations}

\label{sec:app1}

The Hopf generating functional \citep{HopEarm1952a} is defined according to 

\begin{equation}
\Psi(\mathsbv{Q}(t), \mathsbv{P}) = \mathrm{exp}\left(\imath P_{i}Q_{i}(t)\right),
\end{equation}

\noindent where $\imath=\sqrt{-1}$. The moment $\av{\mathsbv{Q}^{\alpha}}$ can
therefore be generated as

\begin{equation}
\av{\mathsbv{Q}^{\alpha}} = (-\imath)^{|\alpha|}\partial^{\alpha}\av{\Psi}\bigg|_{\mathsbv{P}=\mathsbv{0}},
\end{equation}

\noindent where
$\alpha = (\alpha_{1},\alpha_{2},\ldots,\alpha_{d})$ is a
multi-index, such that
$\mathsbv{Q}^{\alpha}=Q_{1}^{\alpha_{1}}Q_{2}^{\alpha_{2}}\ldots\,Q_{d}^{\alpha_{d}}$
and
$\partial^{\alpha}=\partial_{P_{1}}^{\alpha_{1}}\partial_{P_{2}}^{\alpha_{2}}\ldots\partial_{P_{d}}^{\alpha_{d}}$. A
moment $\av{\mathsbv{Q}^{\alpha}}$ can be decomposed into a sum of products of
cumulants $U_{\beta}$, containing all possible factorisations of
the monomial $\mathsbv{Q}^{\alpha}$:

\begin{equation}
\av{\mathsbv{Q}^{\alpha}}=\sum _{\pi\in \Pi(\alpha)}\prod_{\beta \in \pi}U_{\beta},
\label{eq:U2M}
\end{equation}

\noindent where $\pi$ is a multiset that decomposes a multi-index
into addends. For example, if $\alpha=(2,1,0,0,\dots)$ then
$\pi=\{(2,0,\ldots), (0,1,0,\ldots)\}$ would be one such
decomposition. The multiset $\Pi(\alpha)$ consists of all such
decompositions. For example, if $\alpha=(0,0,4)$, then
$\mathsbv{Q}^{\alpha}=Z^{4}$, and

\begin{equation}
  \begin{aligned}
    \Pi(\alpha) = &\left\{ \left\{(0,0,4)\right\}\right., \\
    & \left\{(0,0,3),(0,0,1)\right\}^{4}, \\
    & \left\{(0,0,2),(0,0,2)\right\}^{3}, \\
    & \left\{(0,0,2),(0,0,1),(0,0,1)\right\}^{6},\\
    & \left.\left\{(0,0,1),(0,0,1),(0,0,1),(0,0,1)\right\} \right\},
  \end{aligned}
  \label{eq:z4parts}
\end{equation}

\noindent in which the exponents denote set multiplicities. In the
example above, the set multiplicities arise from the different
ways that a set consisting of $4$ elements can be
partitioned. According to \eqref{eq:U2M} and \eqref{eq:z4parts},
the moment $\av{Z^{4}}$ can be expressed in terms of cumulants as

\begin{equation}
  \av{Z^{4}}=U_{004}+4U_{003}U_{001}+3U_{002}^{2}+6U_{002}U_{001}^{2}+U_{001}^{4}.
    \label{eq:cumulants_eg2}
\end{equation}

\noindent The decomposition \eqref{eq:U2M} is identical to that
which arises when partial derivatives are applied to composite
functions. Indeed, using $\Psi = \mathrm{exp}(\mathrm{log}(\Psi))$, 

\begin{equation}
\partial^{\alpha}\av{\Psi}\bigg|_{\mathsbv{P}=0}=\sum _{\pi\in \Pi(\alpha)}\prod_{\beta \in \pi}\partial^{\beta}\log \av{\Psi}\bigg|_{\mathsbv{P}=0},
\end{equation}

\noindent which shows the logarithm of the moment generating
function is the cumulant generating function.

}
\rev{
\section{Observed cumulant gradients}

\label{sec:app2}

The gradients used to compute the truncation errors displayed in
figures \ref{fig:11a}-\ref{fig:11b} were obtained from simulations
of the Lorenz equations for $256$ values of $r$ uniformly
distributed over a unit interval centred on $r=28$. An
approximation of the partial derivative of non-zero cumulants up
to order $5$ was obtained by minimising the squared difference
between the straight line $(\partial_{r}J|_{r=28})r+J(0)$ and the data,
which are both displayed in figure \ref{fig:08}. The resulting
gradients are tabulated in tables \ref{tab:error2}-\ref{tab:error5}.

\begin{table}[h]
\renewcommand{\arraystretch}{1.0}
      \rev{
  \begin{center}
    \input{error_order_2}
  \end{center}
  \caption{\rev{The constituent parts of the error
  $(\mathsbv{g}^{(2)}, \mathsbv{u}^{(2)}_{3})$ for truncation at order $N=1$.}}
\label{tab:error2}
}
\end{table}

\begin{table}[h]
\renewcommand{\arraystretch}{1.0}
  \rev{
  \begin{center}
    \input{error_order_3}
  \end{center}
  }
  \caption{\rev{The constituent parts of the error
    $(\mathsbv{g}^{(3)}, \mathsbv{u}^{(3)}_{3})$ for truncation at order $N=2$.}}
  \label{tab:error3}
\end{table}

\begin{table}[h]
\renewcommand{\arraystretch}{1.0}
  \rev{
  \begin{center}
    \input{error_order_4}
  \end{center}
  }
  \caption{\rev{The constituent parts of the error
  $(\mathsbv{g}^{(4)}, \mathsbv{u}^{(4)}_{3})$ for truncation at order $N=3$.}}
  \label{tab:error4}
\end{table}

\begin{table}
\renewcommand{\arraystretch}{1.0}
  \rev{
  \begin{center}
    \input{error_order_5}
  \end{center}
  }
  \caption{\rev{The constituent parts of the error
    $(\mathsbv{g}^{(5)}, \mathsbv{u}^{(5)}_{3})$ for truncation at order $N=4$.}}
  \label{tab:error5}
\end{table}

}

\section*{Acknowledgements}

The author gratefully acknowledges funding from an EPSRC Doctoral
Prize under grant number EP/M507878/1 and an Imperial College
Junior Research Fellowship. The work benefited from
discussions with Davide Lasagna at a SIG meeting for Flow
Modelling, Instability and Control on March 29-30 2017, as part of
the UK Fluids Network (EP/N032861/1). Johanna Mader is thanked for
reading this manuscript and providing the author with useful suggestions. 

\addcontentsline{toc}{section}{References}
\bibliographystyle{main}

\input{main.bbl}
\end{document}

%% file: diag04.tex
\definecolor{cffffff}{RGB}{255,255,255}
\definecolor{ce2e2e2}{RGB}{226,226,226}
\definecolor{cf9f5d5}{RGB}{249,245,213}
\definecolor{cdc0000}{RGB}{220,0,0}

\begin{tikzpicture}[y=0.80pt, x=0.80pt, yscale=-1.200000, xscale=1.200000, inner sep=0pt, outer sep=0pt]
  \path[draw=black,fill=cffffff,line join=miter,line cap=butt,miter
    limit=4.00,even odd rule,line width=0.450pt] (398.2731,203.5411) --
    (407.5916,195.5102) -- (410.6478,199.4754) -- (400.2581,206.2963) --
    (401.3291,207.8088) -- (394.1143,208.8430) -- (397.2866,202.2818) -- cycle;
  \path[fill=black,line join=miter,line cap=butt,fill opacity=0.078,even odd
    rule,line width=0.900pt] (318.4152,226.6926) -- (417.5223,220.3310) --
    (363.2813,157.0497) -- cycle;
  \path[draw=black,line join=miter,line cap=butt,even odd rule,line width=0.960pt]
    (363.4654,209.7845) -- (363.4654,107.8117);
  \path[draw=black,dash pattern=on 6.91pt off 6.91pt,line join=miter,line
    cap=butt,miter limit=4.00,even odd rule,line width=0.576pt]
    (293.1078,236.2238) -- (362.7507,132.1167) -- (445.2506,226.0452) -- cycle;
  \path[draw=black,line join=miter,line cap=butt,even odd rule,line width=0.960pt]
    (266.5833,247.0233) -- (363.3380,209.5944) -- (471.8537,231.8684);
  \path[draw=black,fill=ce2e2e2,miter limit=4.00,line width=0.760pt]
    (342.1682,217.6401) circle (0.0931cm);
  \path[draw=black,fill=ce2e2e2,miter limit=4.00,line width=0.760pt]
    (390.1275,214.9096) circle (0.0931cm);
  \path[draw=black,fill=ce2e2e2,miter limit=4.00,line width=0.760pt]
    (363.6953,183.9378) circle (0.0931cm);
  \path[draw=black,fill=ce2e2e2,miter limit=4.00,line width=0.760pt]
    (417.3198,220.2902) circle (0.0931cm);
  \path[draw=black,fill=ce2e2e2,miter limit=4.00,line width=0.760pt]
    (318.6524,226.7475) circle (0.0931cm);
  \path[draw=black,fill=ce2e2e2,miter limit=4.00,line width=0.760pt]
    (363.7066,157.0638) circle (0.0931cm);
  \path[draw=black,fill=ce2e2e2,miter limit=4.00,line width=0.760pt]
    (363.4390,209.1816) circle (0.0931cm);
  \path[draw=black,fill=cf9f5d5,miter limit=4.00,line width=0.749pt]
    (363.1734,131.9895) circle (0.0917cm);
  \path[draw=black,fill=cf9f5d5,miter limit=4.00,line width=0.749pt]
    (293.3188,236.5421) circle (0.0917cm);
  \path[draw=black,fill=cf9f5d5,miter limit=4.00,line width=0.749pt]
    (445.1045,226.1849) circle (0.0917cm);
  \path[draw=black,fill=ce2e2e2,miter limit=4.00,line width=0.760pt]
    (389.0888,187.6121) circle (0.0931cm);
  \path[draw=black,fill=ce2e2e2,miter limit=4.00,line width=0.760pt]
    (341.3837,190.5521) circle (0.0931cm);
  \path[draw=black,fill=ce2e2e2,miter limit=4.00,line width=0.760pt]
    (368.5006,223.8835) circle (0.0931cm);
  \path[draw=black,fill=cf9f5d5,miter limit=4.00,line width=0.749pt]
    (397.4329,229.5571) circle (0.0917cm);
  \path[draw=black,fill=cf9f5d5,miter limit=4.00,line width=0.749pt]
    (343.5043,233.1286) circle (0.0917cm);
  \path[draw=black,fill=cf9f5d5,miter limit=4.00,line width=0.749pt]
    (415.4522,192.3493) circle (0.0917cm);
  \path[draw=black,fill=cf9f5d5,miter limit=4.00,line width=0.749pt]
    (388.4699,161.5738) circle (0.0917cm);
  \path[draw=black,fill=cf9f5d5,miter limit=4.00,line width=0.749pt]
    (341.7312,164.0342) circle (0.0917cm);
  \path[draw=black,fill=cf9f5d5,miter limit=4.00,line width=0.749pt]
    (318.5819,198.8033) circle (0.0917cm);
  \path[draw=black,fill=cf9f5d5,miter limit=4.00,line width=0.749pt]
    (368.3257,195.9857) circle (0.0917cm);
  \path[fill=black,line join=miter,line cap=butt,line width=0.960pt]
    (253.6952,253.5229) node[above right] (text5271) {$\alpha_{1}$};
  \path[draw=black,line join=miter,line cap=butt,even odd rule,line width=0.960pt]
    (273.6707,242.0523) -- (266.8522,247.1030) -- (275.4385,246.3454);
  \path[draw=black,line join=miter,line cap=butt,even odd rule,line width=0.960pt]
    (360.5439,116.0357) -- (363.3218,107.1969) -- (366.3522,116.5408);
  \path[draw=black,line join=miter,line cap=butt,even odd rule,line width=0.960pt]
    (463.3269,227.6576) -- (472.6708,231.9507) -- (461.3066,232.4558);
  \path[draw=black,line join=miter,line cap=butt,miter limit=4.00,even odd
    rule,line width=0.720pt] (314.9796,136.5958) .. controls (314.9796,136.5958)
    and (332.0260,132.3342) .. (334.8670,143.2249) .. controls (337.7081,154.1156)
    and (333.9200,151.2746) .. (338.1816,157.9037);
  \path[fill=black,line join=miter,line cap=butt,line width=0.900pt]
    (283.7281,140.3838) node[above right] (text5061) {$U_{102}$};
  \path[fill=black,line join=miter,line cap=butt,line width=0.900pt]
    (417.2659,147.9200) node[above right] (text5061-9) {$U_{012}$};
  \path[draw=black,line join=miter,line cap=butt,miter limit=4.00,even odd
    rule,line width=0.720pt] (362.4889,250.2565) .. controls (355.7611,241.2413)
    and (359.2097,232.6217) .. (363.7075,226.3414);
  \path[draw=cdc0000,dash pattern=on 1.80pt off 0.90pt,line join=miter,line
    cap=butt,miter limit=4.00,even odd rule,line width=0.450pt]
    (345.0732,187.7635) -- (383.5423,164.5328);
  \path[draw=cdc0000,dash pattern=on 1.80pt off 0.90pt,line join=miter,line
    cap=butt,miter limit=4.00,even odd rule,line width=0.450pt]
    (388.9589,182.6123) .. controls (388.9589,184.1283) and (388.9876,167.3739) ..
    (388.9876,167.3739);
  \path[draw=cdc0000,dash pattern=on 1.80pt off 0.90pt,line join=miter,line
    cap=butt,miter limit=4.00,even odd rule,line width=0.450pt]
    (368.0265,157.1934) -- (383.0688,160.0345);
  \path[draw=cdc0000,line join=miter,line cap=butt,miter limit=4.00,even odd
    rule,line width=0.450pt] (379.8549,157.7193) -- (383.2031,160.2305) --
    (379.3527,160.9002);
  \path[draw=cdc0000,line join=miter,line cap=butt,miter limit=4.00,even odd
    rule,line width=0.450pt] (379.8549,165.2528) -- (383.5379,164.5832) --
    (381.5290,167.5966);
  \path[draw=cdc0000,line join=miter,line cap=butt,miter limit=4.00,even odd
    rule,line width=0.450pt] (387.5558,170.7774) -- (388.8951,167.4292) --
    (390.5692,170.9448);
  \path[draw=black,line join=miter,line cap=butt,miter limit=4.00,even odd
    rule,line width=0.450pt] (433.0007,182.1595) -- (422.9328,189.5346) --
    (420.0034,185.8920) -- (429.1262,177.8327) -- (427.9926,176.3666) --
    (436.1047,175.7407) -- (434.0392,183.3763) -- cycle;
  \path[fill=black,line join=miter,line cap=butt,line width=0.900pt]
    (413.7801,210.7149) node[above right] (text5061-9-9)
    {${\small\mathrm{known}}$};
  \path[fill=black,line join=miter,line cap=butt,line width=0.900pt]
    (442.6909,182.8160) node[above right] (text5061-9-9-9)
    {${\small\mathrm{unknown}}$};
  \path[fill=black,line join=miter,line cap=butt,line width=0.900pt]
    (414.0965,165.6925) node[above right] (text5061-9-9-9-7)
    {${\small\mathrm{models}\ni\mathrm{optimal\ model}}$};
  \path[fill=black,line join=miter,line cap=butt,line width=0.900pt]
    (365.0517,256.4180) node[above right] (text5061-9-7) {$U_{110}$};
  \path[draw=black,line join=miter,line cap=butt,miter limit=4.00,even odd
    rule,line width=0.720pt] (390.1664,156.3496) .. controls (393.3792,143.9863)
    and (405.9223,139.7193) .. (411.4743,142.1444);
  \path[draw=black,line join=miter,line cap=butt,miter limit=4.00,even odd
    rule,line width=0.720pt] (412.4278,165.7817) -- (390.9040,175.2975);
  \path[fill=black,line join=miter,line cap=butt,line width=0.900pt]
    (472.7679,164.0810) node[above right] (text1132) {$$};
  \path[fill=black,line join=miter,line cap=butt,line width=0.960pt]
    (460.1366,247.0430) node[above right] (text5271-5) {$\alpha_{2}$};
  \path[fill=black,line join=miter,line cap=butt,line width=0.960pt]
    (357.6813,98.7171) node[above right] (text5271-5-3) {$\alpha_{3}$};

\end{tikzpicture}

%% file: diag03.tex
\definecolor{cd2b400}{RGB}{210,180,0}
\definecolor{c0000ff}{RGB}{0,0,255}
\definecolor{cdb0000}{RGB}{219,0,0}
\definecolor{cffffff}{RGB}{255,255,255}
\definecolor{cffff00}{RGB}{255,255,0}
\definecolor{ce60000}{RGB}{230,0,0}
\definecolor{c6d5f41}{RGB}{109,95,65}

\begin{tikzpicture}[y=0.80pt, x=0.80pt, yscale=-1.000000, xscale=1.000000, inner sep=0pt, outer sep=0pt]
  \path[draw=black,fill=black,line join=miter,line cap=butt,fill
    opacity=0.080,even odd rule,line width=0.455pt] (372.5286,142.5148) ..
    controls (372.5286,142.5148) and (399.4864,129.1844) .. (445.7759,132.4328) ..
    controls (492.0654,135.6812) and (491.1907,130.3787) .. (491.1907,130.3787) ..
    controls (491.1907,130.3787) and (508.2039,143.4778) .. (536.8891,148.9246) ..
    controls (579.4447,157.0052) and (608.6730,152.1421) .. (608.6730,152.1421) ..
    controls (608.6730,152.1421) and (607.9137,173.0138) .. (558.3160,192.3598) ..
    controls (473.9499,225.2676) and (406.7058,207.3090) .. (406.7058,207.3090) ..
    controls (406.7058,207.3090) and (405.2566,163.6771) .. (372.5286,142.5148) --
    cycle;
  \path[draw=black,fill=cd2b400,dash pattern=on 5.46pt off 5.46pt,line
    join=miter,line cap=butt,miter limit=4.00,fill opacity=0.169,even odd
    rule,line width=0.455pt] (392.3152,148.2729) .. controls (392.3152,148.2729)
    and (408.4944,117.4660) .. (454.7840,120.7143) .. controls (501.0735,123.9627)
    and (522.5940,129.5707) .. (522.5940,129.5707) .. controls (522.5940,129.5707)
    and (535.5876,161.0454) .. (564.2727,166.4923) .. controls (606.8284,174.5729)
    and (628.5916,193.2536) .. (628.5916,193.2536) .. controls (628.5916,193.2536)
    and (614.6745,183.6544) .. (563.8786,199.5912) .. controls (475.6034,227.2869)
    and (434.4873,191.3356) .. (434.6637,191.5708) .. controls (434.6637,191.5708)
    and (446.7615,167.4149) .. (392.3152,148.2729) -- cycle;
  \path[draw=black,fill=c0000ff,dash pattern=on 1.50pt off 1.50pt,line
    join=miter,line cap=butt,miter limit=4.00,fill opacity=0.077,even odd
    rule,line width=0.375pt] (440.3794,166.9635) .. controls (449.8075,162.5367)
    and (459.0033,155.7870) .. (468.6637,153.6836) .. controls (483.0457,155.5221)
    and (489.2658,158.3809) .. (498.8764,160.8159) .. controls (488.2411,164.6753)
    and (479.0388,169.4900) .. (471.7502,175.5804) .. controls (461.6362,169.9646)
    and (450.8996,169.3295) .. (440.3794,166.9635) -- cycle;
  \path[fill=cdb0000,miter limit=4.00,line width=1.471pt] (471.0013,163.3353)
    circle (0.1234cm);
  \path[fill=cffffff,line join=miter,line cap=butt,fill opacity=0.330,even odd
    rule,line width=0.750pt] (394.3097,184.8735) -- (417.7472,189.6759) --
    (456.3729,170.5847) -- (438.8250,166.9727) .. controls (452.1286,159.4781) and
    (453.9328,158.2862) .. (458.8502,155.9968) -- (406.1671,129.6657) --
    (367.1317,133.9470) -- cycle;
  \begin{scope}[cm={{0.66466,0.0,0.0,0.66466,(146.73753,-138.23368)}},miter limit=4.00,line width=0.790pt]
    \path[draw=black,line join=miter,line cap=butt,miter limit=4.00,even odd
      rule,line width=0.790pt] (483.5909,454.1180) -- (243.9745,455.4413);
  \end{scope}
  \begin{scope}[cm={{0.60393,0.0,0.0,0.82271,(166.15308,-200.31101)}},miter limit=4.00,line width=0.745pt]
    \path[draw=black,dash pattern=on 5.96pt off 5.96pt,line join=miter,line
      cap=butt,miter limit=4.00,even odd rule,line width=0.745pt]
      (505.0882,441.8203) -- (272.2785,353.6837);
  \end{scope}
  \path[draw=black,line join=miter,line cap=butt,miter limit=4.00,even odd
    rule,line width=0.433pt] (366.2960,202.1417) -- (366.2960,244.5112) --
    (402.5151,230.1602);
  \path[draw=black,line join=miter,line cap=butt,miter limit=4.00,even odd
    rule,line width=0.433pt] (365.9543,244.8528) -- (420.9663,255.4452);
  \path[draw=black,line join=miter,line cap=butt,even odd rule,line width=0.541pt]
    (364.6476,208.4428) -- (366.2181,202.0401) -- (368.1509,208.3220);
  \path[draw=black,line join=miter,line cap=butt,even odd rule,line width=0.541pt]
    (396.2986,230.9126) -- (403.0637,229.9462) -- (398.8355,233.2079);
  \path[draw=black,line join=miter,line cap=butt,even odd rule,line width=0.541pt]
    (415.8691,252.5368) -- (421.3053,255.5569) -- (413.6946,255.5569);
  \path[fill=black,line join=miter,line cap=butt,line width=0.800pt]
    (527.9198,125.4150) node[above right] (text5071) {$J_{\mu}$};
  \path[fill=black,line join=miter,line cap=butt,line width=0.800pt]
    (556.9864,146.3054) node[above right] (text5071-8) {$J$};
  \path[fill=black,line join=miter,line cap=butt,line width=0.935pt]
    (313.6569,144.1938) node[above right] (text5071-8-1) {$\nabla J_{*}$};
  \path[fill=black,line join=miter,line cap=butt,line width=0.935pt]
    (327.9476,179.6581) node[above right] (text5071-8-1-1) {$\nabla J$};
  \path[fill=black,line join=miter,line cap=butt,line width=0.800pt]
    (222.8571,140.9336) node[above right] (text5162) {$$};
  \path[fill=black,line join=miter,line cap=butt,line width=0.800pt]
    (424.0353,253.7099) node[above right] (text5071-8-1-1-0) {$m_{1}$};
  \path[fill=black,line join=miter,line cap=butt,line width=0.800pt]
    (403.5755,228.0385) node[above right] (text5071-8-1-1-0-4) {$m_{2}$};
  \path[fill=black,line join=miter,line cap=butt,line width=0.800pt]
    (357.0182,207.1298) node[above right] (text5071-8-1-1-0-4-2) {$J$};
  \path[draw=black,line join=miter,line cap=butt,miter limit=4.00,even odd
    rule,line width=0.374pt] (470.6272,163.0370) -- (355.9334,157.0415);
  \path[draw=black,line join=miter,line cap=butt,miter limit=4.00,even odd
    rule,line width=0.525pt] (471.1929,163.1780) -- (352.8597,145.0018);
  \path[draw=black,fill=cffff00,line join=round,line cap=round,miter
    limit=4.00,line width=0.468pt] (302.0634,169.4906) -- (298.0370,166.5789) --
    (294.0612,169.5869) -- (295.5861,164.8578) -- (291.4968,162.0061) --
    (296.4657,161.9950) -- (297.9141,157.2246) -- (299.4601,161.9469) --
    (304.4447,161.8503) -- (300.4313,164.7799) -- cycle;
  \path[draw=black,line join=miter,line cap=butt,miter limit=4.00,even odd
    rule,line width=0.374pt] (470.2096,163.0370) -- (340.4831,129.4814);
  \path[draw=black,line join=miter,line cap=butt,miter limit=4.00,even odd
    rule,line width=0.374pt] (469.7920,163.4545) -- (333.8018,111.1080);
  \path[fill=black,line join=miter,line cap=butt,line width=0.800pt]
    (543.1612,237.4229) node[above right] (text5071-8-1-15) {$\mathrm{optimal\
    model}$};
  \path[draw=black,line join=miter,line cap=butt,miter limit=4.00,even odd
    rule,line width=0.374pt] (469.5832,163.2457) -- (373.6804,172.2831);
  \path[draw=ce60000,dash pattern=on 7.48pt off 7.48pt,line join=miter,line
    cap=butt,miter limit=4.00,even odd rule,line width=0.935pt] (329.0357,92.3351)
    .. controls (329.0357,92.3351) and (339.0750,176.7828) .. (414.6645,189.1842);
  \path[draw=black,line join=miter,line cap=butt,miter limit=4.00,even odd
    rule,line width=0.374pt] (470.0008,163.4545) -- (395.8120,183.3489);
  \path[draw=black,fill=c6d5f41,line join=round,line cap=round,miter
    limit=4.00,fill opacity=0.294,line width=0.468pt] (349.2264,148.5114) --
    (345.2000,145.5997) -- (341.2242,148.6077) -- (342.7491,143.8786) --
    (338.6597,141.0269) -- (343.6286,141.0158) -- (345.0770,136.2454) --
    (346.6231,140.9676) -- (351.6076,140.8710) -- (347.5942,143.8007) -- cycle;
  \path[draw=black,fill=black,line join=miter,line cap=butt,miter limit=4.00,even
    odd rule,line width=0.702pt] (317.3167,166.7109) -- (308.2865,164.6766) --
    (317.3650,161.8784) -- cycle;
  \begin{scope}[cm={{0.60393,0.0,0.0,0.82271,(305.72027,-127.44605)}},miter limit=4.00,line width=0.532pt]
    \path[draw=black,line join=miter,line cap=butt,miter limit=4.00,even odd
      rule,line width=0.532pt] (185.6182,385.3343) -- (273.9914,353.2530);
  \end{scope}
  \path[draw=black,fill=black,line join=miter,line cap=butt,miter limit=4.00,even
    odd rule,line width=0.702pt] (361.6832,148.6811) -- (353.2094,144.9561) --
    (362.6556,143.9472) -- cycle;
  \path[draw=black,fill=black,line join=miter,line cap=butt,miter limit=4.00,even
    odd rule,line width=0.702pt] (337.0506,96.4519) -- (330.1288,90.3062) --
    (339.4289,92.2449) -- cycle;
  \path[draw=black,line join=miter,line cap=butt,miter limit=4.00,even odd
    rule,line width=0.375pt] (578.9062,198.9024) .. controls (578.9062,198.9024)
    and (601.3393,191.2015) .. (623.7723,196.2238) -- (614.3973,191.2015);
  \path[draw=black,line join=miter,line cap=butt,miter limit=4.00,even odd
    rule,line width=0.375pt] (578.2020,203.7422) .. controls (578.2020,203.7422)
    and (600.6351,196.0413) .. (623.0681,201.0636) -- (613.6931,196.0413);
  \path[draw=black,line join=miter,line cap=butt,miter limit=4.00,even odd
    rule,line width=0.375pt] (577.5324,208.0949) .. controls (577.5324,208.0949)
    and (599.9654,200.3940) .. (622.3985,205.4163) -- (613.0235,200.3940);
  \path[draw=black,line join=miter,line cap=butt,miter limit=4.00,even odd
    rule,line width=0.400pt] (492.8571,165.7551) .. controls (492.8571,165.7551)
    and (511.7010,168.8561) .. (516.9643,193.8801) .. controls (522.0472,218.0465)
    and (538.3929,234.0586) .. (538.3929,234.0586);
  \path[fill=black,line join=miter,line cap=butt,line width=0.800pt]
    (595.3260,218.3682) node[above right] (text5071-8-1-15-4) {$\mathrm{models}$};
  \path[fill=black,line join=miter,line cap=butt,line width=0.935pt]
    (356.8625,99.2712) node[above right] (text5071-8-1-6) {$\nabla J_{\mu}$};
  \path[fill=black,line join=miter,line cap=butt,line width=0.800pt]
    (505.0280,162.3332) node[above right] (text5071-9) {$J_{*}$};
  \path[fill=black,line join=miter,line cap=butt,line width=0.935pt]
    (334.9025,78.5123) node[above right] (text5071-8-1-6-7) {$(b)$};
  \path[fill=black,line join=miter,line cap=butt,line width=0.935pt]
    (141.6730,77.9330) node[above right] (text5071-8-1-6-7-3) {$(a)$};
  \path[draw=black,fill=black,line join=miter,line cap=butt,fill
    opacity=0.080,even odd rule,line width=0.270pt] (132.9796,149.4983) ..
    controls (132.9796,149.4983) and (146.1970,152.2149) .. (173.6339,149.6671) ..
    controls (198.7537,147.3344) and (207.7044,147.5003) .. (207.7044,147.5003) ..
    controls (207.7044,147.5003) and (208.1487,157.0591) .. (223.9500,164.1952) ..
    controls (244.0115,173.2551) and (250.9505,187.4137) .. (250.9505,187.4137) ..
    controls (250.9505,187.4137) and (234.7721,181.4464) .. (211.5777,182.4115) ..
    controls (180.6427,183.6986) and (157.5361,193.1825) .. (157.5361,193.1825) ..
    controls (157.5361,193.1825) and (145.3114,166.7998) .. (132.9796,149.4983) --
    cycle;
  \path[fill=cdb0000,miter limit=4.00,line width=1.020pt] (192.3482,164.6431)
    circle (0.0856cm);
  \path[draw=black,line join=miter,line cap=butt,miter limit=4.00,even odd
    rule,line width=0.450pt] (191.9944,164.8496) -- (211.8414,102.0860);
  \path[draw=black,fill=black,line join=miter,line cap=butt,miter limit=4.00,even
    odd rule,line width=0.942pt] (208.4015,107.5996) -- (211.5853,101.8435) --
    (211.6552,108.6125) -- cycle;
  \path[draw=black,line join=miter,line cap=butt,miter limit=4.00,even odd
    rule,line width=0.450pt] (191.5260,164.7506) -- (175.7894,132.2729);
  \path[fill=black,line join=miter,line cap=butt,line width=0.800pt]
    (155.6043,124.8996) node[above right] (text5071-8-1-1-0-4-7)
    {$\boldsymbol{u}_{i}^{(N+1)}$};
  \path[rotate=7.11657,fill=black,line join=miter,line cap=butt,miter
    limit=4.00,fill opacity=0.139,line width=0.383pt] (211.4123,141.3219) ellipse
    (0.6436cm and 0.2675cm);
  \path[draw=black,dash pattern=on 2.70pt off 2.70pt,line join=miter,line
    cap=butt,miter limit=4.00,even odd rule,line width=0.450pt]
    (175.7894,129.5943) -- (200.2314,137.2952);
  \path[draw=black,line join=miter,line cap=butt,miter limit=4.00,even odd
    rule,line width=0.300pt] (197.2165,136.8325) -- (195.9143,140.6206) --
    (199.1105,141.5676);
  \path[draw=black,line join=miter,line cap=butt,miter limit=4.00,even odd
    rule,line width=0.450pt] (176.6265,138.1323) -- (175.7894,132.1055) --
    (180.1421,136.6256);
  \path[draw=black,line join=miter,line cap=butt,miter limit=4.00,even odd
    rule,line width=0.300pt] (203.3933,151.3299) .. controls (203.3933,151.3299)
    and (208.4156,136.5977) .. (219.4647,136.2629) .. controls (230.5138,135.9281)
    and (239.2192,129.5665) .. (239.2192,129.5665);
  \path[fill=black,line join=miter,line cap=butt,line width=0.800pt]
    (228.6791,123.7434) node[above right] (text5071-9-7) {$\mathrm{Null\ space}$};
  \path[fill=black,line join=miter,line cap=butt,line width=0.800pt]
    (202.9348,92.6903) node[above right] (text5071-8-1-1-0-4-7-4)
    {$\boldsymbol{g}^{(N+1)}$};
  \path[draw=black,line join=miter,line cap=butt,miter limit=4.00,even odd
    rule,line width=0.300pt] (197.0936,153.2050) .. controls (197.0936,153.2050)
    and (205.1267,158.0068) .. (210.7974,167.4956) .. controls (215.6909,175.6840)
    and (213.9791,197.7329) .. (213.9791,197.7329);
  \path[fill=black,line join=miter,line cap=butt,line width=0.800pt]
    (207.3088,210.5971) node[above right] (text5071-9-7-1) {$\mathrm{Error}$};

\end{tikzpicture}

%% file: fig_11a.pgf
\begingroup%
\makeatletter%
\begin{pgfpicture}%
\pgfpathrectangle{\pgfpointorigin}{\pgfqpoint{6.088280in}{2.536783in}}%
\pgfusepath{use as bounding box, clip}%
\begin{pgfscope}%
\pgfsetbuttcap%
\pgfsetmiterjoin%
\definecolor{currentfill}{rgb}{1.000000,1.000000,1.000000}%
\pgfsetfillcolor{currentfill}%
\pgfsetlinewidth{0.000000pt}%
\definecolor{currentstroke}{rgb}{1.000000,1.000000,1.000000}%
\pgfsetstrokecolor{currentstroke}%
\pgfsetdash{}{0pt}%
\pgfpathmoveto{\pgfqpoint{0.000000in}{0.000000in}}%
\pgfpathlineto{\pgfqpoint{6.088280in}{0.000000in}}%
\pgfpathlineto{\pgfqpoint{6.088280in}{2.536783in}}%
\pgfpathlineto{\pgfqpoint{0.000000in}{2.536783in}}%
\pgfpathclose%
\pgfusepath{fill}%
\end{pgfscope}%
\begin{pgfscope}%
\pgfsetbuttcap%
\pgfsetmiterjoin%
\definecolor{currentfill}{rgb}{1.000000,1.000000,1.000000}%
\pgfsetfillcolor{currentfill}%
\pgfsetlinewidth{0.000000pt}%
\definecolor{currentstroke}{rgb}{0.000000,0.000000,0.000000}%
\pgfsetstrokecolor{currentstroke}%
\pgfsetstrokeopacity{0.000000}%
\pgfsetdash{}{0pt}%
\pgfpathmoveto{\pgfqpoint{0.654385in}{0.777990in}}%
\pgfpathlineto{\pgfqpoint{2.458063in}{0.777990in}}%
\pgfpathlineto{\pgfqpoint{2.458063in}{2.237794in}}%
\pgfpathlineto{\pgfqpoint{0.654385in}{2.237794in}}%
\pgfpathclose%
\pgfusepath{fill}%
\end{pgfscope}%
\begin{pgfscope}%
\pgfpathrectangle{\pgfqpoint{0.654385in}{0.777990in}}{\pgfqpoint{1.803678in}{1.459805in}} %
\pgfusepath{clip}%
\pgfsetbuttcap%
\pgfsetmiterjoin%
\definecolor{currentfill}{rgb}{1.000000,0.600000,0.000000}%
\pgfsetfillcolor{currentfill}%
\pgfsetlinewidth{0.401500pt}%
\definecolor{currentstroke}{rgb}{0.000000,0.000000,0.000000}%
\pgfsetstrokecolor{currentstroke}%
\pgfsetdash{}{0pt}%
\pgfpathmoveto{\pgfqpoint{0.822671in}{1.507892in}}%
\pgfpathlineto{\pgfqpoint{0.995271in}{1.507892in}}%
\pgfpathlineto{\pgfqpoint{0.995271in}{1.507892in}}%
\pgfpathlineto{\pgfqpoint{0.822671in}{1.507892in}}%
\pgfpathclose%
\pgfusepath{stroke,fill}%
\end{pgfscope}%
\begin{pgfscope}%
\pgfpathrectangle{\pgfqpoint{0.654385in}{0.777990in}}{\pgfqpoint{1.803678in}{1.459805in}} %
\pgfusepath{clip}%
\pgfsetbuttcap%
\pgfsetmiterjoin%
\definecolor{currentfill}{rgb}{1.000000,0.600000,0.000000}%
\pgfsetfillcolor{currentfill}%
\pgfsetlinewidth{0.401500pt}%
\definecolor{currentstroke}{rgb}{0.000000,0.000000,0.000000}%
\pgfsetstrokecolor{currentstroke}%
\pgfsetdash{}{0pt}%
\pgfpathmoveto{\pgfqpoint{1.254173in}{1.507892in}}%
\pgfpathlineto{\pgfqpoint{1.426773in}{1.507892in}}%
\pgfpathlineto{\pgfqpoint{1.426773in}{0.899640in}}%
\pgfpathlineto{\pgfqpoint{1.254173in}{0.899640in}}%
\pgfpathclose%
\pgfusepath{stroke,fill}%
\end{pgfscope}%
\begin{pgfscope}%
\pgfpathrectangle{\pgfqpoint{0.654385in}{0.777990in}}{\pgfqpoint{1.803678in}{1.459805in}} %
\pgfusepath{clip}%
\pgfsetbuttcap%
\pgfsetmiterjoin%
\definecolor{currentfill}{rgb}{1.000000,0.600000,0.000000}%
\pgfsetfillcolor{currentfill}%
\pgfsetlinewidth{0.401500pt}%
\definecolor{currentstroke}{rgb}{0.000000,0.000000,0.000000}%
\pgfsetstrokecolor{currentstroke}%
\pgfsetdash{}{0pt}%
\pgfpathmoveto{\pgfqpoint{1.685674in}{1.507892in}}%
\pgfpathlineto{\pgfqpoint{1.858275in}{1.507892in}}%
\pgfpathlineto{\pgfqpoint{1.858275in}{1.507892in}}%
\pgfpathlineto{\pgfqpoint{1.685674in}{1.507892in}}%
\pgfpathclose%
\pgfusepath{stroke,fill}%
\end{pgfscope}%
\begin{pgfscope}%
\pgfpathrectangle{\pgfqpoint{0.654385in}{0.777990in}}{\pgfqpoint{1.803678in}{1.459805in}} %
\pgfusepath{clip}%
\pgfsetbuttcap%
\pgfsetmiterjoin%
\definecolor{currentfill}{rgb}{1.000000,0.600000,0.000000}%
\pgfsetfillcolor{currentfill}%
\pgfsetlinewidth{0.401500pt}%
\definecolor{currentstroke}{rgb}{0.000000,0.000000,0.000000}%
\pgfsetstrokecolor{currentstroke}%
\pgfsetdash{}{0pt}%
\pgfpathmoveto{\pgfqpoint{2.117176in}{1.507892in}}%
\pgfpathlineto{\pgfqpoint{2.289777in}{1.507892in}}%
\pgfpathlineto{\pgfqpoint{2.289777in}{1.507892in}}%
\pgfpathlineto{\pgfqpoint{2.117176in}{1.507892in}}%
\pgfpathclose%
\pgfusepath{stroke,fill}%
\end{pgfscope}%
\begin{pgfscope}%
\pgfsetbuttcap%
\pgfsetroundjoin%
\definecolor{currentfill}{rgb}{0.000000,0.000000,0.000000}%
\pgfsetfillcolor{currentfill}%
\pgfsetlinewidth{0.803000pt}%
\definecolor{currentstroke}{rgb}{0.000000,0.000000,0.000000}%
\pgfsetstrokecolor{currentstroke}%
\pgfsetdash{}{0pt}%
\pgfsys@defobject{currentmarker}{\pgfqpoint{0.000000in}{-0.048611in}}{\pgfqpoint{0.000000in}{0.000000in}}{%
\pgfpathmoveto{\pgfqpoint{0.000000in}{0.000000in}}%
\pgfpathlineto{\pgfqpoint{0.000000in}{-0.048611in}}%
\pgfusepath{stroke,fill}%
}%
\begin{pgfscope}%
\pgfsys@transformshift{0.908971in}{0.777990in}%
\pgfsys@useobject{currentmarker}{}%
\end{pgfscope}%
\end{pgfscope}%
\begin{pgfscope}%
\pgftext[x=0.937128in,y=0.298211in,left,base,rotate=90.000000]{\rmfamily\fontsize{8.000000}{9.600000}\selectfont \(\displaystyle (2,0,0)\)}%
\end{pgfscope}%
\begin{pgfscope}%
\pgfsetbuttcap%
\pgfsetroundjoin%
\definecolor{currentfill}{rgb}{0.000000,0.000000,0.000000}%
\pgfsetfillcolor{currentfill}%
\pgfsetlinewidth{0.803000pt}%
\definecolor{currentstroke}{rgb}{0.000000,0.000000,0.000000}%
\pgfsetstrokecolor{currentstroke}%
\pgfsetdash{}{0pt}%
\pgfsys@defobject{currentmarker}{\pgfqpoint{0.000000in}{-0.048611in}}{\pgfqpoint{0.000000in}{0.000000in}}{%
\pgfpathmoveto{\pgfqpoint{0.000000in}{0.000000in}}%
\pgfpathlineto{\pgfqpoint{0.000000in}{-0.048611in}}%
\pgfusepath{stroke,fill}%
}%
\begin{pgfscope}%
\pgfsys@transformshift{1.340473in}{0.777990in}%
\pgfsys@useobject{currentmarker}{}%
\end{pgfscope}%
\end{pgfscope}%
\begin{pgfscope}%
\pgftext[x=1.368630in,y=0.298211in,left,base,rotate=90.000000]{\rmfamily\fontsize{8.000000}{9.600000}\selectfont \(\displaystyle (1,1,0)\)}%
\end{pgfscope}%
\begin{pgfscope}%
\pgfsetbuttcap%
\pgfsetroundjoin%
\definecolor{currentfill}{rgb}{0.000000,0.000000,0.000000}%
\pgfsetfillcolor{currentfill}%
\pgfsetlinewidth{0.803000pt}%
\definecolor{currentstroke}{rgb}{0.000000,0.000000,0.000000}%
\pgfsetstrokecolor{currentstroke}%
\pgfsetdash{}{0pt}%
\pgfsys@defobject{currentmarker}{\pgfqpoint{0.000000in}{-0.048611in}}{\pgfqpoint{0.000000in}{0.000000in}}{%
\pgfpathmoveto{\pgfqpoint{0.000000in}{0.000000in}}%
\pgfpathlineto{\pgfqpoint{0.000000in}{-0.048611in}}%
\pgfusepath{stroke,fill}%
}%
\begin{pgfscope}%
\pgfsys@transformshift{1.771975in}{0.777990in}%
\pgfsys@useobject{currentmarker}{}%
\end{pgfscope}%
\end{pgfscope}%
\begin{pgfscope}%
\pgftext[x=1.800132in,y=0.298211in,left,base,rotate=90.000000]{\rmfamily\fontsize{8.000000}{9.600000}\selectfont \(\displaystyle (0,2,0)\)}%
\end{pgfscope}%
\begin{pgfscope}%
\pgfsetbuttcap%
\pgfsetroundjoin%
\definecolor{currentfill}{rgb}{0.000000,0.000000,0.000000}%
\pgfsetfillcolor{currentfill}%
\pgfsetlinewidth{0.803000pt}%
\definecolor{currentstroke}{rgb}{0.000000,0.000000,0.000000}%
\pgfsetstrokecolor{currentstroke}%
\pgfsetdash{}{0pt}%
\pgfsys@defobject{currentmarker}{\pgfqpoint{0.000000in}{-0.048611in}}{\pgfqpoint{0.000000in}{0.000000in}}{%
\pgfpathmoveto{\pgfqpoint{0.000000in}{0.000000in}}%
\pgfpathlineto{\pgfqpoint{0.000000in}{-0.048611in}}%
\pgfusepath{stroke,fill}%
}%
\begin{pgfscope}%
\pgfsys@transformshift{2.203477in}{0.777990in}%
\pgfsys@useobject{currentmarker}{}%
\end{pgfscope}%
\end{pgfscope}%
\begin{pgfscope}%
\pgftext[x=2.231633in,y=0.298211in,left,base,rotate=90.000000]{\rmfamily\fontsize{8.000000}{9.600000}\selectfont \(\displaystyle (0,0,2)\)}%
\end{pgfscope}%
\begin{pgfscope}%
\pgftext[x=1.556224in,y=0.242655in,,top]{\rmfamily\fontsize{10.000000}{12.000000}\selectfont \(\displaystyle \alpha\)}%
\end{pgfscope}%
\begin{pgfscope}%
\pgfsetbuttcap%
\pgfsetroundjoin%
\definecolor{currentfill}{rgb}{0.000000,0.000000,0.000000}%
\pgfsetfillcolor{currentfill}%
\pgfsetlinewidth{0.803000pt}%
\definecolor{currentstroke}{rgb}{0.000000,0.000000,0.000000}%
\pgfsetstrokecolor{currentstroke}%
\pgfsetdash{}{0pt}%
\pgfsys@defobject{currentmarker}{\pgfqpoint{-0.048611in}{0.000000in}}{\pgfqpoint{0.000000in}{0.000000in}}{%
\pgfpathmoveto{\pgfqpoint{0.000000in}{0.000000in}}%
\pgfpathlineto{\pgfqpoint{-0.048611in}{0.000000in}}%
\pgfusepath{stroke,fill}%
}%
\begin{pgfscope}%
\pgfsys@transformshift{0.654385in}{0.859277in}%
\pgfsys@useobject{currentmarker}{}%
\end{pgfscope}%
\end{pgfscope}%
\begin{pgfscope}%
\pgftext[x=0.335909in,y=0.821015in,left,base]{\rmfamily\fontsize{8.000000}{9.600000}\selectfont \(\displaystyle -0.4\)}%
\end{pgfscope}%
\begin{pgfscope}%
\pgfsetbuttcap%
\pgfsetroundjoin%
\definecolor{currentfill}{rgb}{0.000000,0.000000,0.000000}%
\pgfsetfillcolor{currentfill}%
\pgfsetlinewidth{0.803000pt}%
\definecolor{currentstroke}{rgb}{0.000000,0.000000,0.000000}%
\pgfsetstrokecolor{currentstroke}%
\pgfsetdash{}{0pt}%
\pgfsys@defobject{currentmarker}{\pgfqpoint{-0.048611in}{0.000000in}}{\pgfqpoint{0.000000in}{0.000000in}}{%
\pgfpathmoveto{\pgfqpoint{0.000000in}{0.000000in}}%
\pgfpathlineto{\pgfqpoint{-0.048611in}{0.000000in}}%
\pgfusepath{stroke,fill}%
}%
\begin{pgfscope}%
\pgfsys@transformshift{0.654385in}{1.183585in}%
\pgfsys@useobject{currentmarker}{}%
\end{pgfscope}%
\end{pgfscope}%
\begin{pgfscope}%
\pgftext[x=0.335909in,y=1.145322in,left,base]{\rmfamily\fontsize{8.000000}{9.600000}\selectfont \(\displaystyle -0.2\)}%
\end{pgfscope}%
\begin{pgfscope}%
\pgfsetbuttcap%
\pgfsetroundjoin%
\definecolor{currentfill}{rgb}{0.000000,0.000000,0.000000}%
\pgfsetfillcolor{currentfill}%
\pgfsetlinewidth{0.803000pt}%
\definecolor{currentstroke}{rgb}{0.000000,0.000000,0.000000}%
\pgfsetstrokecolor{currentstroke}%
\pgfsetdash{}{0pt}%
\pgfsys@defobject{currentmarker}{\pgfqpoint{-0.048611in}{0.000000in}}{\pgfqpoint{0.000000in}{0.000000in}}{%
\pgfpathmoveto{\pgfqpoint{0.000000in}{0.000000in}}%
\pgfpathlineto{\pgfqpoint{-0.048611in}{0.000000in}}%
\pgfusepath{stroke,fill}%
}%
\begin{pgfscope}%
\pgfsys@transformshift{0.654385in}{1.507892in}%
\pgfsys@useobject{currentmarker}{}%
\end{pgfscope}%
\end{pgfscope}%
\begin{pgfscope}%
\pgftext[x=0.406312in,y=1.469630in,left,base]{\rmfamily\fontsize{8.000000}{9.600000}\selectfont \(\displaystyle 0.0\)}%
\end{pgfscope}%
\begin{pgfscope}%
\pgfsetbuttcap%
\pgfsetroundjoin%
\definecolor{currentfill}{rgb}{0.000000,0.000000,0.000000}%
\pgfsetfillcolor{currentfill}%
\pgfsetlinewidth{0.803000pt}%
\definecolor{currentstroke}{rgb}{0.000000,0.000000,0.000000}%
\pgfsetstrokecolor{currentstroke}%
\pgfsetdash{}{0pt}%
\pgfsys@defobject{currentmarker}{\pgfqpoint{-0.048611in}{0.000000in}}{\pgfqpoint{0.000000in}{0.000000in}}{%
\pgfpathmoveto{\pgfqpoint{0.000000in}{0.000000in}}%
\pgfpathlineto{\pgfqpoint{-0.048611in}{0.000000in}}%
\pgfusepath{stroke,fill}%
}%
\begin{pgfscope}%
\pgfsys@transformshift{0.654385in}{1.832200in}%
\pgfsys@useobject{currentmarker}{}%
\end{pgfscope}%
\end{pgfscope}%
\begin{pgfscope}%
\pgftext[x=0.406312in,y=1.793937in,left,base]{\rmfamily\fontsize{8.000000}{9.600000}\selectfont \(\displaystyle 0.2\)}%
\end{pgfscope}%
\begin{pgfscope}%
\pgfsetbuttcap%
\pgfsetroundjoin%
\definecolor{currentfill}{rgb}{0.000000,0.000000,0.000000}%
\pgfsetfillcolor{currentfill}%
\pgfsetlinewidth{0.803000pt}%
\definecolor{currentstroke}{rgb}{0.000000,0.000000,0.000000}%
\pgfsetstrokecolor{currentstroke}%
\pgfsetdash{}{0pt}%
\pgfsys@defobject{currentmarker}{\pgfqpoint{-0.048611in}{0.000000in}}{\pgfqpoint{0.000000in}{0.000000in}}{%
\pgfpathmoveto{\pgfqpoint{0.000000in}{0.000000in}}%
\pgfpathlineto{\pgfqpoint{-0.048611in}{0.000000in}}%
\pgfusepath{stroke,fill}%
}%
\begin{pgfscope}%
\pgfsys@transformshift{0.654385in}{2.156507in}%
\pgfsys@useobject{currentmarker}{}%
\end{pgfscope}%
\end{pgfscope}%
\begin{pgfscope}%
\pgftext[x=0.406312in,y=2.118245in,left,base]{\rmfamily\fontsize{8.000000}{9.600000}\selectfont \(\displaystyle 0.4\)}%
\end{pgfscope}%
\begin{pgfscope}%
\pgftext[x=0.280353in,y=1.507892in,,bottom,rotate=90.000000]{\rmfamily\fontsize{10.000000}{12.000000}\selectfont \(\displaystyle \boldsymbol{g}^{(2)}\)}%
\end{pgfscope}%
\begin{pgfscope}%
\pgfsetrectcap%
\pgfsetmiterjoin%
\pgfsetlinewidth{0.803000pt}%
\definecolor{currentstroke}{rgb}{0.000000,0.000000,0.000000}%
\pgfsetstrokecolor{currentstroke}%
\pgfsetdash{}{0pt}%
\pgfpathmoveto{\pgfqpoint{0.654385in}{0.777990in}}%
\pgfpathlineto{\pgfqpoint{0.654385in}{2.237794in}}%
\pgfusepath{stroke}%
\end{pgfscope}%
\begin{pgfscope}%
\pgfsetrectcap%
\pgfsetmiterjoin%
\pgfsetlinewidth{0.803000pt}%
\definecolor{currentstroke}{rgb}{0.000000,0.000000,0.000000}%
\pgfsetstrokecolor{currentstroke}%
\pgfsetdash{}{0pt}%
\pgfpathmoveto{\pgfqpoint{2.458063in}{0.777990in}}%
\pgfpathlineto{\pgfqpoint{2.458063in}{2.237794in}}%
\pgfusepath{stroke}%
\end{pgfscope}%
\begin{pgfscope}%
\pgfsetrectcap%
\pgfsetmiterjoin%
\pgfsetlinewidth{0.803000pt}%
\definecolor{currentstroke}{rgb}{0.000000,0.000000,0.000000}%
\pgfsetstrokecolor{currentstroke}%
\pgfsetdash{}{0pt}%
\pgfpathmoveto{\pgfqpoint{0.654385in}{0.777990in}}%
\pgfpathlineto{\pgfqpoint{2.458063in}{0.777990in}}%
\pgfusepath{stroke}%
\end{pgfscope}%
\begin{pgfscope}%
\pgfsetrectcap%
\pgfsetmiterjoin%
\pgfsetlinewidth{0.803000pt}%
\definecolor{currentstroke}{rgb}{0.000000,0.000000,0.000000}%
\pgfsetstrokecolor{currentstroke}%
\pgfsetdash{}{0pt}%
\pgfpathmoveto{\pgfqpoint{0.654385in}{2.237794in}}%
\pgfpathlineto{\pgfqpoint{2.458063in}{2.237794in}}%
\pgfusepath{stroke}%
\end{pgfscope}%
\begin{pgfscope}%
\pgfpathrectangle{\pgfqpoint{0.654385in}{0.777990in}}{\pgfqpoint{1.803678in}{1.459805in}} %
\pgfusepath{clip}%
\pgfsetbuttcap%
\pgfsetmiterjoin%
\definecolor{currentfill}{rgb}{0.960000,0.960000,0.960000}%
\pgfsetfillcolor{currentfill}%
\pgfsetfillopacity{0.600000}%
\pgfsetlinewidth{0.401500pt}%
\definecolor{currentstroke}{rgb}{0.000000,0.000000,0.000000}%
\pgfsetstrokecolor{currentstroke}%
\pgfsetstrokeopacity{0.600000}%
\pgfsetdash{}{0pt}%
\pgfpathmoveto{\pgfqpoint{0.736370in}{1.507892in}}%
\pgfpathlineto{\pgfqpoint{1.081572in}{1.507892in}}%
\pgfpathlineto{\pgfqpoint{1.081572in}{1.919830in}}%
\pgfpathlineto{\pgfqpoint{0.736370in}{1.919830in}}%
\pgfpathclose%
\pgfusepath{stroke,fill}%
\end{pgfscope}%
\begin{pgfscope}%
\pgfpathrectangle{\pgfqpoint{0.654385in}{0.777990in}}{\pgfqpoint{1.803678in}{1.459805in}} %
\pgfusepath{clip}%
\pgfsetbuttcap%
\pgfsetmiterjoin%
\definecolor{currentfill}{rgb}{0.960000,0.960000,0.960000}%
\pgfsetfillcolor{currentfill}%
\pgfsetfillopacity{0.600000}%
\pgfsetlinewidth{0.401500pt}%
\definecolor{currentstroke}{rgb}{0.000000,0.000000,0.000000}%
\pgfsetstrokecolor{currentstroke}%
\pgfsetstrokeopacity{0.600000}%
\pgfsetdash{}{0pt}%
\pgfpathmoveto{\pgfqpoint{1.167872in}{1.507892in}}%
\pgfpathlineto{\pgfqpoint{1.513074in}{1.507892in}}%
\pgfpathlineto{\pgfqpoint{1.513074in}{1.919455in}}%
\pgfpathlineto{\pgfqpoint{1.167872in}{1.919455in}}%
\pgfpathclose%
\pgfusepath{stroke,fill}%
\end{pgfscope}%
\begin{pgfscope}%
\pgfpathrectangle{\pgfqpoint{0.654385in}{0.777990in}}{\pgfqpoint{1.803678in}{1.459805in}} %
\pgfusepath{clip}%
\pgfsetbuttcap%
\pgfsetmiterjoin%
\definecolor{currentfill}{rgb}{0.960000,0.960000,0.960000}%
\pgfsetfillcolor{currentfill}%
\pgfsetfillopacity{0.600000}%
\pgfsetlinewidth{0.401500pt}%
\definecolor{currentstroke}{rgb}{0.000000,0.000000,0.000000}%
\pgfsetstrokecolor{currentstroke}%
\pgfsetstrokeopacity{0.600000}%
\pgfsetdash{}{0pt}%
\pgfpathmoveto{\pgfqpoint{1.599374in}{1.507892in}}%
\pgfpathlineto{\pgfqpoint{1.944576in}{1.507892in}}%
\pgfpathlineto{\pgfqpoint{1.944576in}{2.116144in}}%
\pgfpathlineto{\pgfqpoint{1.599374in}{2.116144in}}%
\pgfpathclose%
\pgfusepath{stroke,fill}%
\end{pgfscope}%
\begin{pgfscope}%
\pgfpathrectangle{\pgfqpoint{0.654385in}{0.777990in}}{\pgfqpoint{1.803678in}{1.459805in}} %
\pgfusepath{clip}%
\pgfsetbuttcap%
\pgfsetmiterjoin%
\definecolor{currentfill}{rgb}{0.960000,0.960000,0.960000}%
\pgfsetfillcolor{currentfill}%
\pgfsetfillopacity{0.600000}%
\pgfsetlinewidth{0.401500pt}%
\definecolor{currentstroke}{rgb}{0.000000,0.000000,0.000000}%
\pgfsetstrokecolor{currentstroke}%
\pgfsetstrokeopacity{0.600000}%
\pgfsetdash{}{0pt}%
\pgfpathmoveto{\pgfqpoint{2.030876in}{1.507892in}}%
\pgfpathlineto{\pgfqpoint{2.376077in}{1.507892in}}%
\pgfpathlineto{\pgfqpoint{2.376077in}{1.957053in}}%
\pgfpathlineto{\pgfqpoint{2.030876in}{1.957053in}}%
\pgfpathclose%
\pgfusepath{stroke,fill}%
\end{pgfscope}%
\begin{pgfscope}%
\pgfsetbuttcap%
\pgfsetroundjoin%
\definecolor{currentfill}{rgb}{0.000000,0.000000,0.000000}%
\pgfsetfillcolor{currentfill}%
\pgfsetlinewidth{0.803000pt}%
\definecolor{currentstroke}{rgb}{0.000000,0.000000,0.000000}%
\pgfsetstrokecolor{currentstroke}%
\pgfsetdash{}{0pt}%
\pgfsys@defobject{currentmarker}{\pgfqpoint{0.000000in}{0.000000in}}{\pgfqpoint{0.048611in}{0.000000in}}{%
\pgfpathmoveto{\pgfqpoint{0.000000in}{0.000000in}}%
\pgfpathlineto{\pgfqpoint{0.048611in}{0.000000in}}%
\pgfusepath{stroke,fill}%
}%
\begin{pgfscope}%
\pgfsys@transformshift{2.458063in}{0.892387in}%
\pgfsys@useobject{currentmarker}{}%
\end{pgfscope}%
\end{pgfscope}%
\begin{pgfscope}%
\pgftext[x=2.555285in,y=0.854125in,left,base]{\rmfamily\fontsize{8.000000}{9.600000}\selectfont \(\displaystyle -4\)}%
\end{pgfscope}%
\begin{pgfscope}%
\pgfsetbuttcap%
\pgfsetroundjoin%
\definecolor{currentfill}{rgb}{0.000000,0.000000,0.000000}%
\pgfsetfillcolor{currentfill}%
\pgfsetlinewidth{0.803000pt}%
\definecolor{currentstroke}{rgb}{0.000000,0.000000,0.000000}%
\pgfsetstrokecolor{currentstroke}%
\pgfsetdash{}{0pt}%
\pgfsys@defobject{currentmarker}{\pgfqpoint{0.000000in}{0.000000in}}{\pgfqpoint{0.048611in}{0.000000in}}{%
\pgfpathmoveto{\pgfqpoint{0.000000in}{0.000000in}}%
\pgfpathlineto{\pgfqpoint{0.048611in}{0.000000in}}%
\pgfusepath{stroke,fill}%
}%
\begin{pgfscope}%
\pgfsys@transformshift{2.458063in}{1.200140in}%
\pgfsys@useobject{currentmarker}{}%
\end{pgfscope}%
\end{pgfscope}%
\begin{pgfscope}%
\pgftext[x=2.555285in,y=1.161877in,left,base]{\rmfamily\fontsize{8.000000}{9.600000}\selectfont \(\displaystyle -2\)}%
\end{pgfscope}%
\begin{pgfscope}%
\pgfsetbuttcap%
\pgfsetroundjoin%
\definecolor{currentfill}{rgb}{0.000000,0.000000,0.000000}%
\pgfsetfillcolor{currentfill}%
\pgfsetlinewidth{0.803000pt}%
\definecolor{currentstroke}{rgb}{0.000000,0.000000,0.000000}%
\pgfsetstrokecolor{currentstroke}%
\pgfsetdash{}{0pt}%
\pgfsys@defobject{currentmarker}{\pgfqpoint{0.000000in}{0.000000in}}{\pgfqpoint{0.048611in}{0.000000in}}{%
\pgfpathmoveto{\pgfqpoint{0.000000in}{0.000000in}}%
\pgfpathlineto{\pgfqpoint{0.048611in}{0.000000in}}%
\pgfusepath{stroke,fill}%
}%
\begin{pgfscope}%
\pgfsys@transformshift{2.458063in}{1.507892in}%
\pgfsys@useobject{currentmarker}{}%
\end{pgfscope}%
\end{pgfscope}%
\begin{pgfscope}%
\pgftext[x=2.555285in,y=1.469630in,left,base]{\rmfamily\fontsize{8.000000}{9.600000}\selectfont \(\displaystyle 0\)}%
\end{pgfscope}%
\begin{pgfscope}%
\pgfsetbuttcap%
\pgfsetroundjoin%
\definecolor{currentfill}{rgb}{0.000000,0.000000,0.000000}%
\pgfsetfillcolor{currentfill}%
\pgfsetlinewidth{0.803000pt}%
\definecolor{currentstroke}{rgb}{0.000000,0.000000,0.000000}%
\pgfsetstrokecolor{currentstroke}%
\pgfsetdash{}{0pt}%
\pgfsys@defobject{currentmarker}{\pgfqpoint{0.000000in}{0.000000in}}{\pgfqpoint{0.048611in}{0.000000in}}{%
\pgfpathmoveto{\pgfqpoint{0.000000in}{0.000000in}}%
\pgfpathlineto{\pgfqpoint{0.048611in}{0.000000in}}%
\pgfusepath{stroke,fill}%
}%
\begin{pgfscope}%
\pgfsys@transformshift{2.458063in}{1.815645in}%
\pgfsys@useobject{currentmarker}{}%
\end{pgfscope}%
\end{pgfscope}%
\begin{pgfscope}%
\pgftext[x=2.555285in,y=1.777383in,left,base]{\rmfamily\fontsize{8.000000}{9.600000}\selectfont \(\displaystyle 2\)}%
\end{pgfscope}%
\begin{pgfscope}%
\pgfsetbuttcap%
\pgfsetroundjoin%
\definecolor{currentfill}{rgb}{0.000000,0.000000,0.000000}%
\pgfsetfillcolor{currentfill}%
\pgfsetlinewidth{0.803000pt}%
\definecolor{currentstroke}{rgb}{0.000000,0.000000,0.000000}%
\pgfsetstrokecolor{currentstroke}%
\pgfsetdash{}{0pt}%
\pgfsys@defobject{currentmarker}{\pgfqpoint{0.000000in}{0.000000in}}{\pgfqpoint{0.048611in}{0.000000in}}{%
\pgfpathmoveto{\pgfqpoint{0.000000in}{0.000000in}}%
\pgfpathlineto{\pgfqpoint{0.048611in}{0.000000in}}%
\pgfusepath{stroke,fill}%
}%
\begin{pgfscope}%
\pgfsys@transformshift{2.458063in}{2.123397in}%
\pgfsys@useobject{currentmarker}{}%
\end{pgfscope}%
\end{pgfscope}%
\begin{pgfscope}%
\pgftext[x=2.555285in,y=2.085135in,left,base]{\rmfamily\fontsize{8.000000}{9.600000}\selectfont \(\displaystyle 4\)}%
\end{pgfscope}%
\begin{pgfscope}%
\pgftext[x=2.740272in,y=1.507892in,,top,rotate=90.000000]{\rmfamily\fontsize{10.000000}{12.000000}\selectfont \(\displaystyle \boldsymbol{u}_{3}^{(2)}\)}%
\end{pgfscope}%
\begin{pgfscope}%
\pgfsetrectcap%
\pgfsetmiterjoin%
\pgfsetlinewidth{0.803000pt}%
\definecolor{currentstroke}{rgb}{0.000000,0.000000,0.000000}%
\pgfsetstrokecolor{currentstroke}%
\pgfsetdash{}{0pt}%
\pgfpathmoveto{\pgfqpoint{0.654385in}{0.777990in}}%
\pgfpathlineto{\pgfqpoint{0.654385in}{2.237794in}}%
\pgfusepath{stroke}%
\end{pgfscope}%
\begin{pgfscope}%
\pgfsetrectcap%
\pgfsetmiterjoin%
\pgfsetlinewidth{0.803000pt}%
\definecolor{currentstroke}{rgb}{0.000000,0.000000,0.000000}%
\pgfsetstrokecolor{currentstroke}%
\pgfsetdash{}{0pt}%
\pgfpathmoveto{\pgfqpoint{2.458063in}{0.777990in}}%
\pgfpathlineto{\pgfqpoint{2.458063in}{2.237794in}}%
\pgfusepath{stroke}%
\end{pgfscope}%
\begin{pgfscope}%
\pgfsetrectcap%
\pgfsetmiterjoin%
\pgfsetlinewidth{0.803000pt}%
\definecolor{currentstroke}{rgb}{0.000000,0.000000,0.000000}%
\pgfsetstrokecolor{currentstroke}%
\pgfsetdash{}{0pt}%
\pgfpathmoveto{\pgfqpoint{0.654385in}{0.777990in}}%
\pgfpathlineto{\pgfqpoint{2.458063in}{0.777990in}}%
\pgfusepath{stroke}%
\end{pgfscope}%
\begin{pgfscope}%
\pgfsetrectcap%
\pgfsetmiterjoin%
\pgfsetlinewidth{0.803000pt}%
\definecolor{currentstroke}{rgb}{0.000000,0.000000,0.000000}%
\pgfsetstrokecolor{currentstroke}%
\pgfsetdash{}{0pt}%
\pgfpathmoveto{\pgfqpoint{0.654385in}{2.237794in}}%
\pgfpathlineto{\pgfqpoint{2.458063in}{2.237794in}}%
\pgfusepath{stroke}%
\end{pgfscope}%
\begin{pgfscope}%
\pgftext[x=1.556224in,y=2.321128in,,base]{\rmfamily\fontsize{9.600000}{11.520000}\selectfont \(\displaystyle N = 1\)}%
\end{pgfscope}%
\begin{pgfscope}%
\pgfsetbuttcap%
\pgfsetmiterjoin%
\definecolor{currentfill}{rgb}{1.000000,0.600000,0.000000}%
\pgfsetfillcolor{currentfill}%
\pgfsetlinewidth{0.401500pt}%
\definecolor{currentstroke}{rgb}{0.000000,0.000000,0.000000}%
\pgfsetstrokecolor{currentstroke}%
\pgfsetdash{}{0pt}%
\pgfpathmoveto{\pgfqpoint{1.821149in}{1.114377in}}%
\pgfpathlineto{\pgfqpoint{2.043371in}{1.114377in}}%
\pgfpathlineto{\pgfqpoint{2.043371in}{1.192155in}}%
\pgfpathlineto{\pgfqpoint{1.821149in}{1.192155in}}%
\pgfpathclose%
\pgfusepath{stroke,fill}%
\end{pgfscope}%
\begin{pgfscope}%
\pgftext[x=2.132260in,y=1.114377in,left,base]{\rmfamily\fontsize{8.000000}{9.600000}\selectfont \(\displaystyle \boldsymbol{g}^{(2)}\)}%
\end{pgfscope}%
\begin{pgfscope}%
\pgfsetbuttcap%
\pgfsetmiterjoin%
\definecolor{currentfill}{rgb}{0.960000,0.960000,0.960000}%
\pgfsetfillcolor{currentfill}%
\pgfsetfillopacity{0.600000}%
\pgfsetlinewidth{0.401500pt}%
\definecolor{currentstroke}{rgb}{0.000000,0.000000,0.000000}%
\pgfsetstrokecolor{currentstroke}%
\pgfsetstrokeopacity{0.600000}%
\pgfsetdash{}{0pt}%
\pgfpathmoveto{\pgfqpoint{1.821149in}{0.915950in}}%
\pgfpathlineto{\pgfqpoint{2.043371in}{0.915950in}}%
\pgfpathlineto{\pgfqpoint{2.043371in}{0.993727in}}%
\pgfpathlineto{\pgfqpoint{1.821149in}{0.993727in}}%
\pgfpathclose%
\pgfusepath{stroke,fill}%
\end{pgfscope}%
\begin{pgfscope}%
\pgftext[x=2.132260in,y=0.915950in,left,base]{\rmfamily\fontsize{8.000000}{9.600000}\selectfont \(\displaystyle \boldsymbol{u}_{3}^{(2)}\)}%
\end{pgfscope}%
\begin{pgfscope}%
\pgfsetbuttcap%
\pgfsetmiterjoin%
\definecolor{currentfill}{rgb}{1.000000,1.000000,1.000000}%
\pgfsetfillcolor{currentfill}%
\pgfsetlinewidth{0.000000pt}%
\definecolor{currentstroke}{rgb}{0.000000,0.000000,0.000000}%
\pgfsetstrokecolor{currentstroke}%
\pgfsetstrokeopacity{0.000000}%
\pgfsetdash{}{0pt}%
\pgfpathmoveto{\pgfqpoint{3.638525in}{0.777990in}}%
\pgfpathlineto{\pgfqpoint{5.442203in}{0.777990in}}%
\pgfpathlineto{\pgfqpoint{5.442203in}{2.237794in}}%
\pgfpathlineto{\pgfqpoint{3.638525in}{2.237794in}}%
\pgfpathclose%
\pgfusepath{fill}%
\end{pgfscope}%
\begin{pgfscope}%
\pgfpathrectangle{\pgfqpoint{3.638525in}{0.777990in}}{\pgfqpoint{1.803678in}{1.459805in}} %
\pgfusepath{clip}%
\pgfsetbuttcap%
\pgfsetmiterjoin%
\definecolor{currentfill}{rgb}{1.000000,0.600000,0.000000}%
\pgfsetfillcolor{currentfill}%
\pgfsetlinewidth{0.401500pt}%
\definecolor{currentstroke}{rgb}{0.000000,0.000000,0.000000}%
\pgfsetstrokecolor{currentstroke}%
\pgfsetdash{}{0pt}%
\pgfpathmoveto{\pgfqpoint{3.806811in}{1.507892in}}%
\pgfpathlineto{\pgfqpoint{3.979412in}{1.507892in}}%
\pgfpathlineto{\pgfqpoint{3.979412in}{1.568714in}}%
\pgfpathlineto{\pgfqpoint{3.806811in}{1.568714in}}%
\pgfpathclose%
\pgfusepath{stroke,fill}%
\end{pgfscope}%
\begin{pgfscope}%
\pgfpathrectangle{\pgfqpoint{3.638525in}{0.777990in}}{\pgfqpoint{1.803678in}{1.459805in}} %
\pgfusepath{clip}%
\pgfsetbuttcap%
\pgfsetmiterjoin%
\definecolor{currentfill}{rgb}{1.000000,0.600000,0.000000}%
\pgfsetfillcolor{currentfill}%
\pgfsetlinewidth{0.401500pt}%
\definecolor{currentstroke}{rgb}{0.000000,0.000000,0.000000}%
\pgfsetstrokecolor{currentstroke}%
\pgfsetdash{}{0pt}%
\pgfpathmoveto{\pgfqpoint{4.238313in}{1.507892in}}%
\pgfpathlineto{\pgfqpoint{4.410913in}{1.507892in}}%
\pgfpathlineto{\pgfqpoint{4.410913in}{2.116144in}}%
\pgfpathlineto{\pgfqpoint{4.238313in}{2.116144in}}%
\pgfpathclose%
\pgfusepath{stroke,fill}%
\end{pgfscope}%
\begin{pgfscope}%
\pgfpathrectangle{\pgfqpoint{3.638525in}{0.777990in}}{\pgfqpoint{1.803678in}{1.459805in}} %
\pgfusepath{clip}%
\pgfsetbuttcap%
\pgfsetmiterjoin%
\definecolor{currentfill}{rgb}{1.000000,0.600000,0.000000}%
\pgfsetfillcolor{currentfill}%
\pgfsetlinewidth{0.401500pt}%
\definecolor{currentstroke}{rgb}{0.000000,0.000000,0.000000}%
\pgfsetstrokecolor{currentstroke}%
\pgfsetdash{}{0pt}%
\pgfpathmoveto{\pgfqpoint{4.669814in}{1.507892in}}%
\pgfpathlineto{\pgfqpoint{4.842415in}{1.507892in}}%
\pgfpathlineto{\pgfqpoint{4.842415in}{1.507892in}}%
\pgfpathlineto{\pgfqpoint{4.669814in}{1.507892in}}%
\pgfpathclose%
\pgfusepath{stroke,fill}%
\end{pgfscope}%
\begin{pgfscope}%
\pgfpathrectangle{\pgfqpoint{3.638525in}{0.777990in}}{\pgfqpoint{1.803678in}{1.459805in}} %
\pgfusepath{clip}%
\pgfsetbuttcap%
\pgfsetmiterjoin%
\definecolor{currentfill}{rgb}{1.000000,0.600000,0.000000}%
\pgfsetfillcolor{currentfill}%
\pgfsetlinewidth{0.401500pt}%
\definecolor{currentstroke}{rgb}{0.000000,0.000000,0.000000}%
\pgfsetstrokecolor{currentstroke}%
\pgfsetdash{}{0pt}%
\pgfpathmoveto{\pgfqpoint{5.101316in}{1.507892in}}%
\pgfpathlineto{\pgfqpoint{5.273917in}{1.507892in}}%
\pgfpathlineto{\pgfqpoint{5.273917in}{1.507892in}}%
\pgfpathlineto{\pgfqpoint{5.101316in}{1.507892in}}%
\pgfpathclose%
\pgfusepath{stroke,fill}%
\end{pgfscope}%
\begin{pgfscope}%
\pgfsetbuttcap%
\pgfsetroundjoin%
\definecolor{currentfill}{rgb}{0.000000,0.000000,0.000000}%
\pgfsetfillcolor{currentfill}%
\pgfsetlinewidth{0.803000pt}%
\definecolor{currentstroke}{rgb}{0.000000,0.000000,0.000000}%
\pgfsetstrokecolor{currentstroke}%
\pgfsetdash{}{0pt}%
\pgfsys@defobject{currentmarker}{\pgfqpoint{0.000000in}{-0.048611in}}{\pgfqpoint{0.000000in}{0.000000in}}{%
\pgfpathmoveto{\pgfqpoint{0.000000in}{0.000000in}}%
\pgfpathlineto{\pgfqpoint{0.000000in}{-0.048611in}}%
\pgfusepath{stroke,fill}%
}%
\begin{pgfscope}%
\pgfsys@transformshift{3.893111in}{0.777990in}%
\pgfsys@useobject{currentmarker}{}%
\end{pgfscope}%
\end{pgfscope}%
\begin{pgfscope}%
\pgftext[x=3.921268in,y=0.298211in,left,base,rotate=90.000000]{\rmfamily\fontsize{8.000000}{9.600000}\selectfont \(\displaystyle (2,0,1)\)}%
\end{pgfscope}%
\begin{pgfscope}%
\pgfsetbuttcap%
\pgfsetroundjoin%
\definecolor{currentfill}{rgb}{0.000000,0.000000,0.000000}%
\pgfsetfillcolor{currentfill}%
\pgfsetlinewidth{0.803000pt}%
\definecolor{currentstroke}{rgb}{0.000000,0.000000,0.000000}%
\pgfsetstrokecolor{currentstroke}%
\pgfsetdash{}{0pt}%
\pgfsys@defobject{currentmarker}{\pgfqpoint{0.000000in}{-0.048611in}}{\pgfqpoint{0.000000in}{0.000000in}}{%
\pgfpathmoveto{\pgfqpoint{0.000000in}{0.000000in}}%
\pgfpathlineto{\pgfqpoint{0.000000in}{-0.048611in}}%
\pgfusepath{stroke,fill}%
}%
\begin{pgfscope}%
\pgfsys@transformshift{4.324613in}{0.777990in}%
\pgfsys@useobject{currentmarker}{}%
\end{pgfscope}%
\end{pgfscope}%
\begin{pgfscope}%
\pgftext[x=4.352770in,y=0.298211in,left,base,rotate=90.000000]{\rmfamily\fontsize{8.000000}{9.600000}\selectfont \(\displaystyle (1,1,1)\)}%
\end{pgfscope}%
\begin{pgfscope}%
\pgfsetbuttcap%
\pgfsetroundjoin%
\definecolor{currentfill}{rgb}{0.000000,0.000000,0.000000}%
\pgfsetfillcolor{currentfill}%
\pgfsetlinewidth{0.803000pt}%
\definecolor{currentstroke}{rgb}{0.000000,0.000000,0.000000}%
\pgfsetstrokecolor{currentstroke}%
\pgfsetdash{}{0pt}%
\pgfsys@defobject{currentmarker}{\pgfqpoint{0.000000in}{-0.048611in}}{\pgfqpoint{0.000000in}{0.000000in}}{%
\pgfpathmoveto{\pgfqpoint{0.000000in}{0.000000in}}%
\pgfpathlineto{\pgfqpoint{0.000000in}{-0.048611in}}%
\pgfusepath{stroke,fill}%
}%
\begin{pgfscope}%
\pgfsys@transformshift{4.756115in}{0.777990in}%
\pgfsys@useobject{currentmarker}{}%
\end{pgfscope}%
\end{pgfscope}%
\begin{pgfscope}%
\pgftext[x=4.784272in,y=0.298211in,left,base,rotate=90.000000]{\rmfamily\fontsize{8.000000}{9.600000}\selectfont \(\displaystyle (0,2,1)\)}%
\end{pgfscope}%
\begin{pgfscope}%
\pgfsetbuttcap%
\pgfsetroundjoin%
\definecolor{currentfill}{rgb}{0.000000,0.000000,0.000000}%
\pgfsetfillcolor{currentfill}%
\pgfsetlinewidth{0.803000pt}%
\definecolor{currentstroke}{rgb}{0.000000,0.000000,0.000000}%
\pgfsetstrokecolor{currentstroke}%
\pgfsetdash{}{0pt}%
\pgfsys@defobject{currentmarker}{\pgfqpoint{0.000000in}{-0.048611in}}{\pgfqpoint{0.000000in}{0.000000in}}{%
\pgfpathmoveto{\pgfqpoint{0.000000in}{0.000000in}}%
\pgfpathlineto{\pgfqpoint{0.000000in}{-0.048611in}}%
\pgfusepath{stroke,fill}%
}%
\begin{pgfscope}%
\pgfsys@transformshift{5.187617in}{0.777990in}%
\pgfsys@useobject{currentmarker}{}%
\end{pgfscope}%
\end{pgfscope}%
\begin{pgfscope}%
\pgftext[x=5.215773in,y=0.298211in,left,base,rotate=90.000000]{\rmfamily\fontsize{8.000000}{9.600000}\selectfont \(\displaystyle (0,0,3)\)}%
\end{pgfscope}%
\begin{pgfscope}%
\pgftext[x=4.540364in,y=0.242655in,,top]{\rmfamily\fontsize{10.000000}{12.000000}\selectfont \(\displaystyle \alpha\)}%
\end{pgfscope}%
\begin{pgfscope}%
\pgfsetbuttcap%
\pgfsetroundjoin%
\definecolor{currentfill}{rgb}{0.000000,0.000000,0.000000}%
\pgfsetfillcolor{currentfill}%
\pgfsetlinewidth{0.803000pt}%
\definecolor{currentstroke}{rgb}{0.000000,0.000000,0.000000}%
\pgfsetstrokecolor{currentstroke}%
\pgfsetdash{}{0pt}%
\pgfsys@defobject{currentmarker}{\pgfqpoint{-0.048611in}{0.000000in}}{\pgfqpoint{0.000000in}{0.000000in}}{%
\pgfpathmoveto{\pgfqpoint{0.000000in}{0.000000in}}%
\pgfpathlineto{\pgfqpoint{-0.048611in}{0.000000in}}%
\pgfusepath{stroke,fill}%
}%
\begin{pgfscope}%
\pgfsys@transformshift{3.638525in}{0.790416in}%
\pgfsys@useobject{currentmarker}{}%
\end{pgfscope}%
\end{pgfscope}%
\begin{pgfscope}%
\pgftext[x=3.261020in,y=0.752154in,left,base]{\rmfamily\fontsize{8.000000}{9.600000}\selectfont \(\displaystyle -0.02\)}%
\end{pgfscope}%
\begin{pgfscope}%
\pgfsetbuttcap%
\pgfsetroundjoin%
\definecolor{currentfill}{rgb}{0.000000,0.000000,0.000000}%
\pgfsetfillcolor{currentfill}%
\pgfsetlinewidth{0.803000pt}%
\definecolor{currentstroke}{rgb}{0.000000,0.000000,0.000000}%
\pgfsetstrokecolor{currentstroke}%
\pgfsetdash{}{0pt}%
\pgfsys@defobject{currentmarker}{\pgfqpoint{-0.048611in}{0.000000in}}{\pgfqpoint{0.000000in}{0.000000in}}{%
\pgfpathmoveto{\pgfqpoint{0.000000in}{0.000000in}}%
\pgfpathlineto{\pgfqpoint{-0.048611in}{0.000000in}}%
\pgfusepath{stroke,fill}%
}%
\begin{pgfscope}%
\pgfsys@transformshift{3.638525in}{1.149154in}%
\pgfsys@useobject{currentmarker}{}%
\end{pgfscope}%
\end{pgfscope}%
\begin{pgfscope}%
\pgftext[x=3.261020in,y=1.110892in,left,base]{\rmfamily\fontsize{8.000000}{9.600000}\selectfont \(\displaystyle -0.01\)}%
\end{pgfscope}%
\begin{pgfscope}%
\pgfsetbuttcap%
\pgfsetroundjoin%
\definecolor{currentfill}{rgb}{0.000000,0.000000,0.000000}%
\pgfsetfillcolor{currentfill}%
\pgfsetlinewidth{0.803000pt}%
\definecolor{currentstroke}{rgb}{0.000000,0.000000,0.000000}%
\pgfsetstrokecolor{currentstroke}%
\pgfsetdash{}{0pt}%
\pgfsys@defobject{currentmarker}{\pgfqpoint{-0.048611in}{0.000000in}}{\pgfqpoint{0.000000in}{0.000000in}}{%
\pgfpathmoveto{\pgfqpoint{0.000000in}{0.000000in}}%
\pgfpathlineto{\pgfqpoint{-0.048611in}{0.000000in}}%
\pgfusepath{stroke,fill}%
}%
\begin{pgfscope}%
\pgfsys@transformshift{3.638525in}{1.507892in}%
\pgfsys@useobject{currentmarker}{}%
\end{pgfscope}%
\end{pgfscope}%
\begin{pgfscope}%
\pgftext[x=3.331423in,y=1.469630in,left,base]{\rmfamily\fontsize{8.000000}{9.600000}\selectfont \(\displaystyle 0.00\)}%
\end{pgfscope}%
\begin{pgfscope}%
\pgfsetbuttcap%
\pgfsetroundjoin%
\definecolor{currentfill}{rgb}{0.000000,0.000000,0.000000}%
\pgfsetfillcolor{currentfill}%
\pgfsetlinewidth{0.803000pt}%
\definecolor{currentstroke}{rgb}{0.000000,0.000000,0.000000}%
\pgfsetstrokecolor{currentstroke}%
\pgfsetdash{}{0pt}%
\pgfsys@defobject{currentmarker}{\pgfqpoint{-0.048611in}{0.000000in}}{\pgfqpoint{0.000000in}{0.000000in}}{%
\pgfpathmoveto{\pgfqpoint{0.000000in}{0.000000in}}%
\pgfpathlineto{\pgfqpoint{-0.048611in}{0.000000in}}%
\pgfusepath{stroke,fill}%
}%
\begin{pgfscope}%
\pgfsys@transformshift{3.638525in}{1.866630in}%
\pgfsys@useobject{currentmarker}{}%
\end{pgfscope}%
\end{pgfscope}%
\begin{pgfscope}%
\pgftext[x=3.331423in,y=1.828368in,left,base]{\rmfamily\fontsize{8.000000}{9.600000}\selectfont \(\displaystyle 0.01\)}%
\end{pgfscope}%
\begin{pgfscope}%
\pgfsetbuttcap%
\pgfsetroundjoin%
\definecolor{currentfill}{rgb}{0.000000,0.000000,0.000000}%
\pgfsetfillcolor{currentfill}%
\pgfsetlinewidth{0.803000pt}%
\definecolor{currentstroke}{rgb}{0.000000,0.000000,0.000000}%
\pgfsetstrokecolor{currentstroke}%
\pgfsetdash{}{0pt}%
\pgfsys@defobject{currentmarker}{\pgfqpoint{-0.048611in}{0.000000in}}{\pgfqpoint{0.000000in}{0.000000in}}{%
\pgfpathmoveto{\pgfqpoint{0.000000in}{0.000000in}}%
\pgfpathlineto{\pgfqpoint{-0.048611in}{0.000000in}}%
\pgfusepath{stroke,fill}%
}%
\begin{pgfscope}%
\pgfsys@transformshift{3.638525in}{2.225368in}%
\pgfsys@useobject{currentmarker}{}%
\end{pgfscope}%
\end{pgfscope}%
\begin{pgfscope}%
\pgftext[x=3.331423in,y=2.187106in,left,base]{\rmfamily\fontsize{8.000000}{9.600000}\selectfont \(\displaystyle 0.02\)}%
\end{pgfscope}%
\begin{pgfscope}%
\pgftext[x=3.205464in,y=1.507892in,,bottom,rotate=90.000000]{\rmfamily\fontsize{10.000000}{12.000000}\selectfont \(\displaystyle \boldsymbol{g}^{(3)}\)}%
\end{pgfscope}%
\begin{pgfscope}%
\pgfsetrectcap%
\pgfsetmiterjoin%
\pgfsetlinewidth{0.803000pt}%
\definecolor{currentstroke}{rgb}{0.000000,0.000000,0.000000}%
\pgfsetstrokecolor{currentstroke}%
\pgfsetdash{}{0pt}%
\pgfpathmoveto{\pgfqpoint{3.638525in}{0.777990in}}%
\pgfpathlineto{\pgfqpoint{3.638525in}{2.237794in}}%
\pgfusepath{stroke}%
\end{pgfscope}%
\begin{pgfscope}%
\pgfsetrectcap%
\pgfsetmiterjoin%
\pgfsetlinewidth{0.803000pt}%
\definecolor{currentstroke}{rgb}{0.000000,0.000000,0.000000}%
\pgfsetstrokecolor{currentstroke}%
\pgfsetdash{}{0pt}%
\pgfpathmoveto{\pgfqpoint{5.442203in}{0.777990in}}%
\pgfpathlineto{\pgfqpoint{5.442203in}{2.237794in}}%
\pgfusepath{stroke}%
\end{pgfscope}%
\begin{pgfscope}%
\pgfsetrectcap%
\pgfsetmiterjoin%
\pgfsetlinewidth{0.803000pt}%
\definecolor{currentstroke}{rgb}{0.000000,0.000000,0.000000}%
\pgfsetstrokecolor{currentstroke}%
\pgfsetdash{}{0pt}%
\pgfpathmoveto{\pgfqpoint{3.638525in}{0.777990in}}%
\pgfpathlineto{\pgfqpoint{5.442203in}{0.777990in}}%
\pgfusepath{stroke}%
\end{pgfscope}%
\begin{pgfscope}%
\pgfsetrectcap%
\pgfsetmiterjoin%
\pgfsetlinewidth{0.803000pt}%
\definecolor{currentstroke}{rgb}{0.000000,0.000000,0.000000}%
\pgfsetstrokecolor{currentstroke}%
\pgfsetdash{}{0pt}%
\pgfpathmoveto{\pgfqpoint{3.638525in}{2.237794in}}%
\pgfpathlineto{\pgfqpoint{5.442203in}{2.237794in}}%
\pgfusepath{stroke}%
\end{pgfscope}%
\begin{pgfscope}%
\pgfpathrectangle{\pgfqpoint{3.638525in}{0.777990in}}{\pgfqpoint{1.803678in}{1.459805in}} %
\pgfusepath{clip}%
\pgfsetbuttcap%
\pgfsetmiterjoin%
\definecolor{currentfill}{rgb}{0.960000,0.960000,0.960000}%
\pgfsetfillcolor{currentfill}%
\pgfsetfillopacity{0.600000}%
\pgfsetlinewidth{0.401500pt}%
\definecolor{currentstroke}{rgb}{0.000000,0.000000,0.000000}%
\pgfsetstrokecolor{currentstroke}%
\pgfsetstrokeopacity{0.600000}%
\pgfsetdash{}{0pt}%
\pgfpathmoveto{\pgfqpoint{3.720510in}{1.507892in}}%
\pgfpathlineto{\pgfqpoint{4.065712in}{1.507892in}}%
\pgfpathlineto{\pgfqpoint{4.065712in}{2.116144in}}%
\pgfpathlineto{\pgfqpoint{3.720510in}{2.116144in}}%
\pgfpathclose%
\pgfusepath{stroke,fill}%
\end{pgfscope}%
\begin{pgfscope}%
\pgfpathrectangle{\pgfqpoint{3.638525in}{0.777990in}}{\pgfqpoint{1.803678in}{1.459805in}} %
\pgfusepath{clip}%
\pgfsetbuttcap%
\pgfsetmiterjoin%
\definecolor{currentfill}{rgb}{0.960000,0.960000,0.960000}%
\pgfsetfillcolor{currentfill}%
\pgfsetfillopacity{0.600000}%
\pgfsetlinewidth{0.401500pt}%
\definecolor{currentstroke}{rgb}{0.000000,0.000000,0.000000}%
\pgfsetstrokecolor{currentstroke}%
\pgfsetstrokeopacity{0.600000}%
\pgfsetdash{}{0pt}%
\pgfpathmoveto{\pgfqpoint{4.152012in}{1.507892in}}%
\pgfpathlineto{\pgfqpoint{4.497214in}{1.507892in}}%
\pgfpathlineto{\pgfqpoint{4.497214in}{1.724305in}}%
\pgfpathlineto{\pgfqpoint{4.152012in}{1.724305in}}%
\pgfpathclose%
\pgfusepath{stroke,fill}%
\end{pgfscope}%
\begin{pgfscope}%
\pgfpathrectangle{\pgfqpoint{3.638525in}{0.777990in}}{\pgfqpoint{1.803678in}{1.459805in}} %
\pgfusepath{clip}%
\pgfsetbuttcap%
\pgfsetmiterjoin%
\definecolor{currentfill}{rgb}{0.960000,0.960000,0.960000}%
\pgfsetfillcolor{currentfill}%
\pgfsetfillopacity{0.600000}%
\pgfsetlinewidth{0.401500pt}%
\definecolor{currentstroke}{rgb}{0.000000,0.000000,0.000000}%
\pgfsetstrokecolor{currentstroke}%
\pgfsetstrokeopacity{0.600000}%
\pgfsetdash{}{0pt}%
\pgfpathmoveto{\pgfqpoint{4.583514in}{1.507892in}}%
\pgfpathlineto{\pgfqpoint{4.928716in}{1.507892in}}%
\pgfpathlineto{\pgfqpoint{4.928716in}{1.539506in}}%
\pgfpathlineto{\pgfqpoint{4.583514in}{1.539506in}}%
\pgfpathclose%
\pgfusepath{stroke,fill}%
\end{pgfscope}%
\begin{pgfscope}%
\pgfpathrectangle{\pgfqpoint{3.638525in}{0.777990in}}{\pgfqpoint{1.803678in}{1.459805in}} %
\pgfusepath{clip}%
\pgfsetbuttcap%
\pgfsetmiterjoin%
\definecolor{currentfill}{rgb}{0.960000,0.960000,0.960000}%
\pgfsetfillcolor{currentfill}%
\pgfsetfillopacity{0.600000}%
\pgfsetlinewidth{0.401500pt}%
\definecolor{currentstroke}{rgb}{0.000000,0.000000,0.000000}%
\pgfsetstrokecolor{currentstroke}%
\pgfsetstrokeopacity{0.600000}%
\pgfsetdash{}{0pt}%
\pgfpathmoveto{\pgfqpoint{5.015016in}{1.507892in}}%
\pgfpathlineto{\pgfqpoint{5.360217in}{1.507892in}}%
\pgfpathlineto{\pgfqpoint{5.360217in}{1.457614in}}%
\pgfpathlineto{\pgfqpoint{5.015016in}{1.457614in}}%
\pgfpathclose%
\pgfusepath{stroke,fill}%
\end{pgfscope}%
\begin{pgfscope}%
\pgfsetbuttcap%
\pgfsetroundjoin%
\definecolor{currentfill}{rgb}{0.000000,0.000000,0.000000}%
\pgfsetfillcolor{currentfill}%
\pgfsetlinewidth{0.803000pt}%
\definecolor{currentstroke}{rgb}{0.000000,0.000000,0.000000}%
\pgfsetstrokecolor{currentstroke}%
\pgfsetdash{}{0pt}%
\pgfsys@defobject{currentmarker}{\pgfqpoint{0.000000in}{0.000000in}}{\pgfqpoint{0.048611in}{0.000000in}}{%
\pgfpathmoveto{\pgfqpoint{0.000000in}{0.000000in}}%
\pgfpathlineto{\pgfqpoint{0.048611in}{0.000000in}}%
\pgfusepath{stroke,fill}%
}%
\begin{pgfscope}%
\pgfsys@transformshift{5.442203in}{0.952098in}%
\pgfsys@useobject{currentmarker}{}%
\end{pgfscope}%
\end{pgfscope}%
\begin{pgfscope}%
\pgftext[x=5.539425in,y=0.913836in,left,base]{\rmfamily\fontsize{8.000000}{9.600000}\selectfont \(\displaystyle -20\)}%
\end{pgfscope}%
\begin{pgfscope}%
\pgfsetbuttcap%
\pgfsetroundjoin%
\definecolor{currentfill}{rgb}{0.000000,0.000000,0.000000}%
\pgfsetfillcolor{currentfill}%
\pgfsetlinewidth{0.803000pt}%
\definecolor{currentstroke}{rgb}{0.000000,0.000000,0.000000}%
\pgfsetstrokecolor{currentstroke}%
\pgfsetdash{}{0pt}%
\pgfsys@defobject{currentmarker}{\pgfqpoint{0.000000in}{0.000000in}}{\pgfqpoint{0.048611in}{0.000000in}}{%
\pgfpathmoveto{\pgfqpoint{0.000000in}{0.000000in}}%
\pgfpathlineto{\pgfqpoint{0.048611in}{0.000000in}}%
\pgfusepath{stroke,fill}%
}%
\begin{pgfscope}%
\pgfsys@transformshift{5.442203in}{1.229995in}%
\pgfsys@useobject{currentmarker}{}%
\end{pgfscope}%
\end{pgfscope}%
\begin{pgfscope}%
\pgftext[x=5.539425in,y=1.191733in,left,base]{\rmfamily\fontsize{8.000000}{9.600000}\selectfont \(\displaystyle -10\)}%
\end{pgfscope}%
\begin{pgfscope}%
\pgfsetbuttcap%
\pgfsetroundjoin%
\definecolor{currentfill}{rgb}{0.000000,0.000000,0.000000}%
\pgfsetfillcolor{currentfill}%
\pgfsetlinewidth{0.803000pt}%
\definecolor{currentstroke}{rgb}{0.000000,0.000000,0.000000}%
\pgfsetstrokecolor{currentstroke}%
\pgfsetdash{}{0pt}%
\pgfsys@defobject{currentmarker}{\pgfqpoint{0.000000in}{0.000000in}}{\pgfqpoint{0.048611in}{0.000000in}}{%
\pgfpathmoveto{\pgfqpoint{0.000000in}{0.000000in}}%
\pgfpathlineto{\pgfqpoint{0.048611in}{0.000000in}}%
\pgfusepath{stroke,fill}%
}%
\begin{pgfscope}%
\pgfsys@transformshift{5.442203in}{1.507892in}%
\pgfsys@useobject{currentmarker}{}%
\end{pgfscope}%
\end{pgfscope}%
\begin{pgfscope}%
\pgftext[x=5.539425in,y=1.469630in,left,base]{\rmfamily\fontsize{8.000000}{9.600000}\selectfont \(\displaystyle 0\)}%
\end{pgfscope}%
\begin{pgfscope}%
\pgfsetbuttcap%
\pgfsetroundjoin%
\definecolor{currentfill}{rgb}{0.000000,0.000000,0.000000}%
\pgfsetfillcolor{currentfill}%
\pgfsetlinewidth{0.803000pt}%
\definecolor{currentstroke}{rgb}{0.000000,0.000000,0.000000}%
\pgfsetstrokecolor{currentstroke}%
\pgfsetdash{}{0pt}%
\pgfsys@defobject{currentmarker}{\pgfqpoint{0.000000in}{0.000000in}}{\pgfqpoint{0.048611in}{0.000000in}}{%
\pgfpathmoveto{\pgfqpoint{0.000000in}{0.000000in}}%
\pgfpathlineto{\pgfqpoint{0.048611in}{0.000000in}}%
\pgfusepath{stroke,fill}%
}%
\begin{pgfscope}%
\pgfsys@transformshift{5.442203in}{1.785789in}%
\pgfsys@useobject{currentmarker}{}%
\end{pgfscope}%
\end{pgfscope}%
\begin{pgfscope}%
\pgftext[x=5.539425in,y=1.747527in,left,base]{\rmfamily\fontsize{8.000000}{9.600000}\selectfont \(\displaystyle 10\)}%
\end{pgfscope}%
\begin{pgfscope}%
\pgfsetbuttcap%
\pgfsetroundjoin%
\definecolor{currentfill}{rgb}{0.000000,0.000000,0.000000}%
\pgfsetfillcolor{currentfill}%
\pgfsetlinewidth{0.803000pt}%
\definecolor{currentstroke}{rgb}{0.000000,0.000000,0.000000}%
\pgfsetstrokecolor{currentstroke}%
\pgfsetdash{}{0pt}%
\pgfsys@defobject{currentmarker}{\pgfqpoint{0.000000in}{0.000000in}}{\pgfqpoint{0.048611in}{0.000000in}}{%
\pgfpathmoveto{\pgfqpoint{0.000000in}{0.000000in}}%
\pgfpathlineto{\pgfqpoint{0.048611in}{0.000000in}}%
\pgfusepath{stroke,fill}%
}%
\begin{pgfscope}%
\pgfsys@transformshift{5.442203in}{2.063686in}%
\pgfsys@useobject{currentmarker}{}%
\end{pgfscope}%
\end{pgfscope}%
\begin{pgfscope}%
\pgftext[x=5.539425in,y=2.025424in,left,base]{\rmfamily\fontsize{8.000000}{9.600000}\selectfont \(\displaystyle 20\)}%
\end{pgfscope}%
\begin{pgfscope}%
\pgftext[x=5.783441in,y=1.507892in,,top,rotate=90.000000]{\rmfamily\fontsize{10.000000}{12.000000}\selectfont \(\displaystyle \boldsymbol{u}_{3}^{(3)}\)}%
\end{pgfscope}%
\begin{pgfscope}%
\pgfsetrectcap%
\pgfsetmiterjoin%
\pgfsetlinewidth{0.803000pt}%
\definecolor{currentstroke}{rgb}{0.000000,0.000000,0.000000}%
\pgfsetstrokecolor{currentstroke}%
\pgfsetdash{}{0pt}%
\pgfpathmoveto{\pgfqpoint{3.638525in}{0.777990in}}%
\pgfpathlineto{\pgfqpoint{3.638525in}{2.237794in}}%
\pgfusepath{stroke}%
\end{pgfscope}%
\begin{pgfscope}%
\pgfsetrectcap%
\pgfsetmiterjoin%
\pgfsetlinewidth{0.803000pt}%
\definecolor{currentstroke}{rgb}{0.000000,0.000000,0.000000}%
\pgfsetstrokecolor{currentstroke}%
\pgfsetdash{}{0pt}%
\pgfpathmoveto{\pgfqpoint{5.442203in}{0.777990in}}%
\pgfpathlineto{\pgfqpoint{5.442203in}{2.237794in}}%
\pgfusepath{stroke}%
\end{pgfscope}%
\begin{pgfscope}%
\pgfsetrectcap%
\pgfsetmiterjoin%
\pgfsetlinewidth{0.803000pt}%
\definecolor{currentstroke}{rgb}{0.000000,0.000000,0.000000}%
\pgfsetstrokecolor{currentstroke}%
\pgfsetdash{}{0pt}%
\pgfpathmoveto{\pgfqpoint{3.638525in}{0.777990in}}%
\pgfpathlineto{\pgfqpoint{5.442203in}{0.777990in}}%
\pgfusepath{stroke}%
\end{pgfscope}%
\begin{pgfscope}%
\pgfsetrectcap%
\pgfsetmiterjoin%
\pgfsetlinewidth{0.803000pt}%
\definecolor{currentstroke}{rgb}{0.000000,0.000000,0.000000}%
\pgfsetstrokecolor{currentstroke}%
\pgfsetdash{}{0pt}%
\pgfpathmoveto{\pgfqpoint{3.638525in}{2.237794in}}%
\pgfpathlineto{\pgfqpoint{5.442203in}{2.237794in}}%
\pgfusepath{stroke}%
\end{pgfscope}%
\begin{pgfscope}%
\pgftext[x=4.540364in,y=2.321128in,,base]{\rmfamily\fontsize{9.600000}{11.520000}\selectfont \(\displaystyle N = 2\)}%
\end{pgfscope}%
\begin{pgfscope}%
\pgfsetbuttcap%
\pgfsetmiterjoin%
\definecolor{currentfill}{rgb}{1.000000,0.600000,0.000000}%
\pgfsetfillcolor{currentfill}%
\pgfsetlinewidth{0.401500pt}%
\definecolor{currentstroke}{rgb}{0.000000,0.000000,0.000000}%
\pgfsetstrokecolor{currentstroke}%
\pgfsetdash{}{0pt}%
\pgfpathmoveto{\pgfqpoint{4.805289in}{2.026438in}}%
\pgfpathlineto{\pgfqpoint{5.027511in}{2.026438in}}%
\pgfpathlineto{\pgfqpoint{5.027511in}{2.104216in}}%
\pgfpathlineto{\pgfqpoint{4.805289in}{2.104216in}}%
\pgfpathclose%
\pgfusepath{stroke,fill}%
\end{pgfscope}%
\begin{pgfscope}%
\pgftext[x=5.116400in,y=2.026438in,left,base]{\rmfamily\fontsize{8.000000}{9.600000}\selectfont \(\displaystyle \boldsymbol{g}^{(3)}\)}%
\end{pgfscope}%
\begin{pgfscope}%
\pgfsetbuttcap%
\pgfsetmiterjoin%
\definecolor{currentfill}{rgb}{0.960000,0.960000,0.960000}%
\pgfsetfillcolor{currentfill}%
\pgfsetfillopacity{0.600000}%
\pgfsetlinewidth{0.401500pt}%
\definecolor{currentstroke}{rgb}{0.000000,0.000000,0.000000}%
\pgfsetstrokecolor{currentstroke}%
\pgfsetstrokeopacity{0.600000}%
\pgfsetdash{}{0pt}%
\pgfpathmoveto{\pgfqpoint{4.805289in}{1.828011in}}%
\pgfpathlineto{\pgfqpoint{5.027511in}{1.828011in}}%
\pgfpathlineto{\pgfqpoint{5.027511in}{1.905789in}}%
\pgfpathlineto{\pgfqpoint{4.805289in}{1.905789in}}%
\pgfpathclose%
\pgfusepath{stroke,fill}%
\end{pgfscope}%
\begin{pgfscope}%
\pgftext[x=5.116400in,y=1.828011in,left,base]{\rmfamily\fontsize{8.000000}{9.600000}\selectfont \(\displaystyle \boldsymbol{u}_{3}^{(3)}\)}%
\end{pgfscope}%
\end{pgfpicture}%
\makeatother%
\endgroup%

%% file: fig_07.pgf
\begingroup%
\makeatletter%
\begin{pgfpicture}%
\pgfpathrectangle{\pgfpointorigin}{\pgfqpoint{4.151100in}{2.767400in}}%
\pgfusepath{use as bounding box, clip}%
\begin{pgfscope}%
\pgfsetbuttcap%
\pgfsetmiterjoin%
\definecolor{currentfill}{rgb}{1.000000,1.000000,1.000000}%
\pgfsetfillcolor{currentfill}%
\pgfsetlinewidth{0.000000pt}%
\definecolor{currentstroke}{rgb}{1.000000,1.000000,1.000000}%
\pgfsetstrokecolor{currentstroke}%
\pgfsetdash{}{0pt}%
\pgfpathmoveto{\pgfqpoint{0.000000in}{0.000000in}}%
\pgfpathlineto{\pgfqpoint{4.151100in}{0.000000in}}%
\pgfpathlineto{\pgfqpoint{4.151100in}{2.767400in}}%
\pgfpathlineto{\pgfqpoint{0.000000in}{2.767400in}}%
\pgfpathclose%
\pgfusepath{fill}%
\end{pgfscope}%
\begin{pgfscope}%
\pgfsetbuttcap%
\pgfsetmiterjoin%
\definecolor{currentfill}{rgb}{1.000000,1.000000,1.000000}%
\pgfsetfillcolor{currentfill}%
\pgfsetlinewidth{0.000000pt}%
\definecolor{currentstroke}{rgb}{0.000000,0.000000,0.000000}%
\pgfsetstrokecolor{currentstroke}%
\pgfsetstrokeopacity{0.000000}%
\pgfsetdash{}{0pt}%
\pgfpathmoveto{\pgfqpoint{0.640063in}{0.493557in}}%
\pgfpathlineto{\pgfqpoint{3.926160in}{0.493557in}}%
\pgfpathlineto{\pgfqpoint{3.926160in}{2.612400in}}%
\pgfpathlineto{\pgfqpoint{0.640063in}{2.612400in}}%
\pgfpathclose%
\pgfusepath{fill}%
\end{pgfscope}%
\begin{pgfscope}%
\pgfsetbuttcap%
\pgfsetroundjoin%
\definecolor{currentfill}{rgb}{0.000000,0.000000,0.000000}%
\pgfsetfillcolor{currentfill}%
\pgfsetlinewidth{0.803000pt}%
\definecolor{currentstroke}{rgb}{0.000000,0.000000,0.000000}%
\pgfsetstrokecolor{currentstroke}%
\pgfsetdash{}{0pt}%
\pgfsys@defobject{currentmarker}{\pgfqpoint{0.000000in}{-0.048611in}}{\pgfqpoint{0.000000in}{0.000000in}}{%
\pgfpathmoveto{\pgfqpoint{0.000000in}{0.000000in}}%
\pgfpathlineto{\pgfqpoint{0.000000in}{-0.048611in}}%
\pgfusepath{stroke,fill}%
}%
\begin{pgfscope}%
\pgfsys@transformshift{0.640063in}{0.493557in}%
\pgfsys@useobject{currentmarker}{}%
\end{pgfscope}%
\end{pgfscope}%
\begin{pgfscope}%
\pgftext[x=0.640063in,y=0.396335in,,top]{\rmfamily\fontsize{8.000000}{9.600000}\selectfont \(\displaystyle 20.0\)}%
\end{pgfscope}%
\begin{pgfscope}%
\pgfsetbuttcap%
\pgfsetroundjoin%
\definecolor{currentfill}{rgb}{0.000000,0.000000,0.000000}%
\pgfsetfillcolor{currentfill}%
\pgfsetlinewidth{0.803000pt}%
\definecolor{currentstroke}{rgb}{0.000000,0.000000,0.000000}%
\pgfsetstrokecolor{currentstroke}%
\pgfsetdash{}{0pt}%
\pgfsys@defobject{currentmarker}{\pgfqpoint{0.000000in}{-0.048611in}}{\pgfqpoint{0.000000in}{0.000000in}}{%
\pgfpathmoveto{\pgfqpoint{0.000000in}{0.000000in}}%
\pgfpathlineto{\pgfqpoint{0.000000in}{-0.048611in}}%
\pgfusepath{stroke,fill}%
}%
\begin{pgfscope}%
\pgfsys@transformshift{1.050825in}{0.493557in}%
\pgfsys@useobject{currentmarker}{}%
\end{pgfscope}%
\end{pgfscope}%
\begin{pgfscope}%
\pgftext[x=1.050825in,y=0.396335in,,top]{\rmfamily\fontsize{8.000000}{9.600000}\selectfont \(\displaystyle 22.5\)}%
\end{pgfscope}%
\begin{pgfscope}%
\pgfsetbuttcap%
\pgfsetroundjoin%
\definecolor{currentfill}{rgb}{0.000000,0.000000,0.000000}%
\pgfsetfillcolor{currentfill}%
\pgfsetlinewidth{0.803000pt}%
\definecolor{currentstroke}{rgb}{0.000000,0.000000,0.000000}%
\pgfsetstrokecolor{currentstroke}%
\pgfsetdash{}{0pt}%
\pgfsys@defobject{currentmarker}{\pgfqpoint{0.000000in}{-0.048611in}}{\pgfqpoint{0.000000in}{0.000000in}}{%
\pgfpathmoveto{\pgfqpoint{0.000000in}{0.000000in}}%
\pgfpathlineto{\pgfqpoint{0.000000in}{-0.048611in}}%
\pgfusepath{stroke,fill}%
}%
\begin{pgfscope}%
\pgfsys@transformshift{1.461587in}{0.493557in}%
\pgfsys@useobject{currentmarker}{}%
\end{pgfscope}%
\end{pgfscope}%
\begin{pgfscope}%
\pgftext[x=1.461587in,y=0.396335in,,top]{\rmfamily\fontsize{8.000000}{9.600000}\selectfont \(\displaystyle 25.0\)}%
\end{pgfscope}%
\begin{pgfscope}%
\pgfsetbuttcap%
\pgfsetroundjoin%
\definecolor{currentfill}{rgb}{0.000000,0.000000,0.000000}%
\pgfsetfillcolor{currentfill}%
\pgfsetlinewidth{0.803000pt}%
\definecolor{currentstroke}{rgb}{0.000000,0.000000,0.000000}%
\pgfsetstrokecolor{currentstroke}%
\pgfsetdash{}{0pt}%
\pgfsys@defobject{currentmarker}{\pgfqpoint{0.000000in}{-0.048611in}}{\pgfqpoint{0.000000in}{0.000000in}}{%
\pgfpathmoveto{\pgfqpoint{0.000000in}{0.000000in}}%
\pgfpathlineto{\pgfqpoint{0.000000in}{-0.048611in}}%
\pgfusepath{stroke,fill}%
}%
\begin{pgfscope}%
\pgfsys@transformshift{1.872349in}{0.493557in}%
\pgfsys@useobject{currentmarker}{}%
\end{pgfscope}%
\end{pgfscope}%
\begin{pgfscope}%
\pgftext[x=1.872349in,y=0.396335in,,top]{\rmfamily\fontsize{8.000000}{9.600000}\selectfont \(\displaystyle 27.5\)}%
\end{pgfscope}%
\begin{pgfscope}%
\pgfsetbuttcap%
\pgfsetroundjoin%
\definecolor{currentfill}{rgb}{0.000000,0.000000,0.000000}%
\pgfsetfillcolor{currentfill}%
\pgfsetlinewidth{0.803000pt}%
\definecolor{currentstroke}{rgb}{0.000000,0.000000,0.000000}%
\pgfsetstrokecolor{currentstroke}%
\pgfsetdash{}{0pt}%
\pgfsys@defobject{currentmarker}{\pgfqpoint{0.000000in}{-0.048611in}}{\pgfqpoint{0.000000in}{0.000000in}}{%
\pgfpathmoveto{\pgfqpoint{0.000000in}{0.000000in}}%
\pgfpathlineto{\pgfqpoint{0.000000in}{-0.048611in}}%
\pgfusepath{stroke,fill}%
}%
\begin{pgfscope}%
\pgfsys@transformshift{2.283112in}{0.493557in}%
\pgfsys@useobject{currentmarker}{}%
\end{pgfscope}%
\end{pgfscope}%
\begin{pgfscope}%
\pgftext[x=2.283112in,y=0.396335in,,top]{\rmfamily\fontsize{8.000000}{9.600000}\selectfont \(\displaystyle 30.0\)}%
\end{pgfscope}%
\begin{pgfscope}%
\pgfsetbuttcap%
\pgfsetroundjoin%
\definecolor{currentfill}{rgb}{0.000000,0.000000,0.000000}%
\pgfsetfillcolor{currentfill}%
\pgfsetlinewidth{0.803000pt}%
\definecolor{currentstroke}{rgb}{0.000000,0.000000,0.000000}%
\pgfsetstrokecolor{currentstroke}%
\pgfsetdash{}{0pt}%
\pgfsys@defobject{currentmarker}{\pgfqpoint{0.000000in}{-0.048611in}}{\pgfqpoint{0.000000in}{0.000000in}}{%
\pgfpathmoveto{\pgfqpoint{0.000000in}{0.000000in}}%
\pgfpathlineto{\pgfqpoint{0.000000in}{-0.048611in}}%
\pgfusepath{stroke,fill}%
}%
\begin{pgfscope}%
\pgfsys@transformshift{2.693874in}{0.493557in}%
\pgfsys@useobject{currentmarker}{}%
\end{pgfscope}%
\end{pgfscope}%
\begin{pgfscope}%
\pgftext[x=2.693874in,y=0.396335in,,top]{\rmfamily\fontsize{8.000000}{9.600000}\selectfont \(\displaystyle 32.5\)}%
\end{pgfscope}%
\begin{pgfscope}%
\pgfsetbuttcap%
\pgfsetroundjoin%
\definecolor{currentfill}{rgb}{0.000000,0.000000,0.000000}%
\pgfsetfillcolor{currentfill}%
\pgfsetlinewidth{0.803000pt}%
\definecolor{currentstroke}{rgb}{0.000000,0.000000,0.000000}%
\pgfsetstrokecolor{currentstroke}%
\pgfsetdash{}{0pt}%
\pgfsys@defobject{currentmarker}{\pgfqpoint{0.000000in}{-0.048611in}}{\pgfqpoint{0.000000in}{0.000000in}}{%
\pgfpathmoveto{\pgfqpoint{0.000000in}{0.000000in}}%
\pgfpathlineto{\pgfqpoint{0.000000in}{-0.048611in}}%
\pgfusepath{stroke,fill}%
}%
\begin{pgfscope}%
\pgfsys@transformshift{3.104636in}{0.493557in}%
\pgfsys@useobject{currentmarker}{}%
\end{pgfscope}%
\end{pgfscope}%
\begin{pgfscope}%
\pgftext[x=3.104636in,y=0.396335in,,top]{\rmfamily\fontsize{8.000000}{9.600000}\selectfont \(\displaystyle 35.0\)}%
\end{pgfscope}%
\begin{pgfscope}%
\pgfsetbuttcap%
\pgfsetroundjoin%
\definecolor{currentfill}{rgb}{0.000000,0.000000,0.000000}%
\pgfsetfillcolor{currentfill}%
\pgfsetlinewidth{0.803000pt}%
\definecolor{currentstroke}{rgb}{0.000000,0.000000,0.000000}%
\pgfsetstrokecolor{currentstroke}%
\pgfsetdash{}{0pt}%
\pgfsys@defobject{currentmarker}{\pgfqpoint{0.000000in}{-0.048611in}}{\pgfqpoint{0.000000in}{0.000000in}}{%
\pgfpathmoveto{\pgfqpoint{0.000000in}{0.000000in}}%
\pgfpathlineto{\pgfqpoint{0.000000in}{-0.048611in}}%
\pgfusepath{stroke,fill}%
}%
\begin{pgfscope}%
\pgfsys@transformshift{3.515398in}{0.493557in}%
\pgfsys@useobject{currentmarker}{}%
\end{pgfscope}%
\end{pgfscope}%
\begin{pgfscope}%
\pgftext[x=3.515398in,y=0.396335in,,top]{\rmfamily\fontsize{8.000000}{9.600000}\selectfont \(\displaystyle 37.5\)}%
\end{pgfscope}%
\begin{pgfscope}%
\pgfsetbuttcap%
\pgfsetroundjoin%
\definecolor{currentfill}{rgb}{0.000000,0.000000,0.000000}%
\pgfsetfillcolor{currentfill}%
\pgfsetlinewidth{0.803000pt}%
\definecolor{currentstroke}{rgb}{0.000000,0.000000,0.000000}%
\pgfsetstrokecolor{currentstroke}%
\pgfsetdash{}{0pt}%
\pgfsys@defobject{currentmarker}{\pgfqpoint{0.000000in}{-0.048611in}}{\pgfqpoint{0.000000in}{0.000000in}}{%
\pgfpathmoveto{\pgfqpoint{0.000000in}{0.000000in}}%
\pgfpathlineto{\pgfqpoint{0.000000in}{-0.048611in}}%
\pgfusepath{stroke,fill}%
}%
\begin{pgfscope}%
\pgfsys@transformshift{3.926160in}{0.493557in}%
\pgfsys@useobject{currentmarker}{}%
\end{pgfscope}%
\end{pgfscope}%
\begin{pgfscope}%
\pgftext[x=3.926160in,y=0.396335in,,top]{\rmfamily\fontsize{8.000000}{9.600000}\selectfont \(\displaystyle 40.0\)}%
\end{pgfscope}%
\begin{pgfscope}%
\pgftext[x=2.283112in,y=0.242655in,,top]{\rmfamily\fontsize{10.000000}{12.000000}\selectfont \(\displaystyle r\)}%
\end{pgfscope}%
\begin{pgfscope}%
\pgfsetbuttcap%
\pgfsetroundjoin%
\definecolor{currentfill}{rgb}{0.000000,0.000000,0.000000}%
\pgfsetfillcolor{currentfill}%
\pgfsetlinewidth{0.803000pt}%
\definecolor{currentstroke}{rgb}{0.000000,0.000000,0.000000}%
\pgfsetstrokecolor{currentstroke}%
\pgfsetdash{}{0pt}%
\pgfsys@defobject{currentmarker}{\pgfqpoint{-0.048611in}{0.000000in}}{\pgfqpoint{0.000000in}{0.000000in}}{%
\pgfpathmoveto{\pgfqpoint{0.000000in}{0.000000in}}%
\pgfpathlineto{\pgfqpoint{-0.048611in}{0.000000in}}%
\pgfusepath{stroke,fill}%
}%
\begin{pgfscope}%
\pgfsys@transformshift{0.640063in}{0.652470in}%
\pgfsys@useobject{currentmarker}{}%
\end{pgfscope}%
\end{pgfscope}%
\begin{pgfscope}%
\pgftext[x=0.332961in,y=0.614208in,left,base]{\rmfamily\fontsize{8.000000}{9.600000}\selectfont \(\displaystyle -25\)}%
\end{pgfscope}%
\begin{pgfscope}%
\pgfsetbuttcap%
\pgfsetroundjoin%
\definecolor{currentfill}{rgb}{0.000000,0.000000,0.000000}%
\pgfsetfillcolor{currentfill}%
\pgfsetlinewidth{0.803000pt}%
\definecolor{currentstroke}{rgb}{0.000000,0.000000,0.000000}%
\pgfsetstrokecolor{currentstroke}%
\pgfsetdash{}{0pt}%
\pgfsys@defobject{currentmarker}{\pgfqpoint{-0.048611in}{0.000000in}}{\pgfqpoint{0.000000in}{0.000000in}}{%
\pgfpathmoveto{\pgfqpoint{0.000000in}{0.000000in}}%
\pgfpathlineto{\pgfqpoint{-0.048611in}{0.000000in}}%
\pgfusepath{stroke,fill}%
}%
\begin{pgfscope}%
\pgfsys@transformshift{0.640063in}{0.917326in}%
\pgfsys@useobject{currentmarker}{}%
\end{pgfscope}%
\end{pgfscope}%
\begin{pgfscope}%
\pgftext[x=0.483812in,y=0.879063in,left,base]{\rmfamily\fontsize{8.000000}{9.600000}\selectfont \(\displaystyle 0\)}%
\end{pgfscope}%
\begin{pgfscope}%
\pgfsetbuttcap%
\pgfsetroundjoin%
\definecolor{currentfill}{rgb}{0.000000,0.000000,0.000000}%
\pgfsetfillcolor{currentfill}%
\pgfsetlinewidth{0.803000pt}%
\definecolor{currentstroke}{rgb}{0.000000,0.000000,0.000000}%
\pgfsetstrokecolor{currentstroke}%
\pgfsetdash{}{0pt}%
\pgfsys@defobject{currentmarker}{\pgfqpoint{-0.048611in}{0.000000in}}{\pgfqpoint{0.000000in}{0.000000in}}{%
\pgfpathmoveto{\pgfqpoint{0.000000in}{0.000000in}}%
\pgfpathlineto{\pgfqpoint{-0.048611in}{0.000000in}}%
\pgfusepath{stroke,fill}%
}%
\begin{pgfscope}%
\pgfsys@transformshift{0.640063in}{1.182181in}%
\pgfsys@useobject{currentmarker}{}%
\end{pgfscope}%
\end{pgfscope}%
\begin{pgfscope}%
\pgftext[x=0.424783in,y=1.143919in,left,base]{\rmfamily\fontsize{8.000000}{9.600000}\selectfont \(\displaystyle 25\)}%
\end{pgfscope}%
\begin{pgfscope}%
\pgfsetbuttcap%
\pgfsetroundjoin%
\definecolor{currentfill}{rgb}{0.000000,0.000000,0.000000}%
\pgfsetfillcolor{currentfill}%
\pgfsetlinewidth{0.803000pt}%
\definecolor{currentstroke}{rgb}{0.000000,0.000000,0.000000}%
\pgfsetstrokecolor{currentstroke}%
\pgfsetdash{}{0pt}%
\pgfsys@defobject{currentmarker}{\pgfqpoint{-0.048611in}{0.000000in}}{\pgfqpoint{0.000000in}{0.000000in}}{%
\pgfpathmoveto{\pgfqpoint{0.000000in}{0.000000in}}%
\pgfpathlineto{\pgfqpoint{-0.048611in}{0.000000in}}%
\pgfusepath{stroke,fill}%
}%
\begin{pgfscope}%
\pgfsys@transformshift{0.640063in}{1.447036in}%
\pgfsys@useobject{currentmarker}{}%
\end{pgfscope}%
\end{pgfscope}%
\begin{pgfscope}%
\pgftext[x=0.424783in,y=1.408774in,left,base]{\rmfamily\fontsize{8.000000}{9.600000}\selectfont \(\displaystyle 50\)}%
\end{pgfscope}%
\begin{pgfscope}%
\pgfsetbuttcap%
\pgfsetroundjoin%
\definecolor{currentfill}{rgb}{0.000000,0.000000,0.000000}%
\pgfsetfillcolor{currentfill}%
\pgfsetlinewidth{0.803000pt}%
\definecolor{currentstroke}{rgb}{0.000000,0.000000,0.000000}%
\pgfsetstrokecolor{currentstroke}%
\pgfsetdash{}{0pt}%
\pgfsys@defobject{currentmarker}{\pgfqpoint{-0.048611in}{0.000000in}}{\pgfqpoint{0.000000in}{0.000000in}}{%
\pgfpathmoveto{\pgfqpoint{0.000000in}{0.000000in}}%
\pgfpathlineto{\pgfqpoint{-0.048611in}{0.000000in}}%
\pgfusepath{stroke,fill}%
}%
\begin{pgfscope}%
\pgfsys@transformshift{0.640063in}{1.711892in}%
\pgfsys@useobject{currentmarker}{}%
\end{pgfscope}%
\end{pgfscope}%
\begin{pgfscope}%
\pgftext[x=0.424783in,y=1.673630in,left,base]{\rmfamily\fontsize{8.000000}{9.600000}\selectfont \(\displaystyle 75\)}%
\end{pgfscope}%
\begin{pgfscope}%
\pgfsetbuttcap%
\pgfsetroundjoin%
\definecolor{currentfill}{rgb}{0.000000,0.000000,0.000000}%
\pgfsetfillcolor{currentfill}%
\pgfsetlinewidth{0.803000pt}%
\definecolor{currentstroke}{rgb}{0.000000,0.000000,0.000000}%
\pgfsetstrokecolor{currentstroke}%
\pgfsetdash{}{0pt}%
\pgfsys@defobject{currentmarker}{\pgfqpoint{-0.048611in}{0.000000in}}{\pgfqpoint{0.000000in}{0.000000in}}{%
\pgfpathmoveto{\pgfqpoint{0.000000in}{0.000000in}}%
\pgfpathlineto{\pgfqpoint{-0.048611in}{0.000000in}}%
\pgfusepath{stroke,fill}%
}%
\begin{pgfscope}%
\pgfsys@transformshift{0.640063in}{1.976747in}%
\pgfsys@useobject{currentmarker}{}%
\end{pgfscope}%
\end{pgfscope}%
\begin{pgfscope}%
\pgftext[x=0.365755in,y=1.938485in,left,base]{\rmfamily\fontsize{8.000000}{9.600000}\selectfont \(\displaystyle 100\)}%
\end{pgfscope}%
\begin{pgfscope}%
\pgfsetbuttcap%
\pgfsetroundjoin%
\definecolor{currentfill}{rgb}{0.000000,0.000000,0.000000}%
\pgfsetfillcolor{currentfill}%
\pgfsetlinewidth{0.803000pt}%
\definecolor{currentstroke}{rgb}{0.000000,0.000000,0.000000}%
\pgfsetstrokecolor{currentstroke}%
\pgfsetdash{}{0pt}%
\pgfsys@defobject{currentmarker}{\pgfqpoint{-0.048611in}{0.000000in}}{\pgfqpoint{0.000000in}{0.000000in}}{%
\pgfpathmoveto{\pgfqpoint{0.000000in}{0.000000in}}%
\pgfpathlineto{\pgfqpoint{-0.048611in}{0.000000in}}%
\pgfusepath{stroke,fill}%
}%
\begin{pgfscope}%
\pgfsys@transformshift{0.640063in}{2.241603in}%
\pgfsys@useobject{currentmarker}{}%
\end{pgfscope}%
\end{pgfscope}%
\begin{pgfscope}%
\pgftext[x=0.365755in,y=2.203340in,left,base]{\rmfamily\fontsize{8.000000}{9.600000}\selectfont \(\displaystyle 125\)}%
\end{pgfscope}%
\begin{pgfscope}%
\pgfsetbuttcap%
\pgfsetroundjoin%
\definecolor{currentfill}{rgb}{0.000000,0.000000,0.000000}%
\pgfsetfillcolor{currentfill}%
\pgfsetlinewidth{0.803000pt}%
\definecolor{currentstroke}{rgb}{0.000000,0.000000,0.000000}%
\pgfsetstrokecolor{currentstroke}%
\pgfsetdash{}{0pt}%
\pgfsys@defobject{currentmarker}{\pgfqpoint{-0.048611in}{0.000000in}}{\pgfqpoint{0.000000in}{0.000000in}}{%
\pgfpathmoveto{\pgfqpoint{0.000000in}{0.000000in}}%
\pgfpathlineto{\pgfqpoint{-0.048611in}{0.000000in}}%
\pgfusepath{stroke,fill}%
}%
\begin{pgfscope}%
\pgfsys@transformshift{0.640063in}{2.506458in}%
\pgfsys@useobject{currentmarker}{}%
\end{pgfscope}%
\end{pgfscope}%
\begin{pgfscope}%
\pgftext[x=0.365755in,y=2.468196in,left,base]{\rmfamily\fontsize{8.000000}{9.600000}\selectfont \(\displaystyle 150\)}%
\end{pgfscope}%
\begin{pgfscope}%
\pgftext[x=0.277406in,y=1.552979in,,bottom,rotate=90.000000]{\rmfamily\fontsize{10.000000}{12.000000}\selectfont \(\displaystyle (\overline{Z}-\overline{Z}_{*})^{2}\)}%
\end{pgfscope}%
\begin{pgfscope}%
\pgfpathrectangle{\pgfqpoint{0.640063in}{0.493557in}}{\pgfqpoint{3.286097in}{2.118843in}} %
\pgfusepath{clip}%
\pgfsetrectcap%
\pgfsetroundjoin%
\pgfsetlinewidth{1.505625pt}%
\definecolor{currentstroke}{rgb}{0.121569,0.466667,0.705882}%
\pgfsetstrokecolor{currentstroke}%
\pgfsetdash{}{0pt}%
\pgfpathmoveto{\pgfqpoint{0.640063in}{1.136713in}}%
\pgfpathlineto{\pgfqpoint{0.691610in}{1.107506in}}%
\pgfpathlineto{\pgfqpoint{0.743156in}{1.080385in}}%
\pgfpathlineto{\pgfqpoint{0.794703in}{1.055349in}}%
\pgfpathlineto{\pgfqpoint{0.846249in}{1.032398in}}%
\pgfpathlineto{\pgfqpoint{0.897796in}{1.011533in}}%
\pgfpathlineto{\pgfqpoint{0.949343in}{0.992753in}}%
\pgfpathlineto{\pgfqpoint{1.000889in}{0.976059in}}%
\pgfpathlineto{\pgfqpoint{1.052436in}{0.961450in}}%
\pgfpathlineto{\pgfqpoint{1.103983in}{0.948927in}}%
\pgfpathlineto{\pgfqpoint{1.129756in}{0.943455in}}%
\pgfpathlineto{\pgfqpoint{1.142643in}{0.952903in}}%
\pgfpathlineto{\pgfqpoint{1.155529in}{0.940288in}}%
\pgfpathlineto{\pgfqpoint{1.168416in}{0.946326in}}%
\pgfpathlineto{\pgfqpoint{1.181302in}{0.943576in}}%
\pgfpathlineto{\pgfqpoint{1.194189in}{0.939464in}}%
\pgfpathlineto{\pgfqpoint{1.219962in}{0.928482in}}%
\pgfpathlineto{\pgfqpoint{1.232849in}{0.942567in}}%
\pgfpathlineto{\pgfqpoint{1.245736in}{1.015589in}}%
\pgfpathlineto{\pgfqpoint{1.258622in}{1.078895in}}%
\pgfpathlineto{\pgfqpoint{1.271509in}{0.968688in}}%
\pgfpathlineto{\pgfqpoint{1.284396in}{1.081608in}}%
\pgfpathlineto{\pgfqpoint{1.297282in}{1.078529in}}%
\pgfpathlineto{\pgfqpoint{1.323056in}{1.067897in}}%
\pgfpathlineto{\pgfqpoint{1.348829in}{1.054832in}}%
\pgfpathlineto{\pgfqpoint{1.361716in}{1.048819in}}%
\pgfpathlineto{\pgfqpoint{1.374602in}{1.046629in}}%
\pgfpathlineto{\pgfqpoint{1.387489in}{1.038080in}}%
\pgfpathlineto{\pgfqpoint{1.400376in}{1.034797in}}%
\pgfpathlineto{\pgfqpoint{1.413262in}{1.029500in}}%
\pgfpathlineto{\pgfqpoint{1.426149in}{1.022593in}}%
\pgfpathlineto{\pgfqpoint{1.529242in}{0.986435in}}%
\pgfpathlineto{\pgfqpoint{1.580789in}{0.971321in}}%
\pgfpathlineto{\pgfqpoint{1.606562in}{0.964123in}}%
\pgfpathlineto{\pgfqpoint{1.645222in}{0.954048in}}%
\pgfpathlineto{\pgfqpoint{1.696769in}{0.942662in}}%
\pgfpathlineto{\pgfqpoint{1.722542in}{0.938026in}}%
\pgfpathlineto{\pgfqpoint{1.735429in}{0.935054in}}%
\pgfpathlineto{\pgfqpoint{1.748315in}{0.934090in}}%
\pgfpathlineto{\pgfqpoint{1.774089in}{0.929954in}}%
\pgfpathlineto{\pgfqpoint{1.825635in}{0.923868in}}%
\pgfpathlineto{\pgfqpoint{1.851409in}{0.921394in}}%
\pgfpathlineto{\pgfqpoint{1.915842in}{0.917813in}}%
\pgfpathlineto{\pgfqpoint{1.980275in}{0.917619in}}%
\pgfpathlineto{\pgfqpoint{2.031822in}{0.919720in}}%
\pgfpathlineto{\pgfqpoint{2.134915in}{0.930179in}}%
\pgfpathlineto{\pgfqpoint{2.199348in}{0.941260in}}%
\pgfpathlineto{\pgfqpoint{2.263782in}{0.956235in}}%
\pgfpathlineto{\pgfqpoint{2.341102in}{0.977560in}}%
\pgfpathlineto{\pgfqpoint{2.405535in}{0.998991in}}%
\pgfpathlineto{\pgfqpoint{2.482855in}{1.029744in}}%
\pgfpathlineto{\pgfqpoint{2.534401in}{1.051984in}}%
\pgfpathlineto{\pgfqpoint{2.624608in}{1.096464in}}%
\pgfpathlineto{\pgfqpoint{2.637495in}{1.105564in}}%
\pgfpathlineto{\pgfqpoint{2.650381in}{1.111404in}}%
\pgfpathlineto{\pgfqpoint{2.676155in}{1.127270in}}%
\pgfpathlineto{\pgfqpoint{2.689041in}{1.132331in}}%
\pgfpathlineto{\pgfqpoint{2.701928in}{1.141464in}}%
\pgfpathlineto{\pgfqpoint{2.714815in}{1.147483in}}%
\pgfpathlineto{\pgfqpoint{2.727701in}{1.155838in}}%
\pgfpathlineto{\pgfqpoint{2.740588in}{1.166465in}}%
\pgfpathlineto{\pgfqpoint{2.753475in}{1.173573in}}%
\pgfpathlineto{\pgfqpoint{2.766361in}{1.183209in}}%
\pgfpathlineto{\pgfqpoint{2.792135in}{1.195886in}}%
\pgfpathlineto{\pgfqpoint{2.817908in}{1.216089in}}%
\pgfpathlineto{\pgfqpoint{2.830794in}{1.223458in}}%
\pgfpathlineto{\pgfqpoint{2.882341in}{1.261723in}}%
\pgfpathlineto{\pgfqpoint{2.895228in}{1.270024in}}%
\pgfpathlineto{\pgfqpoint{2.908114in}{1.283569in}}%
\pgfpathlineto{\pgfqpoint{2.921001in}{1.290576in}}%
\pgfpathlineto{\pgfqpoint{2.933888in}{1.299615in}}%
\pgfpathlineto{\pgfqpoint{2.946774in}{1.312886in}}%
\pgfpathlineto{\pgfqpoint{2.959661in}{1.322289in}}%
\pgfpathlineto{\pgfqpoint{2.972548in}{1.333776in}}%
\pgfpathlineto{\pgfqpoint{3.011208in}{1.362166in}}%
\pgfpathlineto{\pgfqpoint{3.024094in}{1.375702in}}%
\pgfpathlineto{\pgfqpoint{3.036981in}{1.386974in}}%
\pgfpathlineto{\pgfqpoint{3.049868in}{1.393051in}}%
\pgfpathlineto{\pgfqpoint{3.075641in}{1.421544in}}%
\pgfpathlineto{\pgfqpoint{3.127188in}{1.465826in}}%
\pgfpathlineto{\pgfqpoint{3.152961in}{1.490250in}}%
\pgfpathlineto{\pgfqpoint{3.165848in}{1.500387in}}%
\pgfpathlineto{\pgfqpoint{3.178734in}{1.516062in}}%
\pgfpathlineto{\pgfqpoint{3.191621in}{1.528628in}}%
\pgfpathlineto{\pgfqpoint{3.204508in}{1.543001in}}%
\pgfpathlineto{\pgfqpoint{3.217394in}{1.553689in}}%
\pgfpathlineto{\pgfqpoint{3.230281in}{1.567659in}}%
\pgfpathlineto{\pgfqpoint{3.281827in}{1.616773in}}%
\pgfpathlineto{\pgfqpoint{3.294714in}{1.634407in}}%
\pgfpathlineto{\pgfqpoint{3.307601in}{1.645058in}}%
\pgfpathlineto{\pgfqpoint{3.320487in}{1.658493in}}%
\pgfpathlineto{\pgfqpoint{3.333374in}{1.676301in}}%
\pgfpathlineto{\pgfqpoint{3.346261in}{1.687641in}}%
\pgfpathlineto{\pgfqpoint{3.372034in}{1.717389in}}%
\pgfpathlineto{\pgfqpoint{3.384921in}{1.732924in}}%
\pgfpathlineto{\pgfqpoint{3.397807in}{1.743964in}}%
\pgfpathlineto{\pgfqpoint{3.423581in}{1.775673in}}%
\pgfpathlineto{\pgfqpoint{3.462241in}{1.816056in}}%
\pgfpathlineto{\pgfqpoint{3.475127in}{1.833937in}}%
\pgfpathlineto{\pgfqpoint{3.500901in}{1.864340in}}%
\pgfpathlineto{\pgfqpoint{3.513787in}{1.884276in}}%
\pgfpathlineto{\pgfqpoint{3.539561in}{1.913622in}}%
\pgfpathlineto{\pgfqpoint{3.552447in}{1.930787in}}%
\pgfpathlineto{\pgfqpoint{3.578221in}{1.960877in}}%
\pgfpathlineto{\pgfqpoint{3.603994in}{1.997790in}}%
\pgfpathlineto{\pgfqpoint{3.616881in}{2.009300in}}%
\pgfpathlineto{\pgfqpoint{3.642654in}{2.044369in}}%
\pgfpathlineto{\pgfqpoint{3.655540in}{2.066365in}}%
\pgfpathlineto{\pgfqpoint{3.668427in}{2.077234in}}%
\pgfpathlineto{\pgfqpoint{3.694200in}{2.114728in}}%
\pgfpathlineto{\pgfqpoint{3.707087in}{2.130323in}}%
\pgfpathlineto{\pgfqpoint{3.797294in}{2.255718in}}%
\pgfpathlineto{\pgfqpoint{3.810180in}{2.277955in}}%
\pgfpathlineto{\pgfqpoint{3.835954in}{2.315247in}}%
\pgfpathlineto{\pgfqpoint{3.848840in}{2.330875in}}%
\pgfpathlineto{\pgfqpoint{3.913274in}{2.427792in}}%
\pgfpathlineto{\pgfqpoint{3.926160in}{2.446108in}}%
\pgfpathlineto{\pgfqpoint{3.926160in}{2.446108in}}%
\pgfusepath{stroke}%
\end{pgfscope}%
\begin{pgfscope}%
\pgfpathrectangle{\pgfqpoint{0.640063in}{0.493557in}}{\pgfqpoint{3.286097in}{2.118843in}} %
\pgfusepath{clip}%
\pgfsetbuttcap%
\pgfsetroundjoin%
\definecolor{currentfill}{rgb}{1.000000,1.000000,1.000000}%
\pgfsetfillcolor{currentfill}%
\pgfsetlinewidth{1.003750pt}%
\definecolor{currentstroke}{rgb}{0.000000,0.000000,0.000000}%
\pgfsetstrokecolor{currentstroke}%
\pgfsetdash{}{0pt}%
\pgfsys@defobject{currentmarker}{\pgfqpoint{-0.055556in}{-0.055556in}}{\pgfqpoint{0.055556in}{0.055556in}}{%
\pgfpathmoveto{\pgfqpoint{0.000000in}{-0.055556in}}%
\pgfpathcurveto{\pgfqpoint{0.014734in}{-0.055556in}}{\pgfqpoint{0.028866in}{-0.049702in}}{\pgfqpoint{0.039284in}{-0.039284in}}%
\pgfpathcurveto{\pgfqpoint{0.049702in}{-0.028866in}}{\pgfqpoint{0.055556in}{-0.014734in}}{\pgfqpoint{0.055556in}{0.000000in}}%
\pgfpathcurveto{\pgfqpoint{0.055556in}{0.014734in}}{\pgfqpoint{0.049702in}{0.028866in}}{\pgfqpoint{0.039284in}{0.039284in}}%
\pgfpathcurveto{\pgfqpoint{0.028866in}{0.049702in}}{\pgfqpoint{0.014734in}{0.055556in}}{\pgfqpoint{0.000000in}{0.055556in}}%
\pgfpathcurveto{\pgfqpoint{-0.014734in}{0.055556in}}{\pgfqpoint{-0.028866in}{0.049702in}}{\pgfqpoint{-0.039284in}{0.039284in}}%
\pgfpathcurveto{\pgfqpoint{-0.049702in}{0.028866in}}{\pgfqpoint{-0.055556in}{0.014734in}}{\pgfqpoint{-0.055556in}{0.000000in}}%
\pgfpathcurveto{\pgfqpoint{-0.055556in}{-0.014734in}}{\pgfqpoint{-0.049702in}{-0.028866in}}{\pgfqpoint{-0.039284in}{-0.039284in}}%
\pgfpathcurveto{\pgfqpoint{-0.028866in}{-0.049702in}}{\pgfqpoint{-0.014734in}{-0.055556in}}{\pgfqpoint{0.000000in}{-0.055556in}}%
\pgfpathclose%
\pgfusepath{stroke,fill}%
}%
\begin{pgfscope}%
\pgfsys@transformshift{3.761855in}{2.200497in}%
\pgfsys@useobject{currentmarker}{}%
\end{pgfscope}%
\end{pgfscope}%
\begin{pgfscope}%
\pgfpathrectangle{\pgfqpoint{0.640063in}{0.493557in}}{\pgfqpoint{3.286097in}{2.118843in}} %
\pgfusepath{clip}%
\pgfsetbuttcap%
\pgfsetroundjoin%
\definecolor{currentfill}{rgb}{1.000000,1.000000,1.000000}%
\pgfsetfillcolor{currentfill}%
\pgfsetlinewidth{1.003750pt}%
\definecolor{currentstroke}{rgb}{0.000000,0.000000,0.000000}%
\pgfsetstrokecolor{currentstroke}%
\pgfsetdash{}{0pt}%
\pgfsys@defobject{currentmarker}{\pgfqpoint{-0.055556in}{-0.055556in}}{\pgfqpoint{0.055556in}{0.055556in}}{%
\pgfpathmoveto{\pgfqpoint{0.000000in}{-0.055556in}}%
\pgfpathcurveto{\pgfqpoint{0.014734in}{-0.055556in}}{\pgfqpoint{0.028866in}{-0.049702in}}{\pgfqpoint{0.039284in}{-0.039284in}}%
\pgfpathcurveto{\pgfqpoint{0.049702in}{-0.028866in}}{\pgfqpoint{0.055556in}{-0.014734in}}{\pgfqpoint{0.055556in}{0.000000in}}%
\pgfpathcurveto{\pgfqpoint{0.055556in}{0.014734in}}{\pgfqpoint{0.049702in}{0.028866in}}{\pgfqpoint{0.039284in}{0.039284in}}%
\pgfpathcurveto{\pgfqpoint{0.028866in}{0.049702in}}{\pgfqpoint{0.014734in}{0.055556in}}{\pgfqpoint{0.000000in}{0.055556in}}%
\pgfpathcurveto{\pgfqpoint{-0.014734in}{0.055556in}}{\pgfqpoint{-0.028866in}{0.049702in}}{\pgfqpoint{-0.039284in}{0.039284in}}%
\pgfpathcurveto{\pgfqpoint{-0.049702in}{0.028866in}}{\pgfqpoint{-0.055556in}{0.014734in}}{\pgfqpoint{-0.055556in}{0.000000in}}%
\pgfpathcurveto{\pgfqpoint{-0.055556in}{-0.014734in}}{\pgfqpoint{-0.049702in}{-0.028866in}}{\pgfqpoint{-0.039284in}{-0.039284in}}%
\pgfpathcurveto{\pgfqpoint{-0.028866in}{-0.049702in}}{\pgfqpoint{-0.014734in}{-0.055556in}}{\pgfqpoint{0.000000in}{-0.055556in}}%
\pgfpathclose%
\pgfusepath{stroke,fill}%
}%
\begin{pgfscope}%
\pgfsys@transformshift{3.761855in}{2.200497in}%
\pgfsys@useobject{currentmarker}{}%
\end{pgfscope}%
\begin{pgfscope}%
\pgfsys@transformshift{2.932116in}{1.299615in}%
\pgfsys@useobject{currentmarker}{}%
\end{pgfscope}%
\begin{pgfscope}%
\pgfsys@transformshift{1.890757in}{0.918903in}%
\pgfsys@useobject{currentmarker}{}%
\end{pgfscope}%
\begin{pgfscope}%
\pgfsys@transformshift{1.943771in}{0.917369in}%
\pgfsys@useobject{currentmarker}{}%
\end{pgfscope}%
\begin{pgfscope}%
\pgfsys@transformshift{1.949998in}{0.917326in}%
\pgfsys@useobject{currentmarker}{}%
\end{pgfscope}%
\end{pgfscope}%
\begin{pgfscope}%
\pgfpathrectangle{\pgfqpoint{0.640063in}{0.493557in}}{\pgfqpoint{3.286097in}{2.118843in}} %
\pgfusepath{clip}%
\pgfsetbuttcap%
\pgfsetroundjoin%
\definecolor{currentfill}{rgb}{0.000000,0.000000,0.000000}%
\pgfsetfillcolor{currentfill}%
\pgfsetlinewidth{1.003750pt}%
\definecolor{currentstroke}{rgb}{0.000000,0.000000,0.000000}%
\pgfsetstrokecolor{currentstroke}%
\pgfsetdash{}{0pt}%
\pgfsys@defobject{currentmarker}{\pgfqpoint{-0.069444in}{-0.069444in}}{\pgfqpoint{0.069444in}{0.069444in}}{%
\pgfpathmoveto{\pgfqpoint{-0.069444in}{-0.069444in}}%
\pgfpathlineto{\pgfqpoint{0.069444in}{0.069444in}}%
\pgfpathmoveto{\pgfqpoint{-0.069444in}{0.069444in}}%
\pgfpathlineto{\pgfqpoint{0.069444in}{-0.069444in}}%
\pgfusepath{stroke,fill}%
}%
\begin{pgfscope}%
\pgfsys@transformshift{1.954502in}{0.917326in}%
\pgfsys@useobject{currentmarker}{}%
\end{pgfscope}%
\end{pgfscope}%
\begin{pgfscope}%
\pgfsetrectcap%
\pgfsetmiterjoin%
\pgfsetlinewidth{0.803000pt}%
\definecolor{currentstroke}{rgb}{0.000000,0.000000,0.000000}%
\pgfsetstrokecolor{currentstroke}%
\pgfsetdash{}{0pt}%
\pgfpathmoveto{\pgfqpoint{0.640063in}{0.493557in}}%
\pgfpathlineto{\pgfqpoint{0.640063in}{2.612400in}}%
\pgfusepath{stroke}%
\end{pgfscope}%
\begin{pgfscope}%
\pgfsetrectcap%
\pgfsetmiterjoin%
\pgfsetlinewidth{0.803000pt}%
\definecolor{currentstroke}{rgb}{0.000000,0.000000,0.000000}%
\pgfsetstrokecolor{currentstroke}%
\pgfsetdash{}{0pt}%
\pgfpathmoveto{\pgfqpoint{3.926160in}{0.493557in}}%
\pgfpathlineto{\pgfqpoint{3.926160in}{2.612400in}}%
\pgfusepath{stroke}%
\end{pgfscope}%
\begin{pgfscope}%
\pgfsetrectcap%
\pgfsetmiterjoin%
\pgfsetlinewidth{0.803000pt}%
\definecolor{currentstroke}{rgb}{0.000000,0.000000,0.000000}%
\pgfsetstrokecolor{currentstroke}%
\pgfsetdash{}{0pt}%
\pgfpathmoveto{\pgfqpoint{0.640063in}{0.493557in}}%
\pgfpathlineto{\pgfqpoint{3.926160in}{0.493557in}}%
\pgfusepath{stroke}%
\end{pgfscope}%
\begin{pgfscope}%
\pgfsetrectcap%
\pgfsetmiterjoin%
\pgfsetlinewidth{0.803000pt}%
\definecolor{currentstroke}{rgb}{0.000000,0.000000,0.000000}%
\pgfsetstrokecolor{currentstroke}%
\pgfsetdash{}{0pt}%
\pgfpathmoveto{\pgfqpoint{0.640063in}{2.612400in}}%
\pgfpathlineto{\pgfqpoint{3.926160in}{2.612400in}}%
\pgfusepath{stroke}%
\end{pgfscope}%
\begin{pgfscope}%
\pgftext[x=3.761855in,y=1.988613in,left,base]{\rmfamily\fontsize{8.000000}{9.600000}\selectfont 0}%
\end{pgfscope}%
\begin{pgfscope}%
\pgftext[x=3.334663in,y=2.327628in,left,base]{\rmfamily\fontsize{8.000000}{9.600000}\selectfont \(\displaystyle \mu = 15.50\)}%
\end{pgfscope}%
\begin{pgfscope}%
\pgftext[x=2.932116in,y=1.087730in,left,base]{\rmfamily\fontsize{8.000000}{9.600000}\selectfont 1}%
\end{pgfscope}%
\begin{pgfscope}%
\pgftext[x=2.504923in,y=1.426745in,left,base]{\rmfamily\fontsize{8.000000}{9.600000}\selectfont \(\displaystyle \mu = 13.80\)}%
\end{pgfscope}%
\begin{pgfscope}%
\pgftext[x=1.890757in,y=0.707019in,left,base]{\rmfamily\fontsize{8.000000}{9.600000}\selectfont 2}%
\end{pgfscope}%
\begin{pgfscope}%
\pgftext[x=1.463565in,y=1.046034in,left,base]{\rmfamily\fontsize{8.000000}{9.600000}\selectfont \(\displaystyle \mu =  9.63\)}%
\end{pgfscope}%
\end{pgfpicture}%
\makeatother%
\endgroup%

%% file: fig_l9dJ.pgf
\begingroup%
\makeatletter%
\begin{pgfpicture}%
\pgfpathrectangle{\pgfpointorigin}{\pgfqpoint{6.226650in}{2.594438in}}%
\pgfusepath{use as bounding box, clip}%
\begin{pgfscope}%
\pgfsetbuttcap%
\pgfsetmiterjoin%
\definecolor{currentfill}{rgb}{1.000000,1.000000,1.000000}%
\pgfsetfillcolor{currentfill}%
\pgfsetlinewidth{0.000000pt}%
\definecolor{currentstroke}{rgb}{1.000000,1.000000,1.000000}%
\pgfsetstrokecolor{currentstroke}%
\pgfsetdash{}{0pt}%
\pgfpathmoveto{\pgfqpoint{0.000000in}{0.000000in}}%
\pgfpathlineto{\pgfqpoint{6.226650in}{0.000000in}}%
\pgfpathlineto{\pgfqpoint{6.226650in}{2.594438in}}%
\pgfpathlineto{\pgfqpoint{0.000000in}{2.594438in}}%
\pgfpathclose%
\pgfusepath{fill}%
\end{pgfscope}%
\begin{pgfscope}%
\pgfsetbuttcap%
\pgfsetmiterjoin%
\definecolor{currentfill}{rgb}{1.000000,1.000000,1.000000}%
\pgfsetfillcolor{currentfill}%
\pgfsetlinewidth{0.000000pt}%
\definecolor{currentstroke}{rgb}{0.000000,0.000000,0.000000}%
\pgfsetstrokecolor{currentstroke}%
\pgfsetstrokeopacity{0.000000}%
\pgfsetdash{}{0pt}%
\pgfpathmoveto{\pgfqpoint{0.658967in}{0.493557in}}%
\pgfpathlineto{\pgfqpoint{2.302875in}{0.493557in}}%
\pgfpathlineto{\pgfqpoint{2.302875in}{2.295449in}}%
\pgfpathlineto{\pgfqpoint{0.658967in}{2.295449in}}%
\pgfpathclose%
\pgfusepath{fill}%
\end{pgfscope}%
\begin{pgfscope}%
\pgfsetbuttcap%
\pgfsetroundjoin%
\definecolor{currentfill}{rgb}{0.000000,0.000000,0.000000}%
\pgfsetfillcolor{currentfill}%
\pgfsetlinewidth{0.803000pt}%
\definecolor{currentstroke}{rgb}{0.000000,0.000000,0.000000}%
\pgfsetstrokecolor{currentstroke}%
\pgfsetdash{}{0pt}%
\pgfsys@defobject{currentmarker}{\pgfqpoint{0.000000in}{-0.048611in}}{\pgfqpoint{0.000000in}{0.000000in}}{%
\pgfpathmoveto{\pgfqpoint{0.000000in}{0.000000in}}%
\pgfpathlineto{\pgfqpoint{0.000000in}{-0.048611in}}%
\pgfusepath{stroke,fill}%
}%
\begin{pgfscope}%
\pgfsys@transformshift{1.020279in}{0.493557in}%
\pgfsys@useobject{currentmarker}{}%
\end{pgfscope}%
\end{pgfscope}%
\begin{pgfscope}%
\pgftext[x=1.020279in,y=0.396335in,,top]{\rmfamily\fontsize{8.000000}{9.600000}\selectfont \(\displaystyle 4.0\)}%
\end{pgfscope}%
\begin{pgfscope}%
\pgfsetbuttcap%
\pgfsetroundjoin%
\definecolor{currentfill}{rgb}{0.000000,0.000000,0.000000}%
\pgfsetfillcolor{currentfill}%
\pgfsetlinewidth{0.803000pt}%
\definecolor{currentstroke}{rgb}{0.000000,0.000000,0.000000}%
\pgfsetstrokecolor{currentstroke}%
\pgfsetdash{}{0pt}%
\pgfsys@defobject{currentmarker}{\pgfqpoint{0.000000in}{-0.048611in}}{\pgfqpoint{0.000000in}{0.000000in}}{%
\pgfpathmoveto{\pgfqpoint{0.000000in}{0.000000in}}%
\pgfpathlineto{\pgfqpoint{0.000000in}{-0.048611in}}%
\pgfusepath{stroke,fill}%
}%
\begin{pgfscope}%
\pgfsys@transformshift{1.476420in}{0.493557in}%
\pgfsys@useobject{currentmarker}{}%
\end{pgfscope}%
\end{pgfscope}%
\begin{pgfscope}%
\pgftext[x=1.476420in,y=0.396335in,,top]{\rmfamily\fontsize{8.000000}{9.600000}\selectfont \(\displaystyle 4.5\)}%
\end{pgfscope}%
\begin{pgfscope}%
\pgfsetbuttcap%
\pgfsetroundjoin%
\definecolor{currentfill}{rgb}{0.000000,0.000000,0.000000}%
\pgfsetfillcolor{currentfill}%
\pgfsetlinewidth{0.803000pt}%
\definecolor{currentstroke}{rgb}{0.000000,0.000000,0.000000}%
\pgfsetstrokecolor{currentstroke}%
\pgfsetdash{}{0pt}%
\pgfsys@defobject{currentmarker}{\pgfqpoint{0.000000in}{-0.048611in}}{\pgfqpoint{0.000000in}{0.000000in}}{%
\pgfpathmoveto{\pgfqpoint{0.000000in}{0.000000in}}%
\pgfpathlineto{\pgfqpoint{0.000000in}{-0.048611in}}%
\pgfusepath{stroke,fill}%
}%
\begin{pgfscope}%
\pgfsys@transformshift{1.932561in}{0.493557in}%
\pgfsys@useobject{currentmarker}{}%
\end{pgfscope}%
\end{pgfscope}%
\begin{pgfscope}%
\pgftext[x=1.932561in,y=0.396335in,,top]{\rmfamily\fontsize{8.000000}{9.600000}\selectfont \(\displaystyle 5.0\)}%
\end{pgfscope}%
\begin{pgfscope}%
\pgftext[x=1.480921in,y=0.242655in,,top]{\rmfamily\fontsize{10.000000}{12.000000}\selectfont \(\displaystyle \mu\)}%
\end{pgfscope}%
\begin{pgfscope}%
\pgfsetbuttcap%
\pgfsetroundjoin%
\definecolor{currentfill}{rgb}{0.000000,0.000000,0.000000}%
\pgfsetfillcolor{currentfill}%
\pgfsetlinewidth{0.803000pt}%
\definecolor{currentstroke}{rgb}{0.000000,0.000000,0.000000}%
\pgfsetstrokecolor{currentstroke}%
\pgfsetdash{}{0pt}%
\pgfsys@defobject{currentmarker}{\pgfqpoint{-0.048611in}{0.000000in}}{\pgfqpoint{0.000000in}{0.000000in}}{%
\pgfpathmoveto{\pgfqpoint{0.000000in}{0.000000in}}%
\pgfpathlineto{\pgfqpoint{-0.048611in}{0.000000in}}%
\pgfusepath{stroke,fill}%
}%
\begin{pgfscope}%
\pgfsys@transformshift{0.658967in}{0.693767in}%
\pgfsys@useobject{currentmarker}{}%
\end{pgfscope}%
\end{pgfscope}%
\begin{pgfscope}%
\pgftext[x=0.319072in,y=0.655505in,left,base]{\rmfamily\fontsize{8.000000}{9.600000}\selectfont \(\displaystyle -0.2\)}%
\end{pgfscope}%
\begin{pgfscope}%
\pgfsetbuttcap%
\pgfsetroundjoin%
\definecolor{currentfill}{rgb}{0.000000,0.000000,0.000000}%
\pgfsetfillcolor{currentfill}%
\pgfsetlinewidth{0.803000pt}%
\definecolor{currentstroke}{rgb}{0.000000,0.000000,0.000000}%
\pgfsetstrokecolor{currentstroke}%
\pgfsetdash{}{0pt}%
\pgfsys@defobject{currentmarker}{\pgfqpoint{-0.048611in}{0.000000in}}{\pgfqpoint{0.000000in}{0.000000in}}{%
\pgfpathmoveto{\pgfqpoint{0.000000in}{0.000000in}}%
\pgfpathlineto{\pgfqpoint{-0.048611in}{0.000000in}}%
\pgfusepath{stroke,fill}%
}%
\begin{pgfscope}%
\pgfsys@transformshift{0.658967in}{1.094188in}%
\pgfsys@useobject{currentmarker}{}%
\end{pgfscope}%
\end{pgfscope}%
\begin{pgfscope}%
\pgftext[x=0.410894in,y=1.055925in,left,base]{\rmfamily\fontsize{8.000000}{9.600000}\selectfont \(\displaystyle 0.0\)}%
\end{pgfscope}%
\begin{pgfscope}%
\pgfsetbuttcap%
\pgfsetroundjoin%
\definecolor{currentfill}{rgb}{0.000000,0.000000,0.000000}%
\pgfsetfillcolor{currentfill}%
\pgfsetlinewidth{0.803000pt}%
\definecolor{currentstroke}{rgb}{0.000000,0.000000,0.000000}%
\pgfsetstrokecolor{currentstroke}%
\pgfsetdash{}{0pt}%
\pgfsys@defobject{currentmarker}{\pgfqpoint{-0.048611in}{0.000000in}}{\pgfqpoint{0.000000in}{0.000000in}}{%
\pgfpathmoveto{\pgfqpoint{0.000000in}{0.000000in}}%
\pgfpathlineto{\pgfqpoint{-0.048611in}{0.000000in}}%
\pgfusepath{stroke,fill}%
}%
\begin{pgfscope}%
\pgfsys@transformshift{0.658967in}{1.494608in}%
\pgfsys@useobject{currentmarker}{}%
\end{pgfscope}%
\end{pgfscope}%
\begin{pgfscope}%
\pgftext[x=0.410894in,y=1.456346in,left,base]{\rmfamily\fontsize{8.000000}{9.600000}\selectfont \(\displaystyle 0.2\)}%
\end{pgfscope}%
\begin{pgfscope}%
\pgfsetbuttcap%
\pgfsetroundjoin%
\definecolor{currentfill}{rgb}{0.000000,0.000000,0.000000}%
\pgfsetfillcolor{currentfill}%
\pgfsetlinewidth{0.803000pt}%
\definecolor{currentstroke}{rgb}{0.000000,0.000000,0.000000}%
\pgfsetstrokecolor{currentstroke}%
\pgfsetdash{}{0pt}%
\pgfsys@defobject{currentmarker}{\pgfqpoint{-0.048611in}{0.000000in}}{\pgfqpoint{0.000000in}{0.000000in}}{%
\pgfpathmoveto{\pgfqpoint{0.000000in}{0.000000in}}%
\pgfpathlineto{\pgfqpoint{-0.048611in}{0.000000in}}%
\pgfusepath{stroke,fill}%
}%
\begin{pgfscope}%
\pgfsys@transformshift{0.658967in}{1.895028in}%
\pgfsys@useobject{currentmarker}{}%
\end{pgfscope}%
\end{pgfscope}%
\begin{pgfscope}%
\pgftext[x=0.410894in,y=1.856766in,left,base]{\rmfamily\fontsize{8.000000}{9.600000}\selectfont \(\displaystyle 0.4\)}%
\end{pgfscope}%
\begin{pgfscope}%
\pgfsetbuttcap%
\pgfsetroundjoin%
\definecolor{currentfill}{rgb}{0.000000,0.000000,0.000000}%
\pgfsetfillcolor{currentfill}%
\pgfsetlinewidth{0.803000pt}%
\definecolor{currentstroke}{rgb}{0.000000,0.000000,0.000000}%
\pgfsetstrokecolor{currentstroke}%
\pgfsetdash{}{0pt}%
\pgfsys@defobject{currentmarker}{\pgfqpoint{-0.048611in}{0.000000in}}{\pgfqpoint{0.000000in}{0.000000in}}{%
\pgfpathmoveto{\pgfqpoint{0.000000in}{0.000000in}}%
\pgfpathlineto{\pgfqpoint{-0.048611in}{0.000000in}}%
\pgfusepath{stroke,fill}%
}%
\begin{pgfscope}%
\pgfsys@transformshift{0.658967in}{2.295449in}%
\pgfsys@useobject{currentmarker}{}%
\end{pgfscope}%
\end{pgfscope}%
\begin{pgfscope}%
\pgftext[x=0.410894in,y=2.257186in,left,base]{\rmfamily\fontsize{8.000000}{9.600000}\selectfont \(\displaystyle 0.6\)}%
\end{pgfscope}%
\begin{pgfscope}%
\pgftext[x=0.263516in,y=1.394503in,,bottom,rotate=90.000000]{\rmfamily\fontsize{10.000000}{12.000000}\selectfont \(\displaystyle -\partial_{r}\overline{Q}_{6}\)}%
\end{pgfscope}%
\begin{pgfscope}%
\pgfpathrectangle{\pgfqpoint{0.658967in}{0.493557in}}{\pgfqpoint{1.643908in}{1.801891in}} %
\pgfusepath{clip}%
\pgfsetrectcap%
\pgfsetroundjoin%
\pgfsetlinewidth{1.505625pt}%
\definecolor{currentstroke}{rgb}{0.121569,0.466667,0.705882}%
\pgfsetstrokecolor{currentstroke}%
\pgfsetdash{}{0pt}%
\pgfpathmoveto{\pgfqpoint{1.069944in}{1.289305in}}%
\pgfpathlineto{\pgfqpoint{1.077349in}{1.348421in}}%
\pgfpathlineto{\pgfqpoint{1.085577in}{1.395231in}}%
\pgfpathlineto{\pgfqpoint{1.093805in}{1.429860in}}%
\pgfpathlineto{\pgfqpoint{1.102855in}{1.458959in}}%
\pgfpathlineto{\pgfqpoint{1.112729in}{1.483583in}}%
\pgfpathlineto{\pgfqpoint{1.123425in}{1.504596in}}%
\pgfpathlineto{\pgfqpoint{1.135766in}{1.523852in}}%
\pgfpathlineto{\pgfqpoint{1.148931in}{1.540332in}}%
\pgfpathlineto{\pgfqpoint{1.163741in}{1.555401in}}%
\pgfpathlineto{\pgfqpoint{1.181019in}{1.569757in}}%
\pgfpathlineto{\pgfqpoint{1.200766in}{1.583231in}}%
\pgfpathlineto{\pgfqpoint{1.223804in}{1.596252in}}%
\pgfpathlineto{\pgfqpoint{1.251778in}{1.609438in}}%
\pgfpathlineto{\pgfqpoint{1.286335in}{1.623132in}}%
\pgfpathlineto{\pgfqpoint{1.329119in}{1.637594in}}%
\pgfpathlineto{\pgfqpoint{1.385068in}{1.654053in}}%
\pgfpathlineto{\pgfqpoint{1.464877in}{1.675034in}}%
\pgfpathlineto{\pgfqpoint{1.646711in}{1.719751in}}%
\pgfpathlineto{\pgfqpoint{1.781646in}{1.754297in}}%
\pgfpathlineto{\pgfqpoint{1.882025in}{1.782253in}}%
\pgfpathlineto{\pgfqpoint{1.891898in}{1.785144in}}%
\pgfpathlineto{\pgfqpoint{1.891898in}{1.785144in}}%
\pgfusepath{stroke}%
\end{pgfscope}%
\begin{pgfscope}%
\pgfpathrectangle{\pgfqpoint{0.658967in}{0.493557in}}{\pgfqpoint{1.643908in}{1.801891in}} %
\pgfusepath{clip}%
\pgfsetbuttcap%
\pgfsetroundjoin%
\pgfsetlinewidth{0.501875pt}%
\definecolor{currentstroke}{rgb}{0.000000,0.000000,0.000000}%
\pgfsetstrokecolor{currentstroke}%
\pgfsetdash{{16.000000pt}{4.000000pt}}{0.000000pt}%
\pgfpathmoveto{\pgfqpoint{0.658967in}{1.629571in}}%
\pgfpathlineto{\pgfqpoint{0.716562in}{1.642580in}}%
\pgfpathlineto{\pgfqpoint{0.762637in}{1.655137in}}%
\pgfpathlineto{\pgfqpoint{0.800485in}{1.667706in}}%
\pgfpathlineto{\pgfqpoint{0.830105in}{1.679729in}}%
\pgfpathlineto{\pgfqpoint{0.854788in}{1.691956in}}%
\pgfpathlineto{\pgfqpoint{0.876180in}{1.704951in}}%
\pgfpathlineto{\pgfqpoint{0.894282in}{1.718485in}}%
\pgfpathlineto{\pgfqpoint{0.910737in}{1.733732in}}%
\pgfpathlineto{\pgfqpoint{0.923901in}{1.748831in}}%
\pgfpathlineto{\pgfqpoint{0.935420in}{1.765054in}}%
\pgfpathlineto{\pgfqpoint{0.946939in}{1.785280in}}%
\pgfpathlineto{\pgfqpoint{0.956813in}{1.807190in}}%
\pgfpathlineto{\pgfqpoint{0.966686in}{1.835336in}}%
\pgfpathlineto{\pgfqpoint{0.974914in}{1.865881in}}%
\pgfpathlineto{\pgfqpoint{0.983141in}{1.906333in}}%
\pgfpathlineto{\pgfqpoint{0.989724in}{1.949702in}}%
\pgfpathlineto{\pgfqpoint{0.996306in}{2.008694in}}%
\pgfpathlineto{\pgfqpoint{1.002888in}{2.093858in}}%
\pgfpathlineto{\pgfqpoint{1.007825in}{2.187697in}}%
\pgfpathlineto{\pgfqpoint{1.012048in}{2.305449in}}%
\pgfpathmoveto{\pgfqpoint{1.031217in}{2.305449in}}%
\pgfpathlineto{\pgfqpoint{1.031282in}{0.483557in}}%
\pgfpathmoveto{\pgfqpoint{1.043761in}{0.483557in}}%
\pgfpathlineto{\pgfqpoint{1.047318in}{0.757688in}}%
\pgfpathlineto{\pgfqpoint{1.052255in}{0.975037in}}%
\pgfpathlineto{\pgfqpoint{1.057191in}{1.107023in}}%
\pgfpathlineto{\pgfqpoint{1.063774in}{1.219274in}}%
\pgfpathlineto{\pgfqpoint{1.070356in}{1.293180in}}%
\pgfpathlineto{\pgfqpoint{1.076938in}{1.345641in}}%
\pgfpathlineto{\pgfqpoint{1.085166in}{1.393227in}}%
\pgfpathlineto{\pgfqpoint{1.093393in}{1.428342in}}%
\pgfpathlineto{\pgfqpoint{1.103267in}{1.460114in}}%
\pgfpathlineto{\pgfqpoint{1.113140in}{1.484486in}}%
\pgfpathlineto{\pgfqpoint{1.124659in}{1.506731in}}%
\pgfpathlineto{\pgfqpoint{1.136178in}{1.524423in}}%
\pgfpathlineto{\pgfqpoint{1.149342in}{1.540794in}}%
\pgfpathlineto{\pgfqpoint{1.164152in}{1.555778in}}%
\pgfpathlineto{\pgfqpoint{1.182253in}{1.570679in}}%
\pgfpathlineto{\pgfqpoint{1.202000in}{1.583993in}}%
\pgfpathlineto{\pgfqpoint{1.225038in}{1.596886in}}%
\pgfpathlineto{\pgfqpoint{1.253012in}{1.609969in}}%
\pgfpathlineto{\pgfqpoint{1.287569in}{1.623582in}}%
\pgfpathlineto{\pgfqpoint{1.330353in}{1.637982in}}%
\pgfpathlineto{\pgfqpoint{1.387948in}{1.654850in}}%
\pgfpathlineto{\pgfqpoint{1.470225in}{1.676381in}}%
\pgfpathlineto{\pgfqpoint{1.845411in}{1.771770in}}%
\pgfpathlineto{\pgfqpoint{1.935917in}{1.798392in}}%
\pgfpathlineto{\pgfqpoint{2.016549in}{1.824412in}}%
\pgfpathlineto{\pgfqpoint{2.088953in}{1.850119in}}%
\pgfpathlineto{\pgfqpoint{2.154775in}{1.875867in}}%
\pgfpathlineto{\pgfqpoint{2.214015in}{1.901396in}}%
\pgfpathlineto{\pgfqpoint{2.268319in}{1.927156in}}%
\pgfpathlineto{\pgfqpoint{2.302875in}{1.944927in}}%
\pgfpathlineto{\pgfqpoint{2.302875in}{1.944927in}}%
\pgfusepath{stroke}%
\end{pgfscope}%
\begin{pgfscope}%
\pgfpathrectangle{\pgfqpoint{0.658967in}{0.493557in}}{\pgfqpoint{1.643908in}{1.801891in}} %
\pgfusepath{clip}%
\pgfsetbuttcap%
\pgfsetroundjoin%
\definecolor{currentfill}{rgb}{1.000000,0.000000,0.000000}%
\pgfsetfillcolor{currentfill}%
\pgfsetlinewidth{1.003750pt}%
\definecolor{currentstroke}{rgb}{1.000000,0.000000,0.000000}%
\pgfsetstrokecolor{currentstroke}%
\pgfsetdash{}{0pt}%
\pgfsys@defobject{currentmarker}{\pgfqpoint{-0.027778in}{-0.027778in}}{\pgfqpoint{0.027778in}{0.027778in}}{%
\pgfpathmoveto{\pgfqpoint{0.000000in}{-0.027778in}}%
\pgfpathcurveto{\pgfqpoint{0.007367in}{-0.027778in}}{\pgfqpoint{0.014433in}{-0.024851in}}{\pgfqpoint{0.019642in}{-0.019642in}}%
\pgfpathcurveto{\pgfqpoint{0.024851in}{-0.014433in}}{\pgfqpoint{0.027778in}{-0.007367in}}{\pgfqpoint{0.027778in}{0.000000in}}%
\pgfpathcurveto{\pgfqpoint{0.027778in}{0.007367in}}{\pgfqpoint{0.024851in}{0.014433in}}{\pgfqpoint{0.019642in}{0.019642in}}%
\pgfpathcurveto{\pgfqpoint{0.014433in}{0.024851in}}{\pgfqpoint{0.007367in}{0.027778in}}{\pgfqpoint{0.000000in}{0.027778in}}%
\pgfpathcurveto{\pgfqpoint{-0.007367in}{0.027778in}}{\pgfqpoint{-0.014433in}{0.024851in}}{\pgfqpoint{-0.019642in}{0.019642in}}%
\pgfpathcurveto{\pgfqpoint{-0.024851in}{0.014433in}}{\pgfqpoint{-0.027778in}{0.007367in}}{\pgfqpoint{-0.027778in}{0.000000in}}%
\pgfpathcurveto{\pgfqpoint{-0.027778in}{-0.007367in}}{\pgfqpoint{-0.024851in}{-0.014433in}}{\pgfqpoint{-0.019642in}{-0.019642in}}%
\pgfpathcurveto{\pgfqpoint{-0.014433in}{-0.024851in}}{\pgfqpoint{-0.007367in}{-0.027778in}}{\pgfqpoint{0.000000in}{-0.027778in}}%
\pgfpathclose%
\pgfusepath{stroke,fill}%
}%
\begin{pgfscope}%
\pgfsys@transformshift{1.480921in}{1.679060in}%
\pgfsys@useobject{currentmarker}{}%
\end{pgfscope}%
\end{pgfscope}%
\begin{pgfscope}%
\pgfsetrectcap%
\pgfsetmiterjoin%
\pgfsetlinewidth{0.803000pt}%
\definecolor{currentstroke}{rgb}{0.000000,0.000000,0.000000}%
\pgfsetstrokecolor{currentstroke}%
\pgfsetdash{}{0pt}%
\pgfpathmoveto{\pgfqpoint{0.658967in}{0.493557in}}%
\pgfpathlineto{\pgfqpoint{0.658967in}{2.295449in}}%
\pgfusepath{stroke}%
\end{pgfscope}%
\begin{pgfscope}%
\pgfsetrectcap%
\pgfsetmiterjoin%
\pgfsetlinewidth{0.803000pt}%
\definecolor{currentstroke}{rgb}{0.000000,0.000000,0.000000}%
\pgfsetstrokecolor{currentstroke}%
\pgfsetdash{}{0pt}%
\pgfpathmoveto{\pgfqpoint{2.302875in}{0.493557in}}%
\pgfpathlineto{\pgfqpoint{2.302875in}{2.295449in}}%
\pgfusepath{stroke}%
\end{pgfscope}%
\begin{pgfscope}%
\pgfsetrectcap%
\pgfsetmiterjoin%
\pgfsetlinewidth{0.803000pt}%
\definecolor{currentstroke}{rgb}{0.000000,0.000000,0.000000}%
\pgfsetstrokecolor{currentstroke}%
\pgfsetdash{}{0pt}%
\pgfpathmoveto{\pgfqpoint{0.658967in}{0.493557in}}%
\pgfpathlineto{\pgfqpoint{2.302875in}{0.493557in}}%
\pgfusepath{stroke}%
\end{pgfscope}%
\begin{pgfscope}%
\pgfsetrectcap%
\pgfsetmiterjoin%
\pgfsetlinewidth{0.803000pt}%
\definecolor{currentstroke}{rgb}{0.000000,0.000000,0.000000}%
\pgfsetstrokecolor{currentstroke}%
\pgfsetdash{}{0pt}%
\pgfpathmoveto{\pgfqpoint{0.658967in}{2.295449in}}%
\pgfpathlineto{\pgfqpoint{2.302875in}{2.295449in}}%
\pgfusepath{stroke}%
\end{pgfscope}%
\begin{pgfscope}%
\pgftext[x=1.480921in,y=2.378782in,,base]{\rmfamily\fontsize{9.600000}{11.520000}\selectfont \(\displaystyle r = 27.00\)}%
\end{pgfscope}%
\begin{pgfscope}%
\pgfsetbuttcap%
\pgfsetmiterjoin%
\definecolor{currentfill}{rgb}{1.000000,1.000000,1.000000}%
\pgfsetfillcolor{currentfill}%
\pgfsetlinewidth{0.000000pt}%
\definecolor{currentstroke}{rgb}{0.000000,0.000000,0.000000}%
\pgfsetstrokecolor{currentstroke}%
\pgfsetstrokeopacity{0.000000}%
\pgfsetdash{}{0pt}%
\pgfpathmoveto{\pgfqpoint{2.543355in}{0.493557in}}%
\pgfpathlineto{\pgfqpoint{4.187263in}{0.493557in}}%
\pgfpathlineto{\pgfqpoint{4.187263in}{2.295449in}}%
\pgfpathlineto{\pgfqpoint{2.543355in}{2.295449in}}%
\pgfpathclose%
\pgfusepath{fill}%
\end{pgfscope}%
\begin{pgfscope}%
\pgfsetbuttcap%
\pgfsetroundjoin%
\definecolor{currentfill}{rgb}{0.000000,0.000000,0.000000}%
\pgfsetfillcolor{currentfill}%
\pgfsetlinewidth{0.803000pt}%
\definecolor{currentstroke}{rgb}{0.000000,0.000000,0.000000}%
\pgfsetstrokecolor{currentstroke}%
\pgfsetdash{}{0pt}%
\pgfsys@defobject{currentmarker}{\pgfqpoint{0.000000in}{-0.048611in}}{\pgfqpoint{0.000000in}{0.000000in}}{%
\pgfpathmoveto{\pgfqpoint{0.000000in}{0.000000in}}%
\pgfpathlineto{\pgfqpoint{0.000000in}{-0.048611in}}%
\pgfusepath{stroke,fill}%
}%
\begin{pgfscope}%
\pgfsys@transformshift{2.629429in}{0.493557in}%
\pgfsys@useobject{currentmarker}{}%
\end{pgfscope}%
\end{pgfscope}%
\begin{pgfscope}%
\pgftext[x=2.629429in,y=0.396335in,,top]{\rmfamily\fontsize{8.000000}{9.600000}\selectfont \(\displaystyle 3.0\)}%
\end{pgfscope}%
\begin{pgfscope}%
\pgfsetbuttcap%
\pgfsetroundjoin%
\definecolor{currentfill}{rgb}{0.000000,0.000000,0.000000}%
\pgfsetfillcolor{currentfill}%
\pgfsetlinewidth{0.803000pt}%
\definecolor{currentstroke}{rgb}{0.000000,0.000000,0.000000}%
\pgfsetstrokecolor{currentstroke}%
\pgfsetdash{}{0pt}%
\pgfsys@defobject{currentmarker}{\pgfqpoint{0.000000in}{-0.048611in}}{\pgfqpoint{0.000000in}{0.000000in}}{%
\pgfpathmoveto{\pgfqpoint{0.000000in}{0.000000in}}%
\pgfpathlineto{\pgfqpoint{0.000000in}{-0.048611in}}%
\pgfusepath{stroke,fill}%
}%
\begin{pgfscope}%
\pgfsys@transformshift{3.191744in}{0.493557in}%
\pgfsys@useobject{currentmarker}{}%
\end{pgfscope}%
\end{pgfscope}%
\begin{pgfscope}%
\pgftext[x=3.191744in,y=0.396335in,,top]{\rmfamily\fontsize{8.000000}{9.600000}\selectfont \(\displaystyle 3.5\)}%
\end{pgfscope}%
\begin{pgfscope}%
\pgfsetbuttcap%
\pgfsetroundjoin%
\definecolor{currentfill}{rgb}{0.000000,0.000000,0.000000}%
\pgfsetfillcolor{currentfill}%
\pgfsetlinewidth{0.803000pt}%
\definecolor{currentstroke}{rgb}{0.000000,0.000000,0.000000}%
\pgfsetstrokecolor{currentstroke}%
\pgfsetdash{}{0pt}%
\pgfsys@defobject{currentmarker}{\pgfqpoint{0.000000in}{-0.048611in}}{\pgfqpoint{0.000000in}{0.000000in}}{%
\pgfpathmoveto{\pgfqpoint{0.000000in}{0.000000in}}%
\pgfpathlineto{\pgfqpoint{0.000000in}{-0.048611in}}%
\pgfusepath{stroke,fill}%
}%
\begin{pgfscope}%
\pgfsys@transformshift{3.754059in}{0.493557in}%
\pgfsys@useobject{currentmarker}{}%
\end{pgfscope}%
\end{pgfscope}%
\begin{pgfscope}%
\pgftext[x=3.754059in,y=0.396335in,,top]{\rmfamily\fontsize{8.000000}{9.600000}\selectfont \(\displaystyle 4.0\)}%
\end{pgfscope}%
\begin{pgfscope}%
\pgftext[x=3.365309in,y=0.242655in,,top]{\rmfamily\fontsize{10.000000}{12.000000}\selectfont \(\displaystyle \mu\)}%
\end{pgfscope}%
\begin{pgfscope}%
\pgfpathrectangle{\pgfqpoint{2.543355in}{0.493557in}}{\pgfqpoint{1.643908in}{1.801891in}} %
\pgfusepath{clip}%
\pgfsetrectcap%
\pgfsetroundjoin%
\pgfsetlinewidth{1.505625pt}%
\definecolor{currentstroke}{rgb}{0.121569,0.466667,0.705882}%
\pgfsetstrokecolor{currentstroke}%
\pgfsetdash{}{0pt}%
\pgfpathmoveto{\pgfqpoint{2.954332in}{1.591013in}}%
\pgfpathlineto{\pgfqpoint{3.095027in}{1.605767in}}%
\pgfpathlineto{\pgfqpoint{3.193760in}{1.618211in}}%
\pgfpathlineto{\pgfqpoint{3.263696in}{1.629143in}}%
\pgfpathlineto{\pgfqpoint{3.313885in}{1.639113in}}%
\pgfpathlineto{\pgfqpoint{3.350910in}{1.648594in}}%
\pgfpathlineto{\pgfqpoint{3.379707in}{1.658188in}}%
\pgfpathlineto{\pgfqpoint{3.401922in}{1.667849in}}%
\pgfpathlineto{\pgfqpoint{3.420023in}{1.678124in}}%
\pgfpathlineto{\pgfqpoint{3.434833in}{1.689105in}}%
\pgfpathlineto{\pgfqpoint{3.447175in}{1.701045in}}%
\pgfpathlineto{\pgfqpoint{3.457871in}{1.714589in}}%
\pgfpathlineto{\pgfqpoint{3.466922in}{1.729632in}}%
\pgfpathlineto{\pgfqpoint{3.475149in}{1.747737in}}%
\pgfpathlineto{\pgfqpoint{3.482554in}{1.769706in}}%
\pgfpathlineto{\pgfqpoint{3.489137in}{1.796497in}}%
\pgfpathlineto{\pgfqpoint{3.494896in}{1.829152in}}%
\pgfpathlineto{\pgfqpoint{3.499833in}{1.868575in}}%
\pgfpathlineto{\pgfqpoint{3.504769in}{1.926475in}}%
\pgfpathlineto{\pgfqpoint{3.508883in}{2.000652in}}%
\pgfpathlineto{\pgfqpoint{3.512997in}{2.123029in}}%
\pgfpathlineto{\pgfqpoint{3.516375in}{2.305449in}}%
\pgfpathmoveto{\pgfqpoint{3.525890in}{2.305449in}}%
\pgfpathlineto{\pgfqpoint{3.525933in}{0.483557in}}%
\pgfpathmoveto{\pgfqpoint{3.531051in}{0.483557in}}%
\pgfpathlineto{\pgfqpoint{3.533567in}{0.846319in}}%
\pgfpathlineto{\pgfqpoint{3.536858in}{1.075432in}}%
\pgfpathlineto{\pgfqpoint{3.540972in}{1.224095in}}%
\pgfpathlineto{\pgfqpoint{3.545908in}{1.323215in}}%
\pgfpathlineto{\pgfqpoint{3.551668in}{1.391638in}}%
\pgfpathlineto{\pgfqpoint{3.557427in}{1.435499in}}%
\pgfpathlineto{\pgfqpoint{3.564009in}{1.469720in}}%
\pgfpathlineto{\pgfqpoint{3.571414in}{1.496681in}}%
\pgfpathlineto{\pgfqpoint{3.579642in}{1.518200in}}%
\pgfpathlineto{\pgfqpoint{3.588693in}{1.535626in}}%
\pgfpathlineto{\pgfqpoint{3.598566in}{1.549952in}}%
\pgfpathlineto{\pgfqpoint{3.610085in}{1.562715in}}%
\pgfpathlineto{\pgfqpoint{3.623249in}{1.573939in}}%
\pgfpathlineto{\pgfqpoint{3.638882in}{1.584267in}}%
\pgfpathlineto{\pgfqpoint{3.657806in}{1.593990in}}%
\pgfpathlineto{\pgfqpoint{3.680844in}{1.603249in}}%
\pgfpathlineto{\pgfqpoint{3.710464in}{1.612614in}}%
\pgfpathlineto{\pgfqpoint{3.748311in}{1.622133in}}%
\pgfpathlineto{\pgfqpoint{3.776286in}{1.628090in}}%
\pgfpathlineto{\pgfqpoint{3.776286in}{1.628090in}}%
\pgfusepath{stroke}%
\end{pgfscope}%
\begin{pgfscope}%
\pgfpathrectangle{\pgfqpoint{2.543355in}{0.493557in}}{\pgfqpoint{1.643908in}{1.801891in}} %
\pgfusepath{clip}%
\pgfsetbuttcap%
\pgfsetroundjoin%
\pgfsetlinewidth{0.501875pt}%
\definecolor{currentstroke}{rgb}{0.000000,0.000000,0.000000}%
\pgfsetstrokecolor{currentstroke}%
\pgfsetdash{{16.000000pt}{4.000000pt}}{0.000000pt}%
\pgfpathmoveto{\pgfqpoint{2.543355in}{1.556238in}}%
\pgfpathlineto{\pgfqpoint{2.786897in}{1.575806in}}%
\pgfpathlineto{\pgfqpoint{2.974490in}{1.592987in}}%
\pgfpathlineto{\pgfqpoint{3.109425in}{1.607439in}}%
\pgfpathlineto{\pgfqpoint{3.203222in}{1.619555in}}%
\pgfpathlineto{\pgfqpoint{3.270689in}{1.630395in}}%
\pgfpathlineto{\pgfqpoint{3.318410in}{1.640150in}}%
\pgfpathlineto{\pgfqpoint{3.354613in}{1.649693in}}%
\pgfpathlineto{\pgfqpoint{3.382587in}{1.659304in}}%
\pgfpathlineto{\pgfqpoint{3.403979in}{1.668886in}}%
\pgfpathlineto{\pgfqpoint{3.422080in}{1.679484in}}%
\pgfpathlineto{\pgfqpoint{3.436890in}{1.690886in}}%
\pgfpathlineto{\pgfqpoint{3.448409in}{1.702430in}}%
\pgfpathlineto{\pgfqpoint{3.458282in}{1.715190in}}%
\pgfpathlineto{\pgfqpoint{3.468156in}{1.732030in}}%
\pgfpathlineto{\pgfqpoint{3.476384in}{1.750953in}}%
\pgfpathlineto{\pgfqpoint{3.482966in}{1.771143in}}%
\pgfpathlineto{\pgfqpoint{3.489548in}{1.798489in}}%
\pgfpathlineto{\pgfqpoint{3.496130in}{1.837783in}}%
\pgfpathlineto{\pgfqpoint{3.501067in}{1.880882in}}%
\pgfpathlineto{\pgfqpoint{3.506004in}{1.945474in}}%
\pgfpathlineto{\pgfqpoint{3.510940in}{2.053279in}}%
\pgfpathlineto{\pgfqpoint{3.514231in}{2.176968in}}%
\pgfpathlineto{\pgfqpoint{3.516320in}{2.305449in}}%
\pgfpathmoveto{\pgfqpoint{3.524235in}{2.305449in}}%
\pgfpathlineto{\pgfqpoint{3.524303in}{0.483557in}}%
\pgfpathmoveto{\pgfqpoint{3.531128in}{0.483557in}}%
\pgfpathlineto{\pgfqpoint{3.532332in}{0.702777in}}%
\pgfpathlineto{\pgfqpoint{3.533978in}{0.884782in}}%
\pgfpathlineto{\pgfqpoint{3.537269in}{1.095000in}}%
\pgfpathlineto{\pgfqpoint{3.542206in}{1.254372in}}%
\pgfpathlineto{\pgfqpoint{3.547142in}{1.340932in}}%
\pgfpathlineto{\pgfqpoint{3.552079in}{1.395396in}}%
\pgfpathlineto{\pgfqpoint{3.558661in}{1.442934in}}%
\pgfpathlineto{\pgfqpoint{3.565243in}{1.474895in}}%
\pgfpathlineto{\pgfqpoint{3.573471in}{1.502729in}}%
\pgfpathlineto{\pgfqpoint{3.581699in}{1.522630in}}%
\pgfpathlineto{\pgfqpoint{3.591572in}{1.540218in}}%
\pgfpathlineto{\pgfqpoint{3.601446in}{1.553474in}}%
\pgfpathlineto{\pgfqpoint{3.612965in}{1.565426in}}%
\pgfpathlineto{\pgfqpoint{3.626129in}{1.576047in}}%
\pgfpathlineto{\pgfqpoint{3.642584in}{1.586369in}}%
\pgfpathlineto{\pgfqpoint{3.662331in}{1.595994in}}%
\pgfpathlineto{\pgfqpoint{3.687014in}{1.605392in}}%
\pgfpathlineto{\pgfqpoint{3.718280in}{1.614760in}}%
\pgfpathlineto{\pgfqpoint{3.759419in}{1.624585in}}%
\pgfpathlineto{\pgfqpoint{3.815368in}{1.635451in}}%
\pgfpathlineto{\pgfqpoint{3.894354in}{1.648341in}}%
\pgfpathlineto{\pgfqpoint{4.021062in}{1.666504in}}%
\pgfpathlineto{\pgfqpoint{4.187263in}{1.688982in}}%
\pgfpathlineto{\pgfqpoint{4.187263in}{1.688982in}}%
\pgfusepath{stroke}%
\end{pgfscope}%
\begin{pgfscope}%
\pgfpathrectangle{\pgfqpoint{2.543355in}{0.493557in}}{\pgfqpoint{1.643908in}{1.801891in}} %
\pgfusepath{clip}%
\pgfsetbuttcap%
\pgfsetroundjoin%
\definecolor{currentfill}{rgb}{1.000000,0.000000,0.000000}%
\pgfsetfillcolor{currentfill}%
\pgfsetlinewidth{1.003750pt}%
\definecolor{currentstroke}{rgb}{1.000000,0.000000,0.000000}%
\pgfsetstrokecolor{currentstroke}%
\pgfsetdash{}{0pt}%
\pgfsys@defobject{currentmarker}{\pgfqpoint{-0.027778in}{-0.027778in}}{\pgfqpoint{0.027778in}{0.027778in}}{%
\pgfpathmoveto{\pgfqpoint{0.000000in}{-0.027778in}}%
\pgfpathcurveto{\pgfqpoint{0.007367in}{-0.027778in}}{\pgfqpoint{0.014433in}{-0.024851in}}{\pgfqpoint{0.019642in}{-0.019642in}}%
\pgfpathcurveto{\pgfqpoint{0.024851in}{-0.014433in}}{\pgfqpoint{0.027778in}{-0.007367in}}{\pgfqpoint{0.027778in}{0.000000in}}%
\pgfpathcurveto{\pgfqpoint{0.027778in}{0.007367in}}{\pgfqpoint{0.024851in}{0.014433in}}{\pgfqpoint{0.019642in}{0.019642in}}%
\pgfpathcurveto{\pgfqpoint{0.014433in}{0.024851in}}{\pgfqpoint{0.007367in}{0.027778in}}{\pgfqpoint{0.000000in}{0.027778in}}%
\pgfpathcurveto{\pgfqpoint{-0.007367in}{0.027778in}}{\pgfqpoint{-0.014433in}{0.024851in}}{\pgfqpoint{-0.019642in}{0.019642in}}%
\pgfpathcurveto{\pgfqpoint{-0.024851in}{0.014433in}}{\pgfqpoint{-0.027778in}{0.007367in}}{\pgfqpoint{-0.027778in}{0.000000in}}%
\pgfpathcurveto{\pgfqpoint{-0.027778in}{-0.007367in}}{\pgfqpoint{-0.024851in}{-0.014433in}}{\pgfqpoint{-0.019642in}{-0.019642in}}%
\pgfpathcurveto{\pgfqpoint{-0.014433in}{-0.024851in}}{\pgfqpoint{-0.007367in}{-0.027778in}}{\pgfqpoint{0.000000in}{-0.027778in}}%
\pgfpathclose%
\pgfusepath{stroke,fill}%
}%
\begin{pgfscope}%
\pgfsys@transformshift{3.365309in}{1.653071in}%
\pgfsys@useobject{currentmarker}{}%
\end{pgfscope}%
\end{pgfscope}%
\begin{pgfscope}%
\pgfsetrectcap%
\pgfsetmiterjoin%
\pgfsetlinewidth{0.803000pt}%
\definecolor{currentstroke}{rgb}{0.000000,0.000000,0.000000}%
\pgfsetstrokecolor{currentstroke}%
\pgfsetdash{}{0pt}%
\pgfpathmoveto{\pgfqpoint{2.543355in}{0.493557in}}%
\pgfpathlineto{\pgfqpoint{2.543355in}{2.295449in}}%
\pgfusepath{stroke}%
\end{pgfscope}%
\begin{pgfscope}%
\pgfsetrectcap%
\pgfsetmiterjoin%
\pgfsetlinewidth{0.803000pt}%
\definecolor{currentstroke}{rgb}{0.000000,0.000000,0.000000}%
\pgfsetstrokecolor{currentstroke}%
\pgfsetdash{}{0pt}%
\pgfpathmoveto{\pgfqpoint{4.187263in}{0.493557in}}%
\pgfpathlineto{\pgfqpoint{4.187263in}{2.295449in}}%
\pgfusepath{stroke}%
\end{pgfscope}%
\begin{pgfscope}%
\pgfsetrectcap%
\pgfsetmiterjoin%
\pgfsetlinewidth{0.803000pt}%
\definecolor{currentstroke}{rgb}{0.000000,0.000000,0.000000}%
\pgfsetstrokecolor{currentstroke}%
\pgfsetdash{}{0pt}%
\pgfpathmoveto{\pgfqpoint{2.543355in}{0.493557in}}%
\pgfpathlineto{\pgfqpoint{4.187263in}{0.493557in}}%
\pgfusepath{stroke}%
\end{pgfscope}%
\begin{pgfscope}%
\pgfsetrectcap%
\pgfsetmiterjoin%
\pgfsetlinewidth{0.803000pt}%
\definecolor{currentstroke}{rgb}{0.000000,0.000000,0.000000}%
\pgfsetstrokecolor{currentstroke}%
\pgfsetdash{}{0pt}%
\pgfpathmoveto{\pgfqpoint{2.543355in}{2.295449in}}%
\pgfpathlineto{\pgfqpoint{4.187263in}{2.295449in}}%
\pgfusepath{stroke}%
\end{pgfscope}%
\begin{pgfscope}%
\pgftext[x=3.365309in,y=2.378782in,,base]{\rmfamily\fontsize{9.600000}{11.520000}\selectfont \(\displaystyle r = 28.00\)}%
\end{pgfscope}%
\begin{pgfscope}%
\pgfsetbuttcap%
\pgfsetmiterjoin%
\definecolor{currentfill}{rgb}{1.000000,1.000000,1.000000}%
\pgfsetfillcolor{currentfill}%
\pgfsetlinewidth{0.000000pt}%
\definecolor{currentstroke}{rgb}{0.000000,0.000000,0.000000}%
\pgfsetstrokecolor{currentstroke}%
\pgfsetstrokeopacity{0.000000}%
\pgfsetdash{}{0pt}%
\pgfpathmoveto{\pgfqpoint{4.427742in}{0.493557in}}%
\pgfpathlineto{\pgfqpoint{6.071650in}{0.493557in}}%
\pgfpathlineto{\pgfqpoint{6.071650in}{2.295449in}}%
\pgfpathlineto{\pgfqpoint{4.427742in}{2.295449in}}%
\pgfpathclose%
\pgfusepath{fill}%
\end{pgfscope}%
\begin{pgfscope}%
\pgfsetbuttcap%
\pgfsetroundjoin%
\definecolor{currentfill}{rgb}{0.000000,0.000000,0.000000}%
\pgfsetfillcolor{currentfill}%
\pgfsetlinewidth{0.803000pt}%
\definecolor{currentstroke}{rgb}{0.000000,0.000000,0.000000}%
\pgfsetstrokecolor{currentstroke}%
\pgfsetdash{}{0pt}%
\pgfsys@defobject{currentmarker}{\pgfqpoint{0.000000in}{-0.048611in}}{\pgfqpoint{0.000000in}{0.000000in}}{%
\pgfpathmoveto{\pgfqpoint{0.000000in}{0.000000in}}%
\pgfpathlineto{\pgfqpoint{0.000000in}{-0.048611in}}%
\pgfusepath{stroke,fill}%
}%
\begin{pgfscope}%
\pgfsys@transformshift{4.450282in}{0.493557in}%
\pgfsys@useobject{currentmarker}{}%
\end{pgfscope}%
\end{pgfscope}%
\begin{pgfscope}%
\pgftext[x=4.450282in,y=0.396335in,,top]{\rmfamily\fontsize{8.000000}{9.600000}\selectfont \(\displaystyle 2.00\)}%
\end{pgfscope}%
\begin{pgfscope}%
\pgfsetbuttcap%
\pgfsetroundjoin%
\definecolor{currentfill}{rgb}{0.000000,0.000000,0.000000}%
\pgfsetfillcolor{currentfill}%
\pgfsetlinewidth{0.803000pt}%
\definecolor{currentstroke}{rgb}{0.000000,0.000000,0.000000}%
\pgfsetstrokecolor{currentstroke}%
\pgfsetdash{}{0pt}%
\pgfsys@defobject{currentmarker}{\pgfqpoint{0.000000in}{-0.048611in}}{\pgfqpoint{0.000000in}{0.000000in}}{%
\pgfpathmoveto{\pgfqpoint{0.000000in}{0.000000in}}%
\pgfpathlineto{\pgfqpoint{0.000000in}{-0.048611in}}%
\pgfusepath{stroke,fill}%
}%
\begin{pgfscope}%
\pgfsys@transformshift{4.864076in}{0.493557in}%
\pgfsys@useobject{currentmarker}{}%
\end{pgfscope}%
\end{pgfscope}%
\begin{pgfscope}%
\pgftext[x=4.864076in,y=0.396335in,,top]{\rmfamily\fontsize{8.000000}{9.600000}\selectfont \(\displaystyle 2.25\)}%
\end{pgfscope}%
\begin{pgfscope}%
\pgfsetbuttcap%
\pgfsetroundjoin%
\definecolor{currentfill}{rgb}{0.000000,0.000000,0.000000}%
\pgfsetfillcolor{currentfill}%
\pgfsetlinewidth{0.803000pt}%
\definecolor{currentstroke}{rgb}{0.000000,0.000000,0.000000}%
\pgfsetstrokecolor{currentstroke}%
\pgfsetdash{}{0pt}%
\pgfsys@defobject{currentmarker}{\pgfqpoint{0.000000in}{-0.048611in}}{\pgfqpoint{0.000000in}{0.000000in}}{%
\pgfpathmoveto{\pgfqpoint{0.000000in}{0.000000in}}%
\pgfpathlineto{\pgfqpoint{0.000000in}{-0.048611in}}%
\pgfusepath{stroke,fill}%
}%
\begin{pgfscope}%
\pgfsys@transformshift{5.277871in}{0.493557in}%
\pgfsys@useobject{currentmarker}{}%
\end{pgfscope}%
\end{pgfscope}%
\begin{pgfscope}%
\pgftext[x=5.277871in,y=0.396335in,,top]{\rmfamily\fontsize{8.000000}{9.600000}\selectfont \(\displaystyle 2.50\)}%
\end{pgfscope}%
\begin{pgfscope}%
\pgfsetbuttcap%
\pgfsetroundjoin%
\definecolor{currentfill}{rgb}{0.000000,0.000000,0.000000}%
\pgfsetfillcolor{currentfill}%
\pgfsetlinewidth{0.803000pt}%
\definecolor{currentstroke}{rgb}{0.000000,0.000000,0.000000}%
\pgfsetstrokecolor{currentstroke}%
\pgfsetdash{}{0pt}%
\pgfsys@defobject{currentmarker}{\pgfqpoint{0.000000in}{-0.048611in}}{\pgfqpoint{0.000000in}{0.000000in}}{%
\pgfpathmoveto{\pgfqpoint{0.000000in}{0.000000in}}%
\pgfpathlineto{\pgfqpoint{0.000000in}{-0.048611in}}%
\pgfusepath{stroke,fill}%
}%
\begin{pgfscope}%
\pgfsys@transformshift{5.691665in}{0.493557in}%
\pgfsys@useobject{currentmarker}{}%
\end{pgfscope}%
\end{pgfscope}%
\begin{pgfscope}%
\pgftext[x=5.691665in,y=0.396335in,,top]{\rmfamily\fontsize{8.000000}{9.600000}\selectfont \(\displaystyle 2.75\)}%
\end{pgfscope}%
\begin{pgfscope}%
\pgftext[x=5.249696in,y=0.242655in,,top]{\rmfamily\fontsize{10.000000}{12.000000}\selectfont \(\displaystyle \mu\)}%
\end{pgfscope}%
\begin{pgfscope}%
\pgfpathrectangle{\pgfqpoint{4.427742in}{0.493557in}}{\pgfqpoint{1.643908in}{1.801891in}} %
\pgfusepath{clip}%
\pgfsetrectcap%
\pgfsetroundjoin%
\pgfsetlinewidth{1.505625pt}%
\definecolor{currentstroke}{rgb}{0.121569,0.466667,0.705882}%
\pgfsetstrokecolor{currentstroke}%
\pgfsetdash{}{0pt}%
\pgfpathmoveto{\pgfqpoint{4.838719in}{1.692039in}}%
\pgfpathlineto{\pgfqpoint{4.956376in}{1.699164in}}%
\pgfpathlineto{\pgfqpoint{5.075679in}{1.704155in}}%
\pgfpathlineto{\pgfqpoint{5.194981in}{1.706914in}}%
\pgfpathlineto{\pgfqpoint{5.312639in}{1.707420in}}%
\pgfpathlineto{\pgfqpoint{5.428650in}{1.705698in}}%
\pgfpathlineto{\pgfqpoint{5.541370in}{1.701822in}}%
\pgfpathlineto{\pgfqpoint{5.651623in}{1.695813in}}%
\pgfpathlineto{\pgfqpoint{5.660673in}{1.695218in}}%
\pgfpathlineto{\pgfqpoint{5.660673in}{1.695218in}}%
\pgfusepath{stroke}%
\end{pgfscope}%
\begin{pgfscope}%
\pgfpathrectangle{\pgfqpoint{4.427742in}{0.493557in}}{\pgfqpoint{1.643908in}{1.801891in}} %
\pgfusepath{clip}%
\pgfsetbuttcap%
\pgfsetroundjoin%
\pgfsetlinewidth{0.501875pt}%
\definecolor{currentstroke}{rgb}{0.000000,0.000000,0.000000}%
\pgfsetstrokecolor{currentstroke}%
\pgfsetdash{{16.000000pt}{4.000000pt}}{0.000000pt}%
\pgfpathmoveto{\pgfqpoint{4.427742in}{1.647958in}}%
\pgfpathlineto{\pgfqpoint{4.523184in}{1.661241in}}%
\pgfpathlineto{\pgfqpoint{4.626854in}{1.673394in}}%
\pgfpathlineto{\pgfqpoint{4.737106in}{1.684068in}}%
\pgfpathlineto{\pgfqpoint{4.852295in}{1.692974in}}%
\pgfpathlineto{\pgfqpoint{4.970775in}{1.699885in}}%
\pgfpathlineto{\pgfqpoint{5.089255in}{1.704582in}}%
\pgfpathlineto{\pgfqpoint{5.207734in}{1.707076in}}%
\pgfpathlineto{\pgfqpoint{5.324569in}{1.707346in}}%
\pgfpathlineto{\pgfqpoint{5.439758in}{1.705415in}}%
\pgfpathlineto{\pgfqpoint{5.553301in}{1.701280in}}%
\pgfpathlineto{\pgfqpoint{5.663553in}{1.695025in}}%
\pgfpathlineto{\pgfqpoint{5.770514in}{1.686713in}}%
\pgfpathlineto{\pgfqpoint{5.872538in}{1.676579in}}%
\pgfpathlineto{\pgfqpoint{5.971271in}{1.664567in}}%
\pgfpathlineto{\pgfqpoint{6.066713in}{1.650732in}}%
\pgfpathlineto{\pgfqpoint{6.071650in}{1.649954in}}%
\pgfpathlineto{\pgfqpoint{6.071650in}{1.649954in}}%
\pgfusepath{stroke}%
\end{pgfscope}%
\begin{pgfscope}%
\pgfpathrectangle{\pgfqpoint{4.427742in}{0.493557in}}{\pgfqpoint{1.643908in}{1.801891in}} %
\pgfusepath{clip}%
\pgfsetbuttcap%
\pgfsetroundjoin%
\definecolor{currentfill}{rgb}{1.000000,0.000000,0.000000}%
\pgfsetfillcolor{currentfill}%
\pgfsetlinewidth{1.003750pt}%
\definecolor{currentstroke}{rgb}{1.000000,0.000000,0.000000}%
\pgfsetstrokecolor{currentstroke}%
\pgfsetdash{}{0pt}%
\pgfsys@defobject{currentmarker}{\pgfqpoint{-0.027778in}{-0.027778in}}{\pgfqpoint{0.027778in}{0.027778in}}{%
\pgfpathmoveto{\pgfqpoint{0.000000in}{-0.027778in}}%
\pgfpathcurveto{\pgfqpoint{0.007367in}{-0.027778in}}{\pgfqpoint{0.014433in}{-0.024851in}}{\pgfqpoint{0.019642in}{-0.019642in}}%
\pgfpathcurveto{\pgfqpoint{0.024851in}{-0.014433in}}{\pgfqpoint{0.027778in}{-0.007367in}}{\pgfqpoint{0.027778in}{0.000000in}}%
\pgfpathcurveto{\pgfqpoint{0.027778in}{0.007367in}}{\pgfqpoint{0.024851in}{0.014433in}}{\pgfqpoint{0.019642in}{0.019642in}}%
\pgfpathcurveto{\pgfqpoint{0.014433in}{0.024851in}}{\pgfqpoint{0.007367in}{0.027778in}}{\pgfqpoint{0.000000in}{0.027778in}}%
\pgfpathcurveto{\pgfqpoint{-0.007367in}{0.027778in}}{\pgfqpoint{-0.014433in}{0.024851in}}{\pgfqpoint{-0.019642in}{0.019642in}}%
\pgfpathcurveto{\pgfqpoint{-0.024851in}{0.014433in}}{\pgfqpoint{-0.027778in}{0.007367in}}{\pgfqpoint{-0.027778in}{0.000000in}}%
\pgfpathcurveto{\pgfqpoint{-0.027778in}{-0.007367in}}{\pgfqpoint{-0.024851in}{-0.014433in}}{\pgfqpoint{-0.019642in}{-0.019642in}}%
\pgfpathcurveto{\pgfqpoint{-0.014433in}{-0.024851in}}{\pgfqpoint{-0.007367in}{-0.027778in}}{\pgfqpoint{0.000000in}{-0.027778in}}%
\pgfpathclose%
\pgfusepath{stroke,fill}%
}%
\begin{pgfscope}%
\pgfsys@transformshift{5.249696in}{1.707426in}%
\pgfsys@useobject{currentmarker}{}%
\end{pgfscope}%
\end{pgfscope}%
\begin{pgfscope}%
\pgfsetrectcap%
\pgfsetmiterjoin%
\pgfsetlinewidth{0.803000pt}%
\definecolor{currentstroke}{rgb}{0.000000,0.000000,0.000000}%
\pgfsetstrokecolor{currentstroke}%
\pgfsetdash{}{0pt}%
\pgfpathmoveto{\pgfqpoint{4.427742in}{0.493557in}}%
\pgfpathlineto{\pgfqpoint{4.427742in}{2.295449in}}%
\pgfusepath{stroke}%
\end{pgfscope}%
\begin{pgfscope}%
\pgfsetrectcap%
\pgfsetmiterjoin%
\pgfsetlinewidth{0.803000pt}%
\definecolor{currentstroke}{rgb}{0.000000,0.000000,0.000000}%
\pgfsetstrokecolor{currentstroke}%
\pgfsetdash{}{0pt}%
\pgfpathmoveto{\pgfqpoint{6.071650in}{0.493557in}}%
\pgfpathlineto{\pgfqpoint{6.071650in}{2.295449in}}%
\pgfusepath{stroke}%
\end{pgfscope}%
\begin{pgfscope}%
\pgfsetrectcap%
\pgfsetmiterjoin%
\pgfsetlinewidth{0.803000pt}%
\definecolor{currentstroke}{rgb}{0.000000,0.000000,0.000000}%
\pgfsetstrokecolor{currentstroke}%
\pgfsetdash{}{0pt}%
\pgfpathmoveto{\pgfqpoint{4.427742in}{0.493557in}}%
\pgfpathlineto{\pgfqpoint{6.071650in}{0.493557in}}%
\pgfusepath{stroke}%
\end{pgfscope}%
\begin{pgfscope}%
\pgfsetrectcap%
\pgfsetmiterjoin%
\pgfsetlinewidth{0.803000pt}%
\definecolor{currentstroke}{rgb}{0.000000,0.000000,0.000000}%
\pgfsetstrokecolor{currentstroke}%
\pgfsetdash{}{0pt}%
\pgfpathmoveto{\pgfqpoint{4.427742in}{2.295449in}}%
\pgfpathlineto{\pgfqpoint{6.071650in}{2.295449in}}%
\pgfusepath{stroke}%
\end{pgfscope}%
\begin{pgfscope}%
\pgftext[x=5.249696in,y=2.378782in,,base]{\rmfamily\fontsize{9.600000}{11.520000}\selectfont \(\displaystyle r = 29.00\)}%
\end{pgfscope}%
\end{pgfpicture}%
\makeatother%
\endgroup%

%% file: error_order_2.tex
\begin{tabular}{ccc}\hline
$\alpha$ & $\boldsymbol{g}^{(2)}$ & $\boldsymbol{u}_{3}^{(2)}$ \\ 
\hline
(2, 0, 0) & 0.00e+00 & 2.68e+00 \\ 
(1, 1, 0) & -3.75e-01 & 2.67e+00 \\ 
(0, 2, 0) & 0.00e+00 & 3.95e+00 \\ 
(0, 0, 2) & 0.00e+00 & 2.92e+00 \\ 
\hline
\end{tabular}

%% file: error_order_3.tex
\begin{tabular}{ccc}\hline
$\alpha$ & $\boldsymbol{g}^{(3)}$ & $\boldsymbol{u}_{3}^{(3)}$ \\ 
\hline
(2, 0, 1) & 1.70e-03 & 2.19e+01 \\ 
(1, 1, 1) & 1.70e-02 & 7.79e+00 \\ 
(0, 2, 1) & 0.00e+00 & 1.14e+00 \\ 
(0, 0, 3) & 0.00e+00 & -1.81e+00 \\ 
\hline
\end{tabular}

%% file: error_order_4.tex
\begin{tabular}{ccc}\hline
$\alpha$ & $\boldsymbol{g}^{(4)}$ & $\boldsymbol{u}_{3}^{(4)}$ \\ 
\hline
(4, 0, 0) & 0.00e+00 & -3.33e+02 \\ 
(3, 1, 0) & 2.56e-04 & -3.33e+02 \\ 
(2, 2, 0) & 1.27e-04 & -2.87e+02 \\ 
(2, 0, 2) & -1.27e-04 & 1.33e+02 \\ 
(1, 3, 0) & 2.73e-04 & -2.38e+02 \\ 
(1, 1, 2) & -5.46e-04 & -5.72e+00 \\ 
(0, 4, 0) & 0.00e+00 & -2.33e+02 \\ 
(0, 2, 2) & 0.00e+00 & -6.99e+01 \\ 
(0, 0, 4) & 0.00e+00 & -2.32e+02 \\ 
\hline
\end{tabular}

%% file: error_order_5.tex
\begin{tabular}{ccc}\hline
$\alpha$ & $\boldsymbol{g}^{(5)}$ & $\boldsymbol{u}_{3}^{(5)}$ \\ 
\hline
(4, 0, 1) & 8.20e-06 & -7.16e+03 \\ 
(3, 1, 1) & -1.47e-04 & -5.22e+03 \\ 
(2, 2, 1) & -6.37e-06 & -3.37e+03 \\ 
(2, 0, 3) & -9.74e-06 & -1.18e+03 \\ 
(1, 3, 1) & -5.97e-05 & -1.73e+03 \\ 
(1, 1, 3) & -2.66e-05 & -6.31e+02 \\ 
(0, 4, 1) & 0.00e+00 & -5.73e+02 \\ 
(0, 2, 3) & 0.00e+00 & -2.49e+02 \\ 
(0, 0, 5) & 0.00e+00 & -9.62e+02 \\ 
\hline
\end{tabular}

%% file: main.bbl
\begin{thebibliography}{47}
\expandafter\ifx\csname natexlab\endcsname\relax\def\natexlab#1{#1}\fi

\bibitem[Ait-Chaalal {\em et~al.\/}(2016)Ait-Chaalal, Schneider, Meyer \&
  Marston]{ChaFnjp2016a}
{\sc Ait-Chaalal, F., Schneider, T., Meyer, B. \& Marston, J.~B.} 2016 Cumulant
  expansions for atmospheric flows. {\em New Journal of Physics\/} .

\bibitem[Allawala \& Marston(2016)]{AllApre2016a}
{\sc Allawala, A. \& Marston, J.~B.} 2016 Statistics of the stochastically
  forced {L}orenz attractor by the {F}okker-{P}lanck equation and cumulant
  expansions. {\em Phys. Rev. E\/} {\bf 94}, 052218.

\bibitem[Auerbach {\em et~al.\/}(1987)Auerbach,
  Cvitanovi\ifmmode~\acute{c}\else \'{c}\fi{}, Eckmann, Gunaratne \&
  Procaccia]{AueDprl1987a}
{\sc Auerbach, Ditza, Cvitanovi\ifmmode~\acute{c}\else \'{c}\fi{}, Predrag,
  Eckmann, Jean-Pierre, Gunaratne, Gemunu \& Procaccia, Itamar} 1987 Exploring
  chaotic motion through periodic orbits. {\em Phys. Rev. Lett.\/} {\bf 58},
  2387--2389.

\bibitem[Blonigan \& Wang(2014)]{BloPcsf2014a}
{\sc Blonigan, P.~J. \& Wang, Q.} 2014 Least squares shadowing sensitivity
  analysis of a modified {K}uramoto-{S}ivashinsky equation. {\em Chaos,
  Solitons \& Fractals\/} {\bf 64}, 16--25, nonequilibrium Statistical
  Mechanics: Fluctuations and Response.

\bibitem[Bohr {\em et~al.\/}(2005)Bohr, Jensen, Paladin \&
  Vulpiani]{BohTboo2005a}
{\sc Bohr, T., Jensen, M.H., Paladin, G. \& Vulpiani, A.} 2005 {\em Dynamical
  Systems Approach to Turbulence\/}. Cambridge University Press.

\bibitem[Cacuci(2003)]{CacDboo2003a}
{\sc Cacuci, D.G.} 2003 {\em Sensitivity \& Uncertainty Analysis, Volume 1:
  Theory\/}. CRC Press.

\bibitem[Cooper \& Haynes(2011)]{CooFjas2011a}
{\sc Cooper, F.~C. \& Haynes, P.~H.} 2011 Climate sensitivity via a
  nonparametric fluctuation-dissipation theorem. {\em Journal of the
  Atmospheric Sciences\/} {\bf 68}~(5), 937--953.

\bibitem[Dimet \& Talagrand(1986)]{DimFtel1986a}
{\sc Dimet, F.~Le \& Talagrand, O.} 1986 Variational algorithms for analysis
  and assimilation of meteorological observations: theoretical aspects. {\em
  Tellus A: Dynamic Meteorology and Oceanography\/} {\bf 38}~(2), 97--110.

\bibitem[Eckmann \& Ruelle(2004)]{EckJrmp1985a}
{\sc Eckmann, J.-P. \& Ruelle, D.} 2004 {\em Ergodic theory of chaos and
  strange attractors\/}, pp. 273--312. New York, NY: Springer New York.

\bibitem[Eyink {\em et~al.\/}(2004)Eyink, Haine \& Lea]{EyiGnon2004a}
{\sc Eyink, G.~L., Haine, T. W.~N. \& Lea, D.~J.} 2004 Ruelle's linear response
  formula, ensemble adjoint schemes and {L}\'{e}vy flights. {\em
  Nonlinearity\/} {\bf 17}~(5), 1867.

\bibitem[Farrell \& Ioannou(2014)]{FarBarx2014a}
{\sc Farrell, B.~F. \& Ioannou, P.~J.} 2014 Statistical state dynamics: a new
  perspective on turbulence in shear flow. {\em arXiv\/} {\bf 1412.8290v1}.

\bibitem[Farrell {\em et~al.\/}(2016)Farrell, Ioannou, Jim{\' e}nez,
  Constantinou, Lozano-Dur{\' a}n \& Nikolaidis]{FarBjfm2016a}
{\sc Farrell, B.~F., Ioannou, P.~J., Jim{\' e}nez, J., Constantinou, N.~C.,
  Lozano-Dur{\' a}n, A. \& Nikolaidis, M.-A.} 2016 A statistical state
  dynamics-based study of the structure and mechanism of large-scale motions in
  plane {P}oiseuille flow. {\em Journal of Fluid Mechanics\/} {\bf 809},
  290--315.

\bibitem[Farrell {\em et~al.\/}(2014)Farrell, Cotter \& Funke]{FarPsam2014a}
{\sc Farrell, P.~E., Cotter, C.~J. \& Funke, S.~W.} 2014 A framework for the
  automation of generalized stability theory. {\em SIAM Journal on Scientific
  Computing\/} {\bf 36}~(1), C25--C48.

\bibitem[Foures {\em et~al.\/}(2014)Foures, Caulfield \& Schmid]{FouDjfm2014a}
{\sc Foures, D.P.G., Caulfield, C.P. \& Schmid, P.J.} 2014 Optimal mixing in
  two-dimensional plane poiseuille flow at finite {P}\'{e}clet number. {\em
  Journal of Fluid Mechanics\/} {\bf 748}, 241--277.

\bibitem[Frisch(1995)]{FriUboo1995a}
{\sc Frisch, U.} 1995 {\em Turbulence: The Legacy of {A}. {N}. {K}olmogorov\/}.
  Cambridge University Press.

\bibitem[Giles \& Pierce(2000)]{GilMftc2000a}
{\sc Giles, M.~B. \& Pierce, N.~A.} 2000 An introduction to the adjoint
  approach to design. {\em Flow, Turbulence and Combustion\/} .

\bibitem[Hopf(1952)]{HopEarm1952a}
{\sc Hopf, E.} 1952 Statistical hydrodynamics and functional calculus. {\em
  Journal of Rational Mechanics and Analysis\/} {\bf 1}, 87--123.

\bibitem[Jameson(1988)]{JamAjsc1988a}
{\sc Jameson, A.} 1988 Aerodynamic design via control theory. {\em Journal of
  Scientific Computing\/} {\bf 3}~(3), 233--260.

\bibitem[Knobloch(1979)]{KnoEjsp1979a}
{\sc Knobloch, E.} 1979 On the statistical dynamics of the {L}orenz model. {\em
  Journal of Statistical Physics\/} {\bf 20}~(6), 695--709.

\bibitem[Kraichnan(1980)]{KraRmis1980a}
{\sc Kraichnan, R.~H.} 1980 Realizability inequalities and closed moment
  equations. {\em Annals of the New York Academy of Sciences\/} {\bf 357}~(1),
  37--46.

\bibitem[Lasagna(2018)]{LasDsam2018a}
{\sc Lasagna, D.} 2018 Sensitivity analysis of chaotic systems using unstable
  periodic orbits. {\em SIAM Journal on Applied Dynamical Systems\/} {\bf
  17}~(1), 547--580.

\bibitem[Lea {\em et~al.\/}(2000)Lea, Allen \& Haine]{LeaDtel2000a}
{\sc Lea, D., Allen, M. \& Haine, T.} 2000 Sensitivity analysis of the climate
  of a chaotic system. {\em Tellus A\/} {\bf 52}~(5).

\bibitem[Leith \& Kraichnan(1972)]{LeiCjas1972a}
{\sc Leith, C.~E. \& Kraichnan, R.~H.} 1972 Predictability of turbulent flows.
  {\em Journal of the Atmospheric Sciences\/} {\bf 29}~(6), 1041--1058.

\bibitem[Lions(1971)]{LioJboo1971a}
{\sc Lions, J.L.} 1971 {\em Optimal control of systems governed by partial
  differential equations\/}. Springer-Verlag.

\bibitem[Lorenz(1967)]{LorEboo1967a}
{\sc Lorenz, E.N.} 1967 {\em The Nature and Theory of the General Circulation
  of the Atmosphere\/}. World Meteorological Organization.

\bibitem[Lorenz(1963)]{LorEjas1963a}
{\sc Lorenz, E.~N.} 1963 Deterministic nonperiodic flow. {\em J. Atmos. Sci.\/}
  {\bf 20}~(2), 130--141.

\bibitem[Lucas \& Caulfield(2017)]{LucDjfm2017a}
{\sc Lucas, Dan \& Caulfield, C.~P.} 2017 Irreversible mixing by unstable
  periodic orbits in buoyancy dominated stratified turbulence. {\em Journal of
  Fluid Mechanics\/} {\bf 832}.

\bibitem[Luchini \& Bottaro(2014)]{LucPafm2014a}
{\sc Luchini, Paolo \& Bottaro, Alessandro} 2014 Adjoint equations in stability
  analysis. {\em Annual Review of Fluid Mechanics\/} {\bf 46}~(1), 493--517.

\bibitem[Lukacs(1970)]{LukEboo1970a}
{\sc Lukacs, Eugene} 1970 {\em Characteristic functions\/}, 2nd edn. London:
  Griffin.

\bibitem[Marchuk(1995)]{MarGboo1995a}
{\sc Marchuk, G.I.} 1995 {\em Adjoint Equations and Analysis of Complex
  Systems\/}. Springer.

\bibitem[Marconi {\em et~al.\/}(2008)Marconi, Puglisi, Rondoni \&
  Vulpiani]{MarUphr2008a}
{\sc Marconi, U. M.~B., Puglisi, A., Rondoni, L. \& Vulpiani, A.} 2008
  Fluctuation-dissipation: Response theory in statistical physics. {\em Physics
  Reports\/} {\bf 461}, 111--195.

\bibitem[Marston \& Conover(2008)]{MarJjas2008a}
{\sc Marston, J.~B. \& Conover, E.} 2008 Statistics of an unstable barotropic
  jet from a cumulant expansion. {\em Journal of the Atmospheric Sciences\/}
  {\bf 65}~(6), 1955--1966.

\bibitem[Ogura \& Phillips(1962)]{OguYjas1962a}
{\sc Ogura, Y. \& Phillips, N.~A.} 1962 Scale analysis of deep and shallow
  convection in the atmosphere. {\em J. Atmos. Sci.\/} {\bf 19}~(2), 173--179.

\bibitem[Pironneau(1974)]{PirOjfm1974a}
{\sc Pironneau, O.} 1974 On optimum design in fluid mechanics. {\em Journal of
  Fluid Mechanics\/} {\bf 64}, 97--110.

\bibitem[Reiterer {\em et~al.\/}(1998)Reiterer, Lainscsek, Sch{\"u}rrer,
  Letellier \& Maquet]{ReiPjpa1998a}
{\sc Reiterer, P., Lainscsek, C., Sch{\"u}rrer, F., Letellier, C. \& Maquet,
  J.} 1998 A nine-dimensional {L}orenz system to study high-dimensional chaos.
  {\em Journal of Physics A: Mathematical and General\/} {\bf 31}~(34), 7121.

\bibitem[Rothmayer \& Black(1993)]{RotAprs1993a}
{\sc Rothmayer, A.~P. \& Black, D.~W.} 1993 Ensembles of the {L}orenz
  attractor. {\em Proceedings: Mathematical and Physical Sciences\/} {\bf
  441}~(1912), 291--312.

\bibitem[Ruelle(2009)]{RueDnon2009a}
{\sc Ruelle, D.} 2009 A review of linear response theory for general
  differentiable dynamical systems. {\em Nonlinearity\/} {\bf 22}, 855--870.

\bibitem[Russo \& Luchini(2016)]{RusSjfm2016a}
{\sc Russo, S. \& Luchini, P.} 2016 The linear response of turbulent flow to a
  volume force: comparison between eddy-viscosity model and {DNS}. {\em Journal
  of Fluid Mechanics\/} {\bf 790}, 104--127.

\bibitem[Sewell(1987)]{SewMboo1987a}
{\sc Sewell, M.~J.} 1987 {\em Maximum and Minimum Principles\/}. Cambridge
  University Press.

\bibitem[Smale(1967)]{SmaSams1967a}
{\sc Smale, S.} 1967 Differentiable dynamical systems. {\em Bulletin of the
  American Mathematical Society\/} pp. 747--817.

\bibitem[Thuburn(2005)]{ThuJqms2005a}
{\sc Thuburn, J.} 2005 Climate sensitivities via a {F}okker-{P}lanck adjoint
  approach. {\em Quarterly Journal of the Royal Meteorological Society\/} {\bf
  131}~(605), 73--92.

\bibitem[Tobias {\em et~al.\/}(2011)Tobias, Dagon \& Marston]{TobSast2011a}
{\sc Tobias, S.~M., Dagon, K. \& Marston, J.~B.} 2011 Astrophysical fluid
  dynamics via direct statistical simulation. {\em The Astrophysical Journal\/}
  {\bf 727}~(2), 127.

\bibitem[Tobias \& Marston(2013)]{TobSprl2013a}
{\sc Tobias, S.~M. \& Marston, J.~B.} 2013 Direct statistical simulation of
  out-of-equilibrium jets. {\em Phys. Rev. Lett.\/} {\bf 110}, 104502.

\bibitem[Vishnampet {\em et~al.\/}(2015)Vishnampet, Bodony \&
  Freund]{VisRjcp2015a}
{\sc Vishnampet, R., Bodony, D.~J. \& Freund, J.~B.} 2015 A practical
  discrete-adjoint method for high-fidelity compressible turbulence
  simulations. {\em Journal of Computational Physics\/} {\bf 285}, 173--192.

\bibitem[Wang(2013)]{WanQjcp2013a}
{\sc Wang, Q.} 2013 Forward and adjoint sensitivity computation of chaotic
  dynamical systems. {\em Journal of Computational Physics\/} {\bf 235}, 1--13.

\bibitem[Wang {\em et~al.\/}(2014)Wang, Hu \& Blonigan]{WanQjcp2014a}
{\sc Wang, Q., Hu, R. \& Blonigan, P.} 2014 Least squares shadowing sensitivity
  analysis of chaotic limit cycle oscillations. {\em Journal of Computational
  Physics\/} {\bf 267}, 210--224.

\bibitem[Yorke \& Yorke(1979)]{YorJjsp1979a}
{\sc Yorke, James~A. \& Yorke, Ellen~D.} 1979 Metastable chaos: The transition
  to sustained chaotic behavior in the lorenz model. {\em Journal of
  Statistical Physics\/} {\bf 21}~(3), 263--277.

\end{thebibliography}
